\documentclass[a4paper,11pt]{article}
\pdfoutput=1 

\usepackage{jcappub}
\usepackage{amsmath}
\usepackage{amsfonts}
\usepackage{amssymb}
\usepackage{mathtools}
\usepackage{graphics}
\usepackage{tabularx}
\usepackage{adjustbox}
\usepackage{siunitx}
\usepackage{booktabs}
\usepackage{multirow}
\usepackage{citesort}
\usepackage{graphicx}
 \usepackage{url}
\usepackage{soul}
\usepackage{bm}
 \usepackage{xcolor}
\usepackage[utf8]{inputenc}
\usepackage[normalem]{ulem}
\usepackage{hyperref}

\newenvironment{system}%
{\left\lbrace\begin{array}{@{}l@{}}}%
{\end{array}\right.}

\title{Multi-messenger detection prospects of  gamma-ray burst afterglows with  optical jumps}

\author[a]{Ersilia Guarini,}
\author[a]{Irene Tamborra,}
\author[b]{Damien B{\'e}gu{\'e},}
\author[a]{Tetyana Pitik,}
\author[c]{and Jochen Greiner}

\affiliation[a]{Niels Bohr International Academy and DARK, Niels Bohr Institute, University of Copenhagen, Blegdamsvej 17, 2100, Copenhagen, Denmark}
\affiliation[b]{Department of Physics, Bar Ilan University, Ramat-Gan, 52900, Tel Aviv, Israel}
\affiliation[c]{Max-Planck-Institut f\"ur Extraterrestrische Physik, Giessenbachstra{\ss}e 1, 85748, Garching, Germany}

\emailAdd{ersilia.guarini@nbi.ku.dk}
\emailAdd{tamborra@nbi.ku.dk}
\emailAdd{begueda@biu.ac.il}
\emailAdd{tetyana.pitik@nbi.ku.dk}
\emailAdd{jcg@mpe.mpg.de}

\abstract{Some afterglow light curves of gamma-ray bursts (GRBs) exhibit very complex temporal and spectral features, such as a sudden  intensity jump about one hour after the prompt emission in the optical band. 
We assume that this feature is due to  the late collision of two relativistic shells and investigate the corresponding high-energy neutrino emission within a multi-messenger framework,  while contrasting  our findings with the ones from the classic afterglow model. For a constant density circumburst medium, the total number of emitted neutrinos can increase by about an order of magnitude when an optical jump occurs with respect to the self-similar afterglow scenario. By exploring the detection prospects with the IceCube Neutrino Observatory and future radio arrays such as IceCube-Gen2 radio, RNO-G and GRAND200k, as well as the POEMMA spacecraft, we conclude that the detection of neutrinos with IceCube-Gen2 radio could enable us to constrain the fraction of GRB afterglows with a jump as well as the properties of the circumburst medium.
We also investigate the neutrino signal expected for the afterglows of GRB 100621A and a  GRB 130427A-like burst  with an  optical jump. The detection of neutrinos from GRB afterglows  could be  crucial to explore  the yet-to-be unveiled mechanism powering the optical jumps.}

\begin{document}
\maketitle

\section{Introduction}
\label{sec:intro}
Gamma-ray bursts (GRBs) are among the brightest  and most poorly understood transients occurring in our Universe~\cite{Klebesadel:1973iq, Piran:2004ba, Kumar:2014upa}. There are two classes of GRBs; the short ones, lasting less than $2$~s, and the long ones~\cite{Kouveliotou:1993yx, Gehrels:2009qy}. The latter are the focus of this work. They are thought to be harbored within collapsing massive stars~\cite{Woosley:1993wj,MacFadyen:1998vz,Woosley:2006fn}. The isotropic equivalent energy release in gamma-rays spans  $10^{49}$--$10^{55}$~erg and it occurs within a few tens of seconds~\cite{Kumar:2014upa, Atteia:2017dcj}.
The observed spectrum is non-thermal, typically peaking in the $10$--$10^4$~keV energy band~\cite{1993ApJ...413..281B, Gruber:2014iza, vonKienlin:2020xvz}. 

The delayed emission following the prompt phase of GRBs---observed in the X-ray,
optical/infrared, radio and as of recently TeV bands~\cite{Meszaros+1997,MAGIC:2019irs,
Abdalla:2019dlr, HESS:2021dbz}---is the so-called afterglow. It is observed for several weeks after the trigger of the burst and,
in some cases, up to months or even years, making GRBs electromagnetically detectable across all wavebands. The afterglow emission results from  the interaction between the ejecta and the circumburst medium (CBM). The physical mechanism responsible for the multi-wavelength observation is broadly believed to be synchrotron radiation from the relativistic electrons accelerated at the external shock, developing when the  relativistic outflow expands in the CBM~\cite{Meszaros+1997, Waxman:1997if, Waxman:1997ga, Katz:1997nt}.

Observations in the X-ray and optical bands show a rich set of additional features, not described by the simplest afterglow model. 
At X-rays, data from the Gehrels Swift Observatory display a  rapid decline during
the first few hundred seconds~\cite{Barthelmy:2005bu, Lin:2017wpe,OBrien:2006dpq}, strong X-ray flaring during the first few thousand
seconds~\cite{2011MNRAS.410.1064M,2010MNRAS.406.2113C}, and a shallow decay up to
ten-thousand seconds. A canonical view of GRB afterglow is presented in e.g.
\cite{Nousek+2006, Zhang:2005fa}. In the optical band, the forward 
\cite[e.g.,][]{Melandri+2008} and reverse shocks \cite[e.g.,][]{SariPiran1999, 
Japelj+2014} dominate during the first thousand seconds, together with plateaus in the
majority of afterglows, and with X-ray flares, occasionally accompanyied by optical 
flares \cite{Kruehler+2009, 2012ApJ.758.27L}. At later times [i.e., at about
$7$--$10 (1+z)$~days, with $z$ being the redshift], the supernova signal 
emerges~\cite{Bloom:1999aa, refId0} \footnote{It is worth highlighting that we are only listing typical values for all the aforementioned
timescales.}. In this context, one of the biggest surprises was the observation of 
sudden rebrightenings in the afterglow light curve occurring at one to few hours after 
the prompt emission, primarily visible  in the optical band (hereafter called optical 
jump)~\cite{Rana+2009, 2010GCN.11270, Nardini+2011, Greiner:2013dma, Nardini:2013aea}. These optical jumps are very rare, as opposed to e.g. X-ray flares 
occurring in about 50\% of all GRB afterglows. The optical jump can be very large in 
amplitude ($>$1 mag) and is typically  brighter than the one observed in X-rays.
So far, about 10 out of 146 GRBs with well sampled optical light curves collected 
between February 1997 and November 2011 have displayed an optical 
jump~\cite{2013ApJ.774.13L}; for half of these, the brightness at the jump peak is 
comparable to the peak of the afterglow associated to the forward shock.

Several theoretical models attempt to explain such optical
jumps. For instance, they might be due to CBM inhomogeneities generated by 
anisotropic wind ejection of the GRB progenitor or interstellar 
turbulence~\cite{Lazzati:2002ri, Wang:2000fd}; however, numerical simulations of 
spherical explosions exhibit rather regular features and, in addition, density 
fluctuations of the CBM cannot give rise to significant time variability in the 
afterglow light curve~\cite{Nakar:2006gn,vanEerten:2008zb, vanEerten:2009mk}.
Alternatively, the late  variability of the afterglow light curve could be explained by
invoking a late energy injection in the first blast wave emitted by the central engine.
In this picture, the central engine undergoes intermittent late explosions, producing 
multiple shells of matter that  propagate and collide with the slower ones previously 
emitted, as proposed in Ref.~\cite{Vlasis:2011yp}. 
The origin of the late time activity of the central engine is 
unclear~\cite{Ioka:2004gy}. For example, it might be related to the disk fragmentation 
due to gravitational instabilities in the outer regions of the disk, with the resulting
fragments being accreted into the central compact object over different timescales, and
causing the observed time variability in the afterglow light curve~\cite{Perna:2005tv}.
Despite the uncertain origin of the central engine late time activity, this model 
predicts that the second blast wave emitted by the central engine injects new energy in
the initially ejected one, causing the observed rebrightening in the light 
curve~\cite{Kumar:1999gi, Granot:2003vj}.  Even though there is to date no 
smoking-gun signature favoring a specific mechanism to explain the 
appearance of optical jumps, the late collision of two relativistic 
shells~\cite{Vlasis:2011yp} is appealing in light of its ability to  successfully fit 
the light curves of some GRBs with optical jumps~\cite{Greiner:2013dma,   
Laskar:2017qrq}.

These peculiar features of the light curve of GRB afterglows raise questions on the possibly related neutrino emission. In fact, GRBs have been proposed as sources of ultra-high energy cosmic rays and high-energy neutrinos~\cite{Waxman:1997ti, Guetta:2003wi, Meszaros:2017fcs}. In the prompt phase, a copious amount of neutrinos could be produced by  photo-hadronic ($ p\gamma $)~\cite{Waxman:1997ti,Guetta:2003wi,Wang:2018xkp} or hadronic interactions ($pp$ or $pn$), the latter being more efficient in the innermost regions  where  the baryon density is large~\cite{Razzaque:2003uv,Murase:2013ffa,Metzger:2011xs,Heinze:2020zqb}. The neutrinos produced during the prompt GRB phase in the optically thin region have TeV--PeV energies, and their spectral distribution strongly depends on the emission mechanism~\cite{Waxman:1997ti,Meszaros:2015krr,Waxman:2015ues,Murase:2015ndr,Pitik:2021xhb,Toma:2010xw,Zhang:2010jt}. 

High energy neutrinos could also be  produced during the afterglow phase through $p \gamma$ interactions in the PeV--EeV energy range. Protons can be accelerated in the blastwave through Fermi acceleration~\cite{Waxman:1995vg,Vietri:1995hs} and interact with the synchrotron photons produced by accelerated electrons. Within the framework of the classic afterglow model, the neutrino emission from  GRB afterglows has been computed by considering the interaction of the GRB blastwave  with the external medium in two possible scenarios: the forward shock one, according to which particles are accelerated at the shock between the blastwave and the CBM~\cite{Waxman:1999ai, Dermer:2000yd, Li:2002dw, Razzaque:2013dsa} and the reverse shock model, that assumes acceleration of particles at the reverse shock  propagating back towards the ejecta~\cite{Murase:2007yt}. 

Since the neutrino production during the GRB afterglow phase strictly depends on the
photon distribution, an increase of the photon flux as observed for late time jumps in
the light curve should result in an increased neutrino flux, potentially detectable by
current and future high energy neutrinos facilities. In fact, optical photons are ideal targets for the production of PeV neutrinos.
The detection prospects with the IceCube Neutrino Observatory, which routinely observes
neutrinos with energies up to a few PeV~\cite{Ahlers:2018fkn, Ahlers_2015, 
IceCube:2020wum, IceCube:2020abv}, of GRB afterglows displaying an optical jump have 
not been investigated up to now. 
In addition, the possibly higher neutrino flux could be  detectable by  upcoming 
detectors, such as IceCube-Gen2 and its radio extension~\cite{IceCube-Gen2:2020qha}, 
the Radio Neutrino Observatory in Greenland (RNO-G)~\cite{RNO-G:2020rmc} and the full 
planned configuration of the Giant Radio Array for Neutrino Detection 
(GRAND200k)~\cite{GRAND:2018iaj}. The orbiting Probe of Extreme Multi-Messenger 
Astrophysics (POEMMA) spacecraft may also have promising perspectives for the detection
of neutrinos from GRB afterglows~\cite{Venters:2019xwi}.

If a jump is observed in the optical light curve of a GRB, what is its signature in neutrinos? Can we use neutrinos to learn more about this enigmatic feature of some GRBs? In this paper, we address these questions and explore the corresponding neutrino detection prospects. Our reference model is the late collision of two relativistic shells~\cite{Vlasis:2011yp,Greiner:2013dma,   Laskar:2017qrq}. Nevertheless, we stress that our goal is not to prove that the shell collision is the main  mechanism  explaining the GRBs light curves displaying jumps; rather, this scenario provides us with the framework within which we aim explore the associated neutrino signal. 

This paper is organized as follows. 
In Sec.~\ref{sec:collision}, we present the theoretical model for the late collision
of two relativistic shells that we consider to be the mechanism responsible for  the
sudden jump in the afterglow light curve. Section~\ref{sec:EM} focuses on the modeling
of the electromagnetic signal from GRB afterglows in the presence of optical jumps,
while Sec.~\ref{sec:p_nu} is centered on the proton distribution in the blastwave and
the resulting neutrino signals. Section~\ref{sec:neutrino_afterglow} presents our
findings on the neutrino and photon signals expected during the GRB afterglow phase,
in the absence as well as in the presence of optical jumps; while 
Sec.~\ref{sec:detection} investigates the neutrino detection prospects in the context
of quasi-diffuse and point source searches. In particular, we discuss the neutrino
detection prospects for the well studied GRB 100621A~\cite{Greiner:2013dma} and a burst with  model parameters
inspired by GRB 130427A~\cite{Panaitescu:2013pga, Perley:2013kwa, DePasquale:2016vau} having a hypothetical optical jump. Finally, our findings are
summarized in Sec.~\ref{sec:conclusions}. The analytical model on the  late collision
and merger of two relativistic shells is detailed in Appendix~\ref{sec:merger_model}, a
discussion on the degeneracies among the parameters of our model  is reported in
Appendix~\ref{A}, while  Appendix~\ref{sec:cooling} focuses on the cooling times of
protons and mesons of our GRB afterglow model.

\section{Modelling of the merger of two relativistic shells}
\label{sec:collision}

In this section, we outline the blastwave physics,  introducing the  scaling relations describing the temporal evolution of the radius and Lorentz factor of the blastwave. By relying on the late activity scenario for the central engine~\cite{Ioka:2004gy, Falcone:2005tp, Romano:2006rd, Zhang:2005fa}, our model on the late collision of two relativistic shells is then presented. 

\subsection{Physics of the blastwave}
\label{sec:Blastwave}

According to the standard picture, the relativistic GRB jet propagates with half
opening angle $\theta_j$ and Lorentz factor $\Gamma \gg 100$~\cite{Gehrels:2009qy}
in the  reference frame of the central engine. As long as $\Gamma^{-1} < \theta_j$,
the emitting region can approximately  be considered spherical. In order to investigate the
afterglow physics, it is useful to introduce the isotropic equivalent energy of the
blastwave, $\tilde{E}_{\rm{iso}}$~\footnote{We adopt three reference frames: the
blastwave comoving frame, the center of explosion (i.e.~the central compact object)
frame, and the observer frame (the Earth). Quantities in these frame are denoted as
$X^\prime$, $\tilde{X}$ and $X$, respectively. Energy, for example, transforms as
$\tilde{E} = (1+z) E =  \mathcal{D} E^\prime$. Here $z$ is the
redshift and $\mathcal{D} = [\Gamma (1- \beta \cos \theta)]^{-1}$ is the Doppler
factor, where $\beta = \sqrt{1-1/ \Gamma^2}$ and $\theta$ is the angle of propagation
of an element of the ejecta relative to the line of sight.}. We  denote with
$\tilde{E}_{k, \rm{iso}}$ the isotropic equivalent kinetic energy of the blastwave,
defined as $\tilde{E}_{k, \rm{iso}} = \tilde{E}_{\rm{iso}} - \tilde{E}_{\gamma,
\rm{iso}}$ and representing the energy content of the outflow after 
$\tilde{E}_{\gamma, \rm{iso}}$ has been released in $\gamma$-rays during the prompt
phase.

Two shocks develop at the interaction front between the burst and the CBM: a reverse shock, that propagates  towards the core of the jet, and a forward shock propagating in the CBM. After the reverse shock crosses the relativistic ejecta, the blastwave enters the so-called Blandford and McKee self-similar regime~\cite{Blandford:1976uq} (dubbed BM hereafter). In the following, we  focus on the BM phase, during which the emission is associated with the forward shock only. The particle density profile of the CBM is assumed to scale as a function of the distance from the central engine as $n \propto R^{-k}$.  In this work, we consider two  CBM scenarios:  a constant density profile resembling the one of the interstellar medium ($k=0$,  ISM) and a stellar  wind one ($k=2$, wind). 

We assume that the ejecta initially have isotropic kinetic energy
$\tilde{E}_{k, \rm{iso}}$ and Lorentz factor $\Gamma_0$.  
Two extreme scenarios for the hydrodynamical evolution of the blastwave can be
described analytically: fully adiabatic and fully
radiative~\cite{Blandford:1976uq,Sari:1997qe}. In the former case, the blastwave does 
not radiate a significant amount of energy while propagating. On the contrary, it 
quickly cools in the latter scenario, radiating all the internal energy released in the
shock while being decelerated by the CBM. Observational evidence 
suggests that GRB afterglow blastwaves are in the adiabatic regime rather than in the 
radiative one~\cite{zhang_2018}. Therefore, in this paper we focus on the adiabatic scenario.

Within the assumption of a thin shell (for which the reverse shock is mildly relativistic at most), if propagation occurs through a CBM with constant density $ n = n_0$, the blastwave starts to be decelerated at~\cite{Blandford:1976uq,zhang_2018}: 
\begin{equation}
T_{\rm{dec}, \rm{ISM}} = \biggl[ \frac{3 \tilde{E}_{k, \rm{iso}} (1+z)^3}{64 \pi n_0 m_p c^5 \Gamma_0^8} \biggr]^{1/3} \ ;
\label{eq:dec_ism}
\end{equation}
while if it occurs in a wind profile, $n = A R^{-2}$, the deceleration occurs at~\cite{Chevalier:1999mi}: 
\begin{equation}
T_{\rm{dec}, \rm{wind}}=\frac{\tilde{E}_{k, \rm{iso}}(1+z)}{16 \pi A m_p c^3 \Gamma_0^4}\ ,
\label{eq:dec_wind}
\end{equation}
where  $A = \dot{M}_w/(4 \pi v_w m_p) = 3.02 \times 10^{35} A_\star \; \rm{cm}^{-1}$, with $A_\star = {\dot{M}_{-5}}/{v_8}$  corresponding to the typical mass loss rate $\dot{M}_{-5} = \dot{M}/(10^{-5}M_{\odot})$~yr$^{-1}$ and  wind velocity $v_8 = v_w/(10^8 \; \rm{cm}~\rm{s}^{-1})$~\cite{Chevalier:1999jy, Razzaque:2013dsa} \footnote{Care should be taken when comparing our definition of the density profile for a wind CBM (which follows the convetion adopted in e.g.~Ref.~\cite{Kumar:2014upa}) with the one often adopted in the literature, i.e.~$\rho = A R^{-2}$, where $A = 5 \times 10^{-11} \; \rm{g} \; \rm{cm}^{-1} \; A_\star$ and $A_\star = {\dot{M}_{-5}}/{v_8}$. The difference between the two definitions  is the normalization in units of proton mass.}. Here $c= 3 \times 10^{10} \; \rm{cm} \; \rm{s}^{-1}$ is the speed of light and $m_p=0.938 \; {\rm{GeV}} \; c^{-2}$ is the proton mass. 

After the deceleration begins, the Lorentz factor of the shell decreases with time as~\cite{Blandford:1976uq, Sari:1997qe, Chevalier:1999mi, Ghisellini:2009rw}: 
\begin{eqnarray}
\Gamma_{\rm{ISM}} &=& \Gamma_0 \biggl(\frac{T_{\rm{dec, ISM}}}{4t}\biggr)^{3/8}  \label{eq:ad_i} \ ;\\
\Gamma_{\rm{wind}} &=& \Gamma_0 \biggl(\frac{T_{\rm{dec, wind}}}{4t}\biggr)^{1/4}  \label{eq:ad_w}\ ; 
\end{eqnarray}
The radius of the blastwave evolves with time according to~\cite{Razzaque:2013dsa}:
\begin{equation}
R= \frac{\zeta \Gamma^2 tc}{(1+z)}\ ,
\label{eq:radius_blastwave}
\end{equation}
where the correction factor $\zeta$ depends on the hydrodynamical evolution of the shock; we assume $\zeta = 8$~\cite{Sari:1997qe, Razzaque:2013dsa, Derishev:2021ivd}. 

In this work we are mainly interested in estimating the neutrino signal, whose accuracy is mainly dominated by other local uncertainties (e.g.~the proton acceleration efficiency and the fraction of the blastwave internal energy that goes into accelerated protons, that we  introduce in Sec.~\ref{sec:protons}). 
Hereafter, we adopt the uniform shell approximation of the BM solution, as presented in this section. This assumption suits our purposes, since the particle density of a BM shell quickly drops outside the region of width $\propto R/ \Gamma^2$ behind the forward shock and thus the corresponding neutrino emission is negligible.

\subsection{Merger of two relativistic shells}
\label{sec:Blastwave_merger}
The late merger of two relativistic shells has been investigated through  hydrodynamical simulations~\cite{Vlasis:2011yp} and applied  to fit the light curve of GRB 100621A~\cite{Greiner:2013dma}.  However, a simplified analytical modeling aiming to estimate the corresponding neutrino signal is presented in this paper for the first time. 
We assume that the first shell is launched by the central engine. At the onset of its deceleration, it is heated up, as its kinetic energy $\tilde{E}_{k, \rm{iso}}$ is converted in internal energy $\tilde{W}$. From now on, we refer to this shell as the ``slow shell.'' Its dynamics is described by the simplified BM solution in the uniform blastwave approximation introduced in Sec.~\ref{sec:Blastwave}, its Lorentz factor $\Gamma$ and radius $R$ evolve by following Eq.~\ref{eq:ad_i} and Eq.~\ref{eq:radius_blastwave}, respectively. 
\begin{figure}[]
\centering
\includegraphics[scale=0.23]{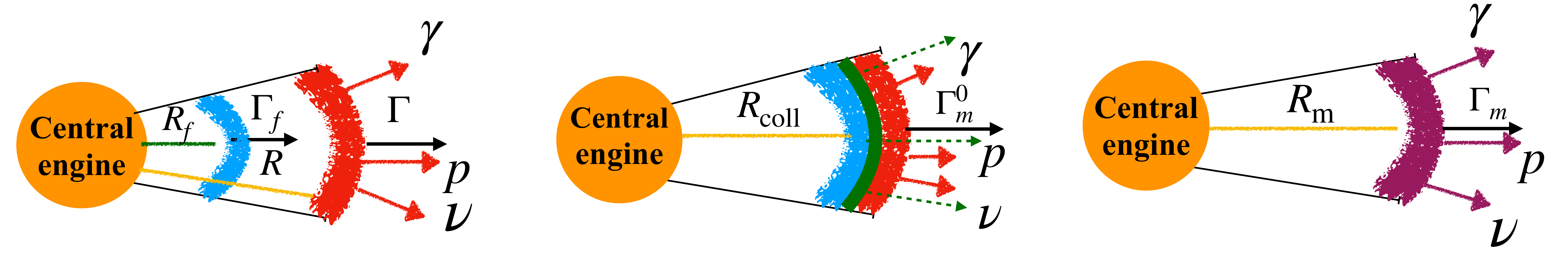}
\caption{Sketch of the collision and merger of two relativistic shells (not in scale). {\it Left panel:} The slow shell (marked in red) is launched by the central engine and decelerated by the interaction with the external medium. A shock develops at the contact surface, leading to the  classic afterglow emission. The fast shell (marked in blue) is launched by the central engine with a temporal delay $\Delta_T$ and propagates freely.  {\it Middle panel:} The fast shell reaches  the slow BM blastwave. Two shocks develop at the collision (marked in green);  the internal energy released in this process is emitted through radiation of secondary particles. {\it Right panel:}   The merged shell (plotted in purple) propagates through the external medium, emitting afterglow radiation.}
\label{fig:snapshot}
\end{figure}

Assuming that the central engine undergoes late activity, a second shell with energy $\tilde{E}_f$ is emitted with a time delay $\Delta_T$ with respect to the slow one, see the left panel of Fig. \ref{fig:snapshot}. We  refer to this second shell as the ``fast shell.'' This shell propagates in an almost empty environment since most of the matter has been  swept up by the slow shell~\cite{Vlasis:2011yp}. Thus, the fast shell moves with a constant Lorentz factor $\Gamma_f$,  eventually reaches the slow shell, and merges with it, as sketched in the middle and right panels of Fig. \ref{fig:snapshot}. Details on the analytical model describing the shell merger and the related conserved quantities are reported in Appendix~\ref{sec:merger_model}.

In order for the collision to happen at a given time $T_{\rm{coll}}$, the slow and  fast shells must be at the same position at $t=T_{\rm{coll}}$: $R(T_{\rm{coll}}) = R_f(T_{\rm{coll}})$. As extensively discussed in Appendix~\ref{A}, this condition gives rise to a degeneracy between $\Gamma_f$ and $\Delta_T$ (see also Appendix C of~\cite{Laskar:2017qrq} for a similar discussion). Indeed, a shell launched with a large delay and large speed could reach the slow shell at the same time of a slower shell launched with a smaller time delay. Understanding this degeneracy among the characteristic shell parameters is important, since $\Gamma_f$ directly affects the dynamics of the collision between the two shells. 

As the slow and fast shells collide, two shocks develop: a reverse shock, propagating
back towards the fast shell, and a forward shock, propagating through  the slow shell.
A detailed modeling of the collision between the fast and the slow shell is not necessary to estimate the production of neutrinos. Therefore, we assume that both the forward and reverse shocks created in
the  shell collision instantly cross the slow (forward shock) and the fast (reverse
shock) shell, which thus merge in a single shell at  $T_{\rm{coll}}$. In other words,
when the collision occurs,  a hot ``merged'' shell instantaneously forms as described in
Appendix~\ref{sec:merger_model}; see the right panel of Fig. \ref{fig:snapshot}.
Despite the simplifying assumption of  instantaneous merger between
the two shells,  our overarching goal of computing  the time-integrated neutrino
event rate is not affected since the neutrino emission during the merger interval is 
overall negligible, see discussion in Sec.~\ref{sec:neutrino_afterglow}.

In order to characterize the properties of the merged shell, we apply the
energy-momentum conservation equations, expanding on the model describing  the
collision of two relativistic shells for the internal shock scenario employed to
model  the prompt phase~\cite{Kobayashi:1997jk, Daigne:1998xc}. The main difference
with respect to the internal shock scenario~\cite{Kobayashi:1997jk, Daigne:1998xc} is that our slow
shell is hot and  is sweeping up material from the external medium. Thus, we need to
include the internal energy of the slow shell and the swept up mass in our calculation. As discussed in Appendix~\ref{sec:merger_model},
the following equations are obtained within the assumption of instantaneous merger.
Therefore,  we evaluate the quantities describing the slow and the
fast shells at time $t = T_{\rm{coll}}$. The initial Lorentz
factor of the merged shell is
\begin{equation}
\Gamma_m^0 \simeq \sqrt{\frac{m_f \Gamma_f + m_{\rm{eff}} \Gamma}{m_f/ \Gamma_f + m_{\rm{eff}}/ \Gamma}} \ ,
\label{eq:gamma_merg_init}
\end{equation}
where $m_f = {\tilde{E}_f}/{(\Gamma_f c^2)}$ is the mass of the fast shell and $m_{\rm{eff}}=m+ \hat{\gamma} W^\prime/ (c^2)$ is the effective mass of the slow shell. Here $\hat{\gamma} = 4/3$ is the adiabatic index in the relativistic limit (which holds since the slow shell is hot) and $m$ is the mass of the slow shell, i.e.~the sum between the initial mass of the ejecta $m_0=\tilde{E}_{\rm{iso}}/(\Gamma_0 c^2)$ and the swept up mass from the CBM up to the radius $R$, 
\begin{equation}
m = m_0+ 4 \pi \int_0^R dR^\prime R^{\prime 2} n(R^\prime) m_p \ .
\label{eq:mass_slow}
\end{equation}
Furthermore, at the collision, the internal energy $\tilde{W}^0_m$ is generated:
\begin{equation}
\tilde{W}^0_m \equiv \Gamma_m^0 W^{\prime 0}_m = \frac{1}{\hat{\gamma}} \left[(m_f \Gamma_f  + m \Gamma ) c^2 - (m + m_f) \Gamma_m^0 c^2 \right] + \Gamma W^\prime \ .
\label{eq:e_m_main}
\end{equation}

In the last stage of evolution, the merged shell moves in the CBM and interacts with it, giving rise to the standard afterglow radiation. Note that another degeneracy occurs. In fact,  the same value of $\Gamma_m^0$ can be obtained for different pairs of $(\tilde{E}_{k, \rm{iso}}, n_0)$ or $(\tilde{E}_{k, \rm{iso}}, A_\star)$. Thus, different initial conditions can lead to the same initial setup of the merged shell, nevertheless as discussed in Appendix~\ref{A} and in Sec.~\ref{sec:EM},  this  degeneracy  is not reflected in the observed photon flux. 

The dynamics of the  slow shell depends on the comoving dynamical time~\cite{Razzaque:2013dsa},
\begin{equation}
t^\prime_{\rm{dyn}} \simeq \frac{R}{8 \Gamma c} \ ,
\label{eq:dynamical_slow}
\end{equation}
and the related comoving width is~\cite{Blandford:1976uq}
\begin{equation}
l^\prime \simeq c t^\prime_{\rm{dyn}} = \frac{R}{8 \Gamma} \ ,
\label{eq:width_slow}
\end{equation}
where the radius $R$ is defined in Eq.~\ref{eq:radius_blastwave}.

The fast shell propagates with constant Lorentz factor $\Gamma_f \gg 1 $,  thus its radius evolves as~\cite{Kumar:2014upa}:
\begin{equation}
R_f = \frac{2 \Gamma_f^2 (t - \Delta_T) c}{(1+ z)} \ .
\label{eq:radius_fast}
\end{equation}
The comoving dynamical time of the fast shell is given by
\begin{equation}
t^\prime_{\rm{dyn}, f} \simeq \frac{R_f}{2 \Gamma_f c} \ ,
\label{eq:dynamical_fast}
\end{equation}
and its comoving width is
\begin{equation}
l^\prime_f \simeq c t^\prime_{\rm{dyn}, f} = \frac{R_f}{2 \Gamma_f} \ .
\label{eq:width_fast}
\end{equation}

The initial width of the  merged shell is approximated as
\begin{equation}
\l^{\prime 0}_m \simeq \Gamma_m^0 \biggl(\frac{l^\prime_f}{\Gamma_f} + \frac{l^\prime}{\Gamma} \biggr) \ ;
\label{eq:width_merger}
\end{equation}
while the dynamical time characterizing the merged shell at the collision is
\begin{equation}
t^{\prime 0}_{\rm{dyn, m}} \simeq \frac{l^{\prime 0}_m}{c} \ ,
\label{eq:dynamical_merg}
\end{equation}
where we have considered the Lorentz transformation for the length between the comoving and laboratory frames.

After a transient phase during which the merged shell relaxes, it is decelerated by the CBM and enters the BM regime. Since  we neglect the time needed by the merged shell to relax soon after the merger, a sharp jump results in the light curve; this treatment is not adequate for realistic fits of the electromagnetic signal, see Sec.~\ref{sec:photon_merger} for a discussion, but such task is beyond the scope of this paper. The semi-analytical treatment of the hydrodynamics of the collision, also taking into account the reverse shock crossing the fast shell was obtained in Ref.~\cite{Kumar:1999gi}; considering such a treatment  would not substantially affect the neutrino signal, since current and future neutrino telescopes may only be sensitive to the time-integrated spectral distribution in the most optimistic scenarios (see Sec.~\ref{sec:detection}). 

Once the merged shell enters the BM regime, its Lorentz factor $\Gamma_m$ evolves as described by  Eqs.~\ref{eq:ad_i}, by replacing $\Gamma_0 \rightarrow \Gamma_m^0$ and using the appropriate deceleration time. Indeed, even though the dynamics of the merged shell resembles the BM solution, there are some important and non trivial precautions to take into account for the definition of its deceleration radius and time , see Appendix \ref{sec:merger_model}. This is due to the fact that the merged shell is already hot and contains swept-up material. 
Once the deceleration time of the merged shell is properly defined, its radius $R_m$ follows Eq.~\ref{eq:radius_blastwave} by replacing $\Gamma \rightarrow \Gamma_m$. 
Finally, the width and dynamical time of the merged shell after its deceleration are given by Eqs.~\ref{eq:dynamical_slow} and \ref{eq:width_slow}, with  $\Gamma \rightarrow \Gamma_m$ and $R \rightarrow R_m$.


\section{Photon energy distribution and light curve} \label{sec:EM}
In this section, we introduce the main ingredients for the modeling of the emission
of  electromagnetic radiation during the classic afterglow and at the shell merger
which produces the optical jump. In the following, we consider a generic shell with
Lorentz factor  $\Gamma$ for the sake of simplicity, however our  treatment holds for
the afterglow generated both by the slow and the merged shell. The proper Lorentz
factor has to be taken into account for each case, i.e.~Eqs.~\ref{eq:ad_i} and
\ref{eq:ad_w} for the slow and the merged shell during the afterglow with the
appropriate initial Lorentz factor and deceleration time, as discussed in
Sec.~\ref{sec:Blastwave_merger}. As for the  collision, the
relevant Lorentz factor is given by Eq.~\ref{eq:gamma_merg_init}.

\subsection{Photon energy distribution during the afterglow} 
\label{sec:photons}
It is assumed that particles undergo Fermi acceleration~\cite{Waxman:1995vg,
Vietri:1995hs, Waxman:1998tn} at the forward shock. The synchrotron radiation
coming from shock accelerated electrons is broadly considered to be the origin
of the observed afterglow light curve~\cite{Sari:1997qe}. For the modeling of the
synchrotron photon spectrum, we follow Refs.~\cite{Sari:1997qe,Chevalier:1999jy, Panaitescu:2000bk}.
The internal energy density of the blastwave is given by the shock jump conditions 
(Eqs.~\ref{eq:gamma} and \ref{eq:en_density}). Therefore, the internal energy density
generated by the forward shock is~\cite{Blandford:1976uq}:
\begin{equation}
w^\prime = 4  m_p c^2 \Gamma (\Gamma -1) n \ ,
\label{eq:densityBlastwave}
\end{equation}
where $n = n_0$ and $n = A R^{-2}$ for the ISM and  wind scenarios, respectively.
A fraction $\epsilon_e$ of this energy  goes into accelerated electrons, a fraction $\epsilon_B$ into magnetic field, while protons receive the fraction $\epsilon_p \lesssim 1-\epsilon_e-\epsilon_B$. Thus, the magnetic field generated by the shock at the BM blastwave front is \begin{equation}
B^\prime = \sqrt{32 \pi m_p c^2 n \Gamma(\Gamma-1)\epsilon_B} \ .
\label{eq:magnetic_field}
\end{equation}

Electrons are expected to be accelerated to a power-law distribution $N_e(\gamma_e) \propto \gamma_e^{-k_e}$, where $k_e$ is the electron spectral index. The resulting electron distribution has three characteristic Lorentz factors:  minimum ($\gamma^\prime_{e, \rm{min}}$),  cooling ($\gamma^\prime_{e, \rm{cool}}$), and  maximum ($\gamma^\prime_{e, \rm{max}}$) ones. The minimum Lorentz factor corresponds to the minimum injection energy of electrons in the blastwave; the cooling Lorentz factor characterizes the energy of electrons that have time to radiate a substantial fraction of their energy in one dynamical time; the maximum Lorentz factor corresponds to the maximum energy that  electrons can achieve in the acceleration region~\cite{Sari:1997qe, zhang_2018}. These characteristic Lorentz factors are  given by~\cite{Sari:1997qe}:
\begin{eqnarray}
 \gamma^\prime_{e, \rm{min}} &=& \epsilon_e \frac{m_p}{m_e} \frac{(k_e - 2)}{(k_e - 1)} (\Gamma - 1)\ , \\ 
 \gamma^\prime_{e, \rm{cool}} &=& \frac{6 \pi m_e c}{\sigma_T B^{\prime 2}} \frac{(1+z)}{t \Gamma}\ , \\
 \gamma^\prime_{e, \rm{max}} &=& \biggl( \frac{6 \pi e}{\sigma_T B^\prime \xi} \biggr)^{1/2}\ ,\label{eq:gammaSat}
\end{eqnarray}
where $\sigma_T = 6.65 \times 10^{-25} \; \rm{cm}^{-2}$ is the Thompson cross section, $\xi$ represents the number of gyroradii needed for accelerating particles, $m_e = 5 \times 10^{-4} \; {\rm{GeV}} \; c^{-2}$ is the electron mass and $e = \sqrt{\alpha \hbar c}$ is the electron charge, where $\alpha \sim 1 / 137$ is the fine-structure constant and $\hbar \simeq 6.58 \times 10^{-25} \; \rm{GeV} \; \rm{s}$ is the reduced Planck constant. We take $\xi \equiv 10$~\cite{Gao:2012ay}.
The three characteristic Lorentz factors result into three observed characteristics break energies $E_{\gamma, \min}, E_{\gamma, \rm{cool}}$ and $E_{\gamma, \rm{max}}$, in the synchrotron photon spectrum at Earth:
\begin{equation}
E_{\gamma} \equiv h \nu_{\gamma} = \frac{3}{2} \frac{B^\prime}{B_Q} m_e c^2 \gamma^{\prime\ 2}_e \frac{\Gamma}{(1+z)}\ ,
\label{eq:synch_energies}
\end{equation}
where $B_Q = 4.41 \times 10^{13}$~G. The electrons are in the ``fast cooling regime'' when $\nu_{\gamma, \rm{min}} > \nu_{\gamma, \rm{cool}}$, while the ``slow cooling regime'' occurs when $\nu_{\gamma, \rm{min}} < \nu_{\gamma, \rm{cool}}$.

For the picture to be complete, the synchrotron self-absorption (SSA) frequency should  be considered as well. However, properly accounting for the SSA requires detailed information about the shell structure and the eventual thermal electron distribution~\cite{Warren:2018lyx}. Since this frequency is expected to be in the radio band~\cite{zhang_2018}, and since its inclusion does not change the results presented herein, we neglect SSA in the rest of this paper. 

We are interested in the comoving photon density in the blastwave  [in units of GeV$^{-1}$~cm$^{-3}$]. In the fast cooling regime, it is given by~\cite{Sari:1997qe, Thomas:2017dft}:
\begin{equation}
    n^\prime_\gamma(E^\prime_\gamma) = A^\prime_{\gamma}
    \begin{system}
    \left( E^\prime_\gamma/E^\prime_{\gamma, \rm{cool}} \right)^{-2/3} \; \; \;\;\; \; \; \; \; \; \;\;\; \; \; \; \; \; \; \; \; \; \; \; \; \; \; \;  \; \; \; \;  \; \; \; \; \; \; \; \; \; \; \; \; \; \; \; \; \; \; \; \;  \; \; \; \; \; E^\prime_\gamma< E^\prime_{\gamma, \rm{cool}}\\
\left( E^\prime_\gamma/ E^\prime_{\gamma, \rm{cool}} \right)^{-3/2} \; \; \; \; \; \;\;\; \; \;\;   \; \; \; \; \; \; \; \; \; \; \; \; \; \; \; \;\;\; \; \; \;  \; \; \;\;\; \; \; \; \; \; \; \; \; \; \; \; \; \ \; \;\; \; \;  \; \;  E^\prime_{\gamma, \rm{cool}} \leq E^\prime_\gamma \leq E^\prime_{\gamma, \rm{min}} \\
\left(E^\prime_{\gamma, \rm{min}} / E^\prime_{\gamma, \rm{cool}} \right)^{-3/2} \left(E^\prime_\gamma / E^\prime_{\gamma, \rm{min}}\right)^{-(k_e+2)/2} {\rm{e}}^{{-\frac{{E'_{\gamma}}}{E'_{\gamma, \rm{max}}}}}\; \; \; \; E^\prime_{\gamma, \rm{min}} < E^\prime_\gamma \leq E^\prime_{\gamma, \rm{max}}
    \end{system}
    \ ;
    \label{eq:lum_fast}
\end{equation}
while, in the slow cooling regime, it is
\begin{equation}
    n^\prime_\gamma(E^\prime_\gamma)  =  A^{\prime}_\gamma
    \begin{system}
\left(E^\prime_\gamma/E^\prime_{\gamma, \min} \right)^{-2/3} \; \;\; \; \; \; \; \; \; \; \;\;\; \; \; \; \; \; \; \; \; \; \; \; \;\; \;\; \; \; \; \; \; \;\; \; \; \;\; \; \; \; \; \; \;  \; \; \;  \; \; \;   \; \; \; \;  \; \; \; \; \;  \; \; \; E^\prime_\gamma < E^\prime_{\gamma, \min}\\
\left(E^\prime_\gamma/ E^\prime_{\gamma, \min}\right)^{-(k_e+1)/2} \; \; \; \; \; \; \; \; \;  \; \;\; \; \;\;\; \; \; \; \; \; \; \; \; \; \; \; \; \; \; \; \; \; \;  \; \; \; \; \; \; \; \; \; \; \; \;  \; \; \; \; \; \; \;  \; \; \; E^\prime_{\gamma, \min} \leq E^\prime_\gamma \leq E^\prime_{\gamma, \rm{cool}} \\
\left(E^\prime_{\gamma, \rm{cool}}/ E^\prime_{\gamma, \rm{min}} \right)^{-(k_e+1)/2} \left(E^\prime_\gamma /E^\prime_{\gamma, \rm{cool}} \right)^{-(k_e+2)/2} {\rm{e}}^{ {-\frac{{E^\prime_{\gamma}}}{E^\prime_{\gamma, \rm{max}}}}} \; \; \;  E^\prime_{\gamma, \rm{cool}} <  E^\prime_\gamma \leq E^\prime_{\gamma, \rm{max}}
\end{system}
\ .
\label{eq:lum_slow}
\end{equation}
Here $E^\prime_\gamma \equiv h \nu^\prime_\gamma$ is the comoving photon energy. The normalization constant is
\begin{equation}
A^{\prime}_\gamma= \frac{1}{2} \frac{L^{\prime}_{\gamma, \rm{max}}}{4 \pi R^2 c \; \min(E^{\prime}_{\gamma, \min}, E^{\prime}_{\gamma, \rm{cool}})} \ ,
\end{equation}
where $L^{\prime}_{\gamma, \rm{max}}= N_e P^\prime_{\max}(\gamma^\prime_{e, \rm{min}}) \phi_e/(E^\prime_{\gamma, \rm{min}})$ is the comoving specific luminosity [in units of s$^{-1}$], and $1/2$ is the geometrical correction coming from the assumption of isotropic synchrotron emission in the comoving frame~\cite{Dermer:2000yd}. 
The number of electrons in the blastwave is $N_e = {4}/{3} \pi n_0 R^3$ in the ISM scenario and $N_e = 4 \pi A R$ in the wind scenario, while $P^\prime_{\max}(\gamma^\prime_{e, \min})$ is the maximum synchrotron power emitted by electrons with Lorentz factor $\gamma^\prime_{e, \min}$ and  defined as
$P^\prime_{\max}(\gamma^\prime_{e, \min}) = c \sigma_T B^{\prime\ 2}\gamma^{\prime\ 2}_{e, \min} /{(6 \pi)}$. Finally, $\phi_e \simeq 0.6$ is a constant depending on the spectral index $k_e$~\cite{Wijers:1998st}; we adopt  $k_e=2.5$, as suggested from simulations of relativistic collisionsless shocks~\cite{Kirk:2000yh,Sironi:2013ri}. This value is also consistent with that obtained from the study of X-ray afterglows, see e.g.~\cite{2010ApJ...716L.135C}. Note that for the classic afterglow, we consider the transition from fast to slow cooling through the time evolution of the blastwave. Indeed, at late times the blastwave is in the slow cooling regime, in agreement with observations (see e.g.~\cite{2011A&A...526A..30G,2011MNRAS.412..561O}).

\subsection{Photon energy distribution during the shell merger}
\label{sec:photon_merger}
When the two shells collide, the internal energy $\tilde{W}^0_m$  is released, see Eq. \ref{eq:e_m_main}\footnote{For the sake of clarity, we denote the physical quantities characteristic of shell collision with the  apex ``$0$'', to distinguish them from the parameters describing the deceleration phase of the merged shell  (marked with the subscript ``$m$''). }.
Assuming that $\epsilon_{B, m}^{0}$ is the fraction of the internal comoving energy density released during the collision and going in magnetic energy density, the comoving magnetic field is 
\begin{equation}
B^{\prime 0}_m= \sqrt{8 \pi \epsilon_{B, m}^{0} w^{\prime 0}_m} \ ,
\label{eq:bfield_merg}
\end{equation}
where the comoving internal energy density is defined as
\begin{equation}
w^{\prime 0}_m = \frac{\tilde{W}^0_m}{\Gamma_m^0 V^{\prime}_m} = \frac{\tilde{W}^0_m}{\Gamma_m^0 4 \pi R(T_{\rm{coll}})^2 l^{\prime 0}_m}\ ,
\label{eq:en_den_merg}
\end{equation}
where $l^{\prime 0}_m$ is given by Eq.~\ref{eq:width_merger} and  $V^{\prime 0}_m = 4 \pi R(T_{\rm{coll}})^2 l^{\prime 0}_m $ is the volume of the merged shell right after its creation.

We assume that, at the collision, electrons are accelerated with the same index  as the one of the particles accelerated at the shock between the slow blastwave and the CBM ($k_e = 2.5$). The fraction $\epsilon_{e, m}^{0}$ of  internal energy density released at the collision goes into accelerated electrons, which  cool through  synchrotron radiation. The characteristic energies of the resulting photon spectrum are ${E}^{m, 0}_{\gamma, \min}$, $E^{ m, 0}_{\gamma, \rm{cool}}$ and $E^{m, 0}_{\gamma, \rm{max}}$ and are defined as in Eq.~\ref{eq:synch_energies} by replacing $\Gamma \rightarrow \Gamma_m^0$, and where the magnetic field is given by  Eq.~\ref{eq:bfield_merg}.

The shell collision and the afterglow are two distinct processes. The former involves a hot and a cold shell, the latter is related to the interaction between the slow, hot shell and the cold CBM. Therefore, the  microphysical parameters $\epsilon_{e, m}^{0}$ and $\epsilon_{B, m}^{0}$ do not need to be the same as $\epsilon_e$ and $\epsilon_B$. Moreover, while observations suggest a slow cooling regime for the classic afterglow at late times, electrons accelerated at the collision might be either in the fast or in the slow cooling regime, depending on the relevant parameters.

If for fixed initial conditions of the colliding shells and collision time the condition $\gamma^{\prime m,0}_{e, \rm{min}} > \gamma^{\prime m, 0}_{e, \rm{cool}}$ is verified, then the spectral energy distribution at the collision is
\begin{equation}
n_{\gamma}^{\prime m,0}(E_{\gamma}^{\prime}) =
A_{\gamma}^{\prime m,0} 
\begin{system}
\left( E^{\prime}_\gamma/E_{\gamma, \rm{cool}}^{\prime m, 0} \right)^{-2/3} \; \; \; \; \; \; \; \; \; \; \; \; \;\; \; \; \; \; \; \; \; \; \; \; \;\; \; \; \;  \; \; \;\; \; \; \; \;\; \; \; \;\; \; \; \;  \; \;  \; \; \;\; \;  \; \; \; \; \; E^{\prime}_\gamma< E_{\gamma, \rm{cool}}^{\prime m, 0}\\
\left( E_\gamma^{\prime}/ E_{\gamma, \rm{cool}}^{\prime m, 0} \right)^{- 3 /2} \; \; \; \; \;  \; \; \; \; \; \; \;\; \; \; \;  \; \;\;\; \;\; \;  \; \; \; \; \; \; \; \; \; \; \;  \; \; \; \; \; \;  \; \; \; \; \; \; \; \; \; \; \; \; \; \; \; \; \; E_{\gamma, \rm{cool}}^{\prime m, 0} \leq E_\gamma^{\prime} \leq E_{\gamma, \rm{min}}^{\prime m ,0} \\
\left( E_{\gamma, \rm{min}}^{\prime m, 0}/ E_{\gamma, \rm{cool}}^{\prime m, 0} \right)^{-3/2} \left( E_\gamma^{\prime} /E^{\prime m, 0}_{\gamma, \rm{min}} \right)^{- (k_e+2)/2} {\rm{e}}^{-\frac{E^\prime_\gamma}{E^{\prime m, 0}_{\gamma, \rm{max}}}} \; \; \; \; E^{\prime m, 0}_{\gamma, \rm{min}} < E^{\prime}_\gamma \leq E^{\prime m, 0}_{\gamma, \rm{max}}
\end{system}
\; ,
\label{eq:lum_fastCOLL}
\end{equation}
where 
\begin{equation}
A_{\gamma}^{\prime m, 0} = \frac{{\epsilon_{e, m}^{0} w^{\prime 0}_m }}{{\int_{\gamma^{\prime m, 0}_{\min}}^{\gamma^{\prime m, 0}_{\rm{sat}}} d E^{\prime}_\gamma n_{\gamma}^{\prime m, 0} (E^{\prime}_\gamma) E^{\prime}_\gamma }} \ .
\end{equation}
If instead $\gamma^{\prime m, 0}_{e, \rm{min}} < \gamma^{\prime m, 0}_{e, \rm{cool}}$, then the photon density is properly described by a slow cooling spectrum
\begin{equation}
    n^{\prime m, 0}_\gamma(E^\prime_\gamma)  =  A^{\prime m, 0}_\gamma
    \begin{system}
\left( E^{\prime m, 0}_\gamma/E^{\prime m, 0}_{\gamma, \min} \right)^{-2/3} \; \;\; \; \; \; \; \; \; \; \;\;\; \; \; \; \; \; \; \; \; \; \; \; \;\; \;\; \; \; \; \; \; \;\; \; \; \;\;  \; \; \;  \; \; \;  \; \; \;   \; \; \; \;  \; \; \; \; \;  \; \; E^\prime_\gamma < E^{\prime m, 0}_{\gamma, \min}\\
\left( E^{\prime m, 0}_\gamma/ E^{\prime m, 0}_{\gamma, \min} \right)^{-(k_e+1)/2} \; \; \; \; \; \; \; \; \;  \; \;\; \; \;\;\; \; \; \; \; \; \; \; \; \; \; \; \; \; \; \; \; \; \;  \; \;  \; \; \; \; \; \; \;  \; \; \; \; \; \; \;  \; \; E^{\prime m, 0}_{\gamma, \min} \leq E^{\prime}_\gamma \leq E^{\prime m, 0}_{\gamma, \rm{cool}} \\
\left( E^{\prime m, 0}_{\gamma, \rm{cool}}/ E^{\prime m, 0}_{\gamma, \rm{min}} \right)^{-(k_e+1)/2} \left(E^\prime_\gamma /E^{\prime m, 0}_{\gamma, \rm{cool}} \right)^{-(k_e+2)/2} {\rm{e}}^{ {-\frac{{E^\prime_{\gamma}}}{E^{\prime m, 0}_{\gamma, \rm{max}}}}} \; \; \;  E^{\prime m, 0}_{\gamma, \rm{cool}} <  E^\prime_\gamma \leq E^{\prime m, 0}_{\gamma, \rm{max}}
\end{system}
\ .
\label{eq:lum_slowCOLL}
\end{equation}
where  
\begin{equation}
A^{\prime m, 0}_{\gamma} = \frac{\epsilon_{e, m}^{ 0} w^{\prime 0}_m}{{\int_{\gamma^{\prime m, 0}_{\min}}^{\gamma^{\prime m, 0}_{\rm{sat}}} d E^{\prime}_\gamma n_{\gamma}^{\prime m, 0} (E^{\prime}_\gamma) E^{\prime}_\gamma }}\left(\frac{\gamma_{e, \rm{cool}}^{\prime m, 0}}{\gamma_{e, \rm{min}}^{\prime m, 0}} \right)^{(-k_e+2)}  \ .
\end{equation}
In the last expression we have taken into account the fact that only a fraction of electrons  radiates. 

\subsection{Light curve}
\label{sec:light_curves}
We now have  all the ingredients for investigating the expected light curve if the merger of two relativistic shells occurs.
We can distinguish between three time windows  in the photon light curve: an ``afterglow phase'' ($T_{\rm{dec}} \leq t < T_{\rm{coll}}$), the ``merging phase'' responsible for  the jump origin ($t = T_{\rm{coll}})$, and a  ``late afterglow phase'' ($t  > T_{{\rm{dec}}, m}$, with $T_{{\rm{dec}}, m}$ given by Eq.~\ref{eq:dec_merg}).

In our simplified model, the  photon lightcurve is a stepwise function obtained as  follows. For $T_{\rm{dec}} \leq t < T_{\rm{coll}}$, the  flux results from the interaction between the slow shell and the external medium. Therefore, it is described by the synchrotron spectrum introduced in Sec.~\ref{sec:photons}. At $t = T_{\rm{coll}}$, the flux undergoes a sharp increase: this is obtained as the sum between the afterglow radiation generated by the slow shell at  $t = T_{\rm{coll}}$ and the synchrotron radiation instantaneously emitted at the collision, see Sec.~\ref{sec:photon_merger} for its description. 
Finally, for $t >T_{{\rm{dec}}, m}$, the radiation comes from the deceleration of the merged shell. Thus, the light curve follows again the predicted broken power-law for the classic afterglow. The relations derived in Sec.~\ref{sec:photons} hold by applying the temporal evolution of the Lorentz factor and the radius of the merged shell as prescribed in Appendix~\ref{sec:merger_model}. 

Let $F^s_\gamma(E_\gamma)$ and $F^m_\gamma(E_\gamma)$ be the photon fluxes at Earth obtained from the photon distributions of the slow and merged shell, respectively, i.e.~Eqs.~\ref{eq:lum_fast}--\ref{eq:lum_slow}, taken with the proper Lorentz factor and radius; $F^{{m, 0}}_\gamma(E_\gamma)$ is instead the photon flux from electrons accelerated at the collision, corresponding to the photon distributions Eqs.~\ref{eq:lum_fastCOLL}--\ref{eq:lum_slowCOLL}. Therefore,  the resulting flux at Earth $F_{\gamma}(E_{\gamma})$ reads as
\begin{equation}
F_{\gamma}(E_{\gamma}) =
\begin{system}
F^s_\gamma(E_\gamma) \; \; \; \; \; \; \; \; \; \; \; \; \; \; \;  \; \; \; \; \;\; \; \; \; \; \; \; \; \; \; \; \; \; \; \; \; \; \;  T_{\rm{dec}} \leq t < T_{\rm{coll}} \\
F^s_\gamma(E_\gamma, t=T_{\rm{coll}}) + F^{m, 0}_\gamma(E_\gamma) \; \; \; \; \; \;  t = T_{\rm{coll}}\;. \\
F^m_\gamma(E_\gamma) \; \; \; \; \; \; \; \; \; \; \; \; \; \; \; \; \; \; \; \; \; \; \; \; \; \; \; \;\; \; \; \; \; \; \;\; \; \;  t \geq T_{\rm{dec, m}}
\end{system}
\end{equation}
This prescription does not aim to fit the afterglow light curves in the presence of a jump. Rather, it is a qualitative parametrization useful for contrasting the neutrino signal in the presence of a jump with the classic afterglow case.

We conclude by observing that we cannot model  the transition phase $T_{\rm{coll}} < t < T_{\rm{dec, m}}$ analytically. Indeed, we should take into account the time needed by the merged shell to relax before starting its deceleration; on the contrary, we are assuming an instantaneous merger. This approximation may lead to overestimate or underestimate the photon flux in the aforementioned time window. Even though this is may be problematic for the electromagnetic signal, it does not affect  the neutrino forecast substantially, as  discussed in Sec.~\ref{sec:neutrino_afterglow}.

\section{Energy distributions of protons and  neutrinos} \label{sec:p_nu}
In this section, the energy distribution of protons is introduced together with the most relevant cooling timescales. The steps followed to compute the neutrino flux expected at Earth are also outlined.

\subsection{Proton energy distribution}
\label{sec:protons}
We assume that protons are Fermi accelerated at the shock front, although  the process responsible for particle acceleration is still subject to debate, see e.g.~Refs.~\cite{Sironi:2013ri,Guo:2014via,Nalewajko:2015gma,Petropoulou:2018bvv,Kilian:2020yyw}.
Accelerated protons have a non-thermal power-law plus exponential cutoff distribution defined in the frame comoving with the blastwave as
\begin{equation}
n^\prime_p(E^\prime_p) = A^\prime_p E^{^\prime -k_p}_p \exp\biggl[-\biggl( \frac{E^\prime_p}{E^\prime_{p, \rm{max}}} \biggr)^{\alpha_p} \biggr] \Theta(E^\prime_p - E^\prime_{p, \rm{min}}) \ ,
\label{eq:proton_distribution}
\end{equation}
where $\Theta$ is the Heaviside function, $\alpha_p = 2$~\cite{Hummer:2010vx} and $k_p$ is the proton spectral index. The proton spectral index resulting from  non-relativistic shock diffusive acceleration theory is expected to be $k_p \simeq 2$~\cite{Matthews:2020lig}, while it is estimated to be  $k_p \simeq 2.2$ from Monte Carlo simulations of ultra-relativistic shocks~\cite{Sironi:2013ri}, assuming isotropic diffusion in the downstream. In this work, we assume $k_p = 2$. The normalization constant   is $A^\prime_p= \epsilon_p w^\prime [\int_{E^\prime_{p, \rm{min}}}^{E^\prime_{p, \rm{max}}} dE^\prime_p  E^\prime_p n^\prime_p(E^\prime_p)]^{-1}$, where $\epsilon_p + \epsilon_e + \epsilon_B \lesssim 1$ and $w^\prime$ is the comoving energy density of the  blastwave. For the slow and merged shells, $w^\prime$ is given by Eq.~\ref{eq:densityBlastwave}, by considering  the Lorentz factor and radius of the respective shell, while the energy density during the merger is given by Eq.~\ref{eq:en_den_merg}. The minimum energy of accelerated protons is $E^\prime_{p, \rm{min}} = \Gamma m_p c^2$~\cite{Dermer:2000yd,Murase:2007yt,Razzaque:2013dsa}. Finally, $E^\prime_{p, \rm{max}}$ is the maximum energy up to which protons can be accelerated in the blastwave and is obtained by the constraint of the Larmor radius  being smaller than the size of the acceleration region or imposing that the acceleration timescale,  
\begin{equation}
t^{^\prime -1}_{p, \rm{acc}} = \frac{c e B^\prime}{\xi E^\prime_p}\ ,
\label{eq:acceleration_time}
\end{equation}
is smaller than the total cooling timescale for protons.
Similarly to the electrons, we assume that $\xi =10$ for protons~\cite{Gao:2012ay}. 

The total cooling timescale for protons, at a fixed time of the evolution of the blastwave, is
\begin{equation}
t^{^\prime -1}_{p, \rm{cool}} = t^{^\prime -1}_{\rm{ad}} + t^{^\prime -1}_{p, \rm{sync}} +t^{^\prime -1}_{p \rm{\gamma}} + t^{^\prime -1}_{pp} + t^{^\prime -1}_{p, \rm{BH}} + t^{^\prime -1}_{p, \rm{IC}}\ ,
\label{eq:total_cooling}
\end{equation}
where $t^{^\prime -1}_{\rm{ad}}$, $t^{^\prime -1}_{p, \rm{sync}}$, $t^{^\prime -1}_{p \gamma}$, $t^{^\prime -1}_{{pp}}$, $t^{^\prime -1}_{p, \rm{BH}}$, $t^{^\prime -1}_{p, \rm{IC}}$ are the adiabatic, synchrotron, photo-hadronic ($p \gamma$), hadronic ($pp$), Bethe-Heitler (BH, $p \gamma \rightarrow p e^+ e^-$) and inverse Compton (IC) cooling timescales, respectively; these are defined as follows~\cite{dermer_book, Gao:2012ay, Razzaque:2005bh}: 
\begin{eqnarray}
 t^{\prime -1}_{\rm{ad}} &=& \frac{8 c \Gamma}{R}\ , \label{eq:adiabatic_time} \\
 t^{\prime -1}_{p, \rm{sync}} &=& \frac{4 \sigma_T m_e^2 E^\prime_p B^{\prime 2}}{3 m_p^4 c^3 8 \pi}\ , \\
 t^{\prime -1}_{p \gamma} &=& \frac{c}{2 \gamma^{\prime 2}_p} \int_{E_{\rm{th}}}^\infty dE^\prime_\gamma \frac{n^\prime_\gamma(E^\prime_\gamma)}{E^{\prime 2}_\gamma} \int_{E_{\rm{th}}}^{2 \gamma^\prime_p E^\prime_\gamma} dE_r E_r \sigma_{p \gamma}(E_r) K_{p \gamma}(E_r)\ ,  \\
 t^{\prime -1}_{{pp}} &=& c n^\prime_p \sigma_{pp} K_{pp}\ ,  \\
 t^{\prime -1}_{p, \rm{BH}} &=& \frac{7 m_e \alpha \sigma_T c}{9 \sqrt{2} \pi m_p \gamma^{\prime 2}_p} \int_{\gamma_p^{\prime -1}}^{\frac{E^\prime_{\gamma, \rm{max}}}{m_e c^2}} d\epsilon^\prime \frac{n^\prime_\gamma(\epsilon^\prime)}{\epsilon^{^\prime 2}} \biggl\{ (2 \gamma^\prime_p \epsilon^\prime)^{3/2} \biggl[\ln(\gamma^\prime_p \epsilon^\prime) -\frac{2}{3} \biggr]+ \frac{2^{5/2}}{3} \biggr\}\ ,  \\
 t^{\prime -1}_{p, \rm{IC}} &=& \frac{3 (m_e c^2)^2 \sigma_T c}{16 \gamma_p^{\prime 2}( \gamma^\prime_p-1) \beta^\prime_p} \int_{E^\prime_{\gamma, \rm{min}}}^{E^\prime_{\gamma, \rm{max}}} \frac{dE^\prime_\gamma}{E_\gamma^{^\prime 2}}F(E^\prime_\gamma, \gamma^\prime_p) n^\prime_\gamma(E^\prime_\gamma)\ , 
\end{eqnarray}
where $\gamma_p = E^\prime_p/m_p c^2$, $\epsilon^\prime = E^\prime_\gamma /m_e c^2$, $E_{\rm{th}}=0.150$~GeV is the threshold for photo-pion production, and   $\beta^\prime_p \approx 1$ for  relativistic particles. The function $F(E^\prime_\gamma, \gamma^\prime_p)$ is given in Ref.~\cite{PhysRev.137.B1306}, with the replacement $m_e \rightarrow m_p$. The cross sections for $p \gamma$ and $pp$ interactions, $\sigma_{p \gamma}$ and $\sigma_{pp}$, are defined following Ref.~\cite{ParticleDataGroup:2020ssz}. The function $K_{p\gamma}(E_r)$ is the $p\gamma$ inelasticity, given by Eq.~9.9 in~\cite{dermer_book}:
\begin{equation}
K_{p\gamma}(E_r) = 
\begin{system}
0.2 \; \; \; \; \; \;  \; \; \;  E_{\rm{th}} < E_r < 1~\rm{GeV}\\
0.6 \; \; \; \; \; \;  \; \; \;  E_r > 1~\rm{GeV}
\end{system} \
\end{equation}
where $E_r = \gamma^\prime_p E^\prime_\gamma (1 - \beta^\prime_p \cos\theta^\prime)$ is the relative energy between a proton with Lorentz factor $\gamma^\prime_p$ and a photon with energy $E^\prime_\gamma$, moving such that they form an angle $\theta^\prime$ in the comoving frame of the blastwave. The comoving proton density in the blastwave, $n^\prime_p$, is obtained from the jump conditions (see Appendix~\ref{sec:merger_model}) and is such that $n^\prime_p = 4 n \Gamma$.  The inelasticity of  $pp$ interactions is $K_{pp} \simeq 0.8$ \cite{Pitik:2021xhb} and $n^\prime_\gamma(E^\prime_\gamma)$ is the photon target for  accelerated protons. 

\subsection{Neutrino energy distribution and flux expected at Earth}
\label{sec:neutrinos}
The blastwave  is rich of photons radiated by shock accelerated electrons, which are  ideal targets for protons co-accelerated at the shock. This results in efficient neutrino production through $p \gamma$ interactions, mostly dominated by the $\Delta^+$ resonance: 
\begin{equation}
   p+ \gamma \longrightarrow \Delta^+ \longrightarrow
   \begin{system}
   n + \pi^+ \; \; \; \; \; \; \; 1/3 \; \rm{\; of\; all\; cases} \\
   p+\pi^0  \; \; \; \; \; \; \; \;  2/3 \; \rm{\; of \; all\; cases }  \ .
   \end{system}
    \label{reaction_channel}
\end{equation} 
Neutral pions decay in two photons: $\pi^0 \longrightarrow  2 \gamma$; while charged pions can produce neutrinos through the decay chain $\pi^+ \longrightarrow \mu^ + + \nu_\mu$, followed by the muon decay $\mu^+ \longrightarrow \bar{\nu}_\mu + \nu_e + e^+$.
Note that, since the number of photons in the blastwave is much larger than the number of protons swept up from the CBM by the blastwave,  we can safely neglect the contribution to the neutrino emission due to $pp$ interactions. Indeed, the cooling timescales satisfy $t^{-1}_{pp} \ll t^{-1}_{p \gamma}$ for typical GRB afterglow parameters, as shown in Appendix~\ref{sec:cooling}. 

In order to compute the neutrino spectral energy distribution resulting from  $p \gamma$ interactions, we rely on the semi-analytic photo-hadronic model described in Ref.~\cite{Hummer:2010vx}. This model is based on SOPHIA~\cite{Mucke:1999yb}, which takes into account the $\Delta^+$ channel in Eq.~\ref{reaction_channel}, as well as the $N$ resonances, the multi-pion and direct-pion production channels. 

The procedure adopted to compute the neutrino energy distribution is the same for all three time windows of our GRB afterglow model, after taking into account the corresponding distributions of  photons and protons.
Given the comoving photon energy distribution, $n^\prime_\gamma(E^\prime_\gamma)$, and the comoving proton energy distribution $n^\prime_p(E^\prime_p)$ [both in units of GeV$^{-1}$~cm$^{-3}$],  the rate of production of secondary particles $l= \pi^{\pm}, \pi^0, K^+$ in the comoving frame [in units of GeV$^{-1}$~cm$^{-3}$~s$^{-1}$] is given by~\cite{Hummer:2010vx}:
\begin{equation}
Q^\prime_l(E^\prime_l) = c \int_{E^\prime_l}^\infty \frac{d E^\prime_p}{E^\prime_p} n^\prime(E^\prime_p) \int_{E_{\rm th}/2 \gamma^\prime_p}^\infty dE^\prime_\gamma n^\prime_\gamma(E^\prime_\gamma) R(x, y)\ ,
\label{eq:rate_secondaries}
\end{equation}
where $x = E^\prime_l/E^\prime_p$ is the fraction of  proton energy that goes into the secondary particles, $y = \gamma^\prime_p E^\prime_l$ and $R(x, y)$ is the response function, which contains  information on the interaction, i.e.~cross section and multiplicity. 

Before decaying, charged mesons undergo energy losses. Their  energy distribution at  decay is approximated by:
\begin{equation}
Q^{^\prime \rm{dec}}_l(E^\prime_l) = Q^\prime_{l}(E^\prime_{l}) \biggr[1 - \exp\biggl(- \frac{t^\prime_{l, \rm{cool}} m_l}{E^\prime_l \tau^\prime_l}\biggr)\biggr] \ ,
\label{eq:decayed_spectrum}
\end{equation}
where  $t^\prime_{l, \rm{cool}}$ is the cooling time scale of the $l$ meson, $m_l$ its mass and $\tau^\prime_l$ its lifetime. Finally, mesons decay and the resulting neutrino comoving spectrum [in units of GeV~cm$^{-3}$~s$^{-1}$] is
\begin{equation}
Q^\prime_{\nu_\alpha}(E^\prime_{\nu}) = \int_{E^\prime_{\nu}}^{\infty} \frac{dE^\prime_l}{E^\prime_l} Q^{^\prime \rm{dec}}_{l}(E^\prime_l) F_{l \rightarrow \nu_\alpha} \biggl(\frac{E^\prime_{\nu}}{E^\prime_l} \biggr)\ ,
\label{eq:rate_neutrini}
\end{equation}
where $\alpha = e, \mu$ is the neutrino flavor at production and $F_{l \rightarrow \nu_\alpha}$ is a function defined as in Ref.~\cite{Lipari:2007su}.
Kaons suffer less from radiative cooling compared to charged pions, due to their larger mass and shorter lifetime. Thus, their contribution to the resulting neutrino spectrum is always sub-leading at lower energies, but may become  dominant at higher energies~\cite{He:2012tq,Asano:2006zzb,Petropoulou:2014lja,Tamborra:2015qza}.

For a  source at redshift $z$, the flux of neutrinos of flavor $\alpha$ expected at Earth [in units of GeV$^{-1}$~cm$^{-2}$~s$^{-1}$] is:
\begin{equation}
\Phi_{\nu_\alpha}(E_{\nu}, z) = \frac{(1+z)^2}{4 \pi d_L^2(z)}  V^\prime_{\rm{shell}} \sum_\beta P_{\nu_\beta \rightarrow \nu_\alpha} (E_{\nu}) Q^\prime_{\nu_\beta}\biggl[\frac{E_{\nu}(1+z)}{\Gamma}\biggr]\ ,
\label{eq:neutrino_flux}
\end{equation}
where $V^\prime_{\rm{shell}}= 4 \pi R^2 l^\prime$ is the volume of the emitting shell~\cite{Baerwald:2011ee} and $l^\prime$ its width. The neutrino oscillation probability $ P_{\nu_\beta \rightarrow \nu_\alpha}(E_{\nu_\alpha})$ is  such that $P_{\nu_\beta \rightarrow \nu_\alpha}= P_{\bar{\nu}_\beta \rightarrow \bar{\nu}_\alpha}$ and is given by~\cite{Anchordoqui:2013dnh}:
\begin{eqnarray}
P_{\nu_e \rightarrow \nu_\mu} &=& P_{\nu_\mu \rightarrow \nu_e} = P_{\nu_e \rightarrow \nu_\tau} = \frac{1}{4} \sin^2 2\theta_{12}\ , \\
P_{\nu_\mu \rightarrow \nu_\mu} &=& P_{\nu_\mu \rightarrow \nu_\tau  }= \frac{1}{8}(4-\sin^2 \theta_{12})\ ,\\
P_{\nu_e \rightarrow \nu_e} &=& 1- \frac{1}{2} \sin^2 2\theta_{12}\ ,
\end{eqnarray}
with $\theta_{12} \simeq 33.5^\circ$~\cite{Esteban:2020cvm}. The luminosity distance  in a standard flat $\Lambda \rm{CDM}$ cosmology is 
\begin{equation}
d_L(z) = (1+z) \frac{c}{H_0} \int_0^z \frac{dz^\prime}{\sqrt{\Omega_\Lambda + \Omega_M(1+z^\prime)^3}}\ ,
\end{equation}
where we adopt $H_0 =  67. 4$~km~s$^{-1}$~Mpc$^{-1}$, $\Omega_M = 0.315$, and $\Omega_\Lambda = 0.685$~\cite{Cahill:2020nhu}.

\section{Afterglow  signals}
\label{sec:neutrino_afterglow}
In this section, we present our findings on the particle distributions  expected at Earth from the GRB afterglow. 
We explore the photon light curve as well as the temporal evolution of the neutrino spectral energy distribution in  three  time windows: the afterglow generated by the first shell launched by the central engine, the time at which the fast shell collides and merges with the slow one, and  the afterglow generated by the merged shell.

\subsection{Particle emission in the absence of a late shell collision} \label{sec:afterglow_nobump}
\label{sec:classic_afterglow}
\begin{table}[b]
\caption{Characteristic parameters assumed for our benchmark GRB afterglow in the ISM  and wind CBM  scenarios.}
\begin{tabular}{cccccccccccc}
\\
\toprule \toprule
& $\tilde{E}_{k, \rm{iso}} \; (\rm{erg})$ & $\Gamma_0$ & $n_0 \; (\rm{cm}^{-3})$ or $A_{\star}$ & $\epsilon_e$ & $\epsilon_B$ & $\epsilon_{e, m}^{ 0}$ & $\epsilon_{B, m}^{0}$ & $T_{\rm{coll}} \; (\rm{s})$ & $\tilde{E}_f$ (erg)  & $k_e$ & $k_p$ \\ \midrule
ISM & $10^{53}$                               & $300$      & $1.0$                               & $0.1$        & $0.1$        & $0.1$          & $0.1$          & $5 \times 10^3 $            & $2 \times 10^{53}$    & $2.5$ & 2   \\  
Wind  & $10^{53}$                               & $100$      & $0.1$                               & $0.1$        & $0.1$        & $0.1$          & $0.1$          & $5 \times 10^3$             & $ 2 \times 10^{53}$  & $2.5$  & 2   \\ \bottomrule \bottomrule
\end{tabular}
\label{tab:benchmark_grb}
\end{table}
We consider a benchmark  GRB with characteristic parameters as in Table~\ref{tab:benchmark_grb} and located at  $z = 1$. 
The chosen value for the isotropic kinetic energy is motivated by  post-Swift observations reporting an average isotropic energy emitted in photons $\tilde{E}_{\gamma, \rm{iso}} =  \mathcal{O}(10^{52})$~erg~\cite{Cenko+2010}
and assuming a conversion efficiency of $\sim 10\% \tilde{E}_{\rm{iso}}$ into gamma-rays, therefore leading to the isotropic kinetic energy $\tilde{E}_{k, \rm{iso}} \sim 10^{53}$~erg.
Moreover, we rely on the standard microphysical parameters reported in Ref.~\cite{Sari:1997qe}.
Since there is no evidence for the values of typical microphysical parameters characteristic of the collision, we fix $\epsilon_{e, m}^{0} = \epsilon_e$ and $\epsilon_{B, m}^{0} = \epsilon_B$.
Finally, as for the CBM densities, we follow Refs.~\cite{Sari:1997qe, Chevalier:1999jy}.
  
Concerning the fast shell, we fix $\Gamma_f$ by taking $\Delta_T \ll T_{\rm{coll}}$, so that $\Gamma_f \simeq 2 \Gamma(T_{\rm{coll}})$ (see Appendix~\ref{A}). Since there are no theoretical constraints on the energy $\tilde{E}_f$, we fix the latter by following Ref.~\cite{Vlasis:2011yp}. We choose $\tilde{E}_f = 2 \tilde{E}_{\rm{iso}}$ relying on the results of ``case 4'' of Ref.~\cite{Vlasis:2011yp}, for which  the strongest rebrightening is obtained. Moreover, we fix $T_{\rm{coll}} = 5 \times 10^3 \; \rm{s}$ both for the ISM and the wind scenarios.
At this time the light curve is decreasing in both scenarios, and it has been chosen consistently with the observation of jumps between a few hundred seconds and $\sim 1$~day after the onset of the burst~\cite{Nardini:2013aea, Greiner:2013dma, 2012ApJ.758.27L, 2013ApJ.774.13L}. 

\begin{figure}[]
\includegraphics[scale=0.38]{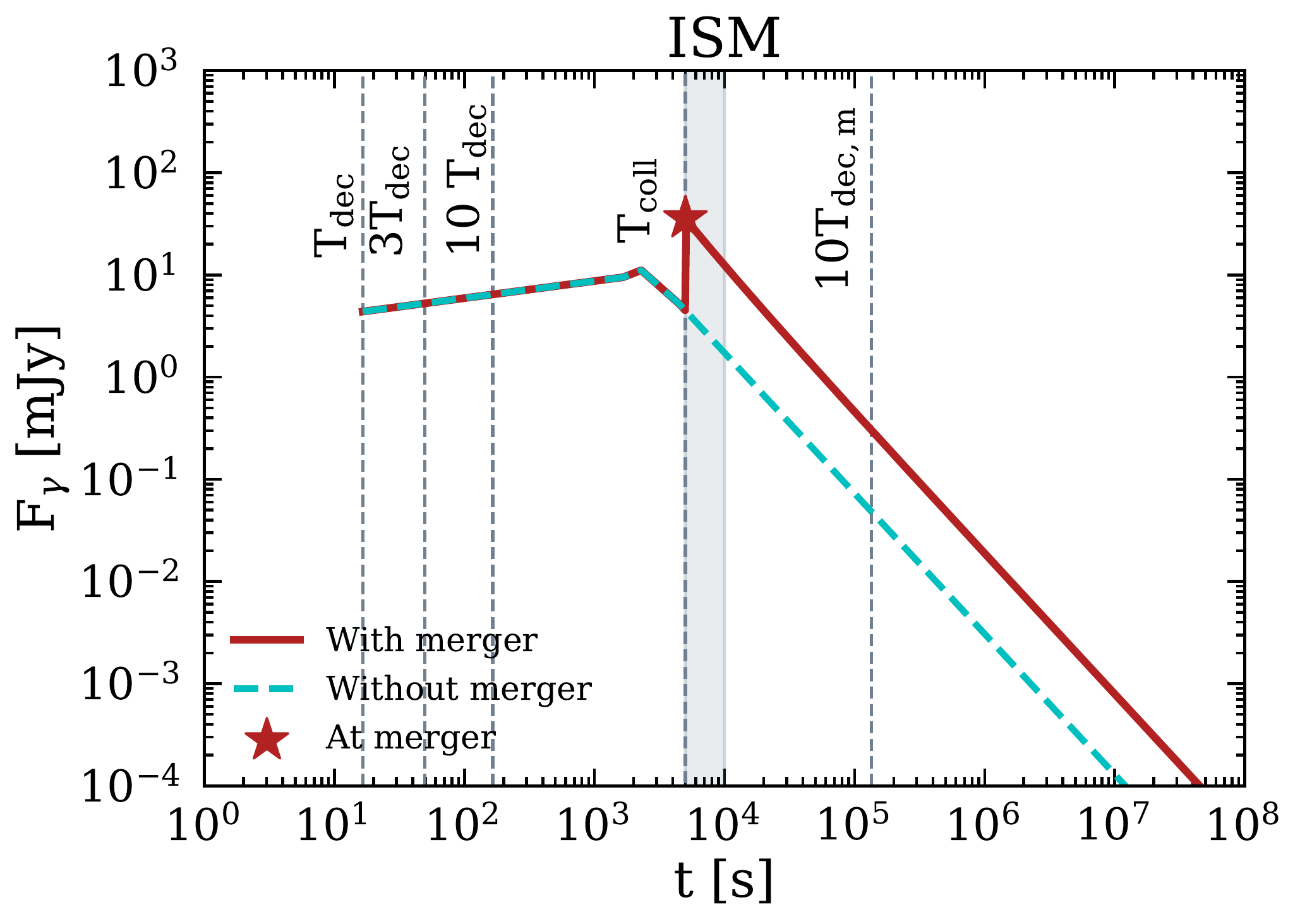}
\includegraphics[scale=0.38]{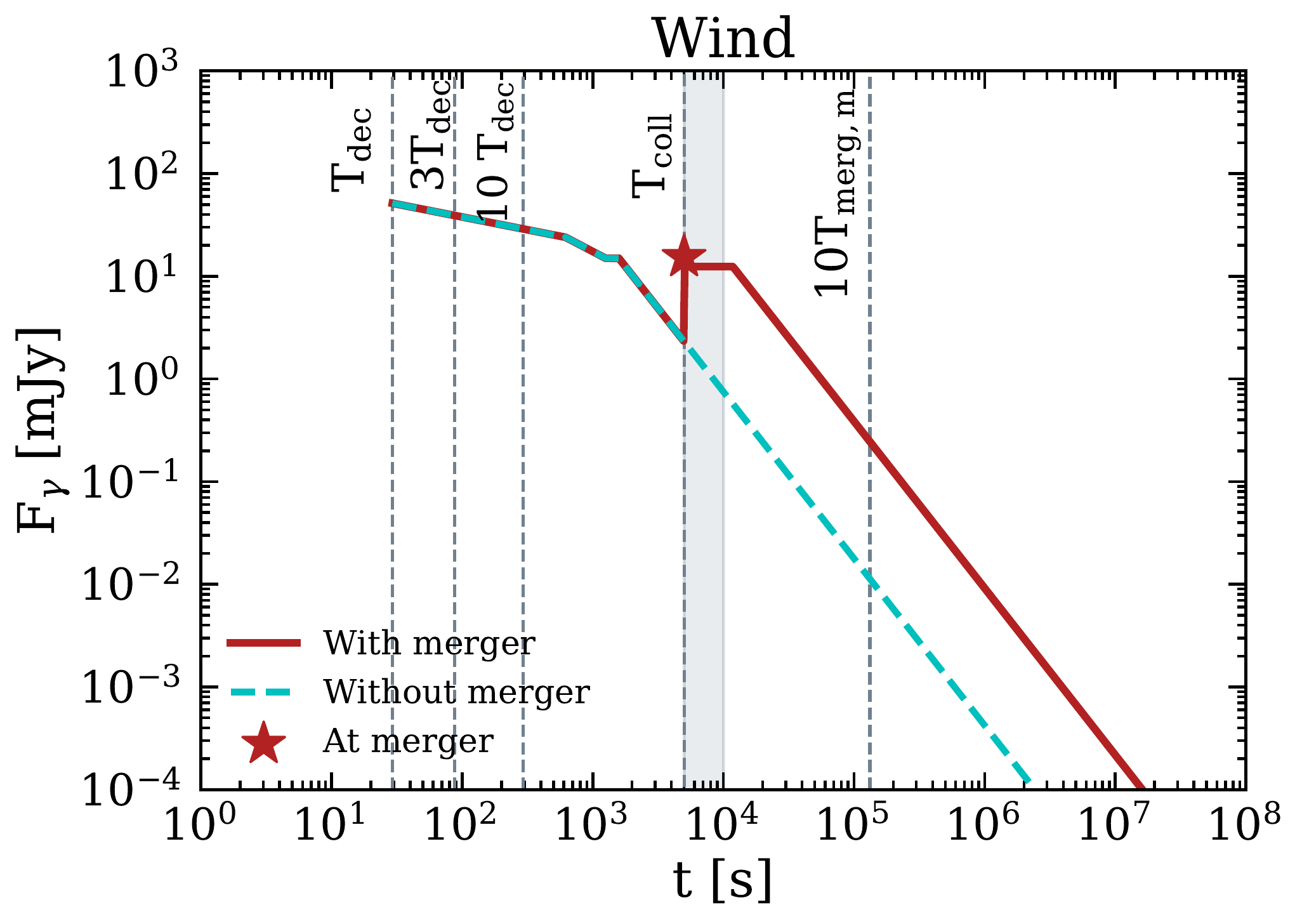}
\caption{Light curves expected at Earth for our benchmark GRB at $z = 1$ for the classic afterglow scenario (cyan dashed line) and in the presence of an optical  jump (brown solid line)  for an observed photon frequency $\nu_{\gamma}= 6 \times 10^{14} \; \rm{Hz}$. The brown star marks the flux generated at $T_{\rm{coll}}$. At the merger and after it, the observed flux is larger than the one expected from the classic afterglow. The gray shadowed region ($T_{\rm{coll}} < t < T_{\rm{dec, m}}$) is  excluded from the computation of the neutrino signal  since we cannot treat  this transition phase analytically (see the main text for details). We assume a photon spectral index $k_e = 2.5$. In order to guide the eye, the vertical grey dashed lines mark the times at which we show snapshots of the spectral energy distribution of photons and neutrinos (see Figs.~\ref{fig:photonISM} and \ref{fig:merger_snapshots}).  These light curves should be considered for illustrative purposes only, since we assume the instantaneous shell collision for simplicity.} \label{fig:light_curve} 
\end{figure}

In the classic afterglow scenario, the time evolution of the photon light curve at Earth, computed as described in Sec.~\ref{sec:photons}, for our benchmark GRB  is shown in Fig.~\ref{fig:light_curve} (cyan dashed line). The light curve is computed for an observed photon frequency $\nu_{\gamma}= 6 \times 10^{14} \; \rm{Hz}$, i.e.~in the optical band. 
For both the ISM and wind scenarios, the breaks in the light curve are determined by the times at which the break frequencies $\nu_{\gamma, \min}$ and $\nu_{\gamma, \rm{cool}}$ cross the observed one $\nu_{\gamma}$, and  $\nu_{\gamma, \min} = \nu_{\gamma, \rm{cool}}$. 

\begin{figure}[]
\includegraphics[scale=0.36]{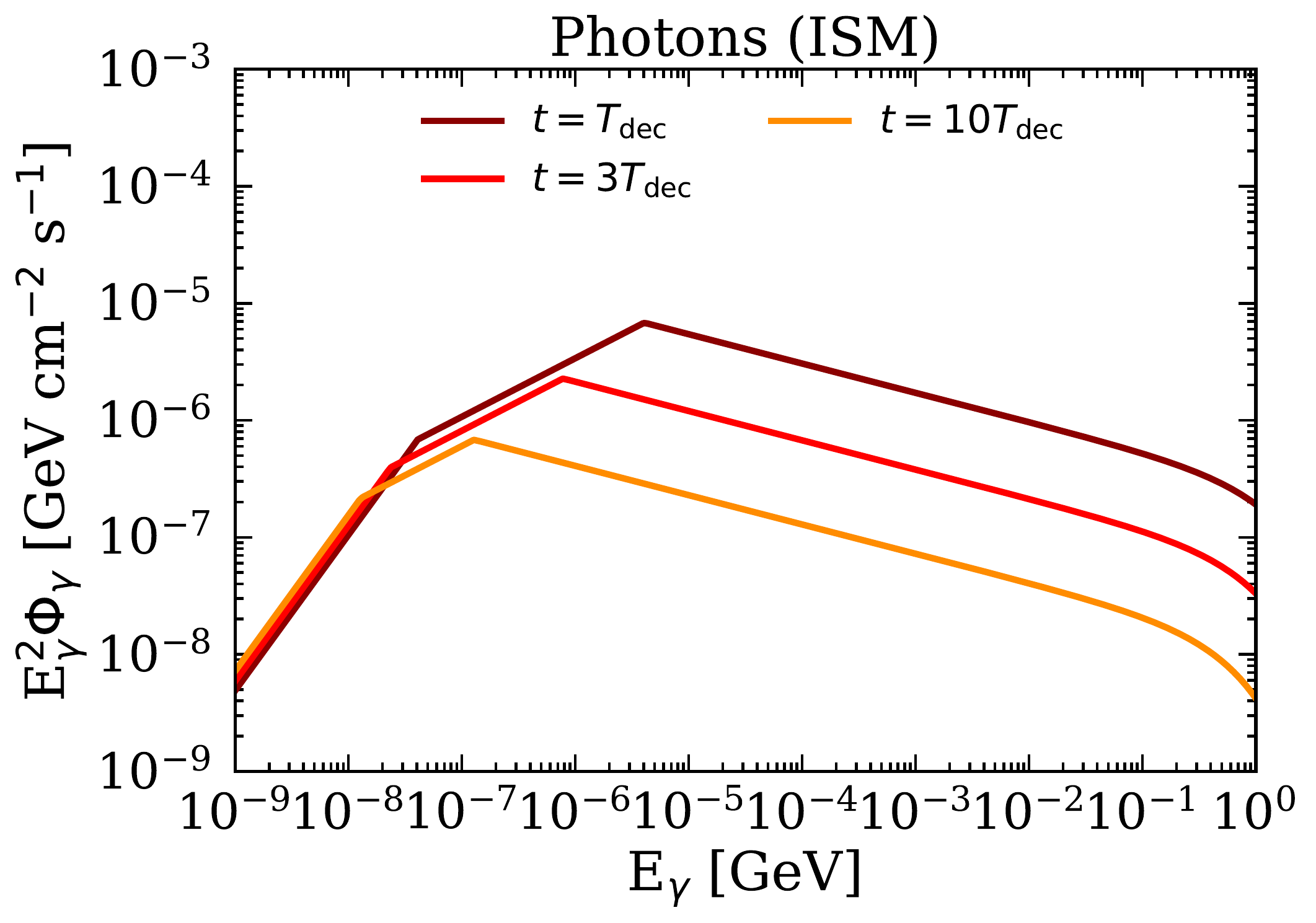}
 \includegraphics[scale=0.36]{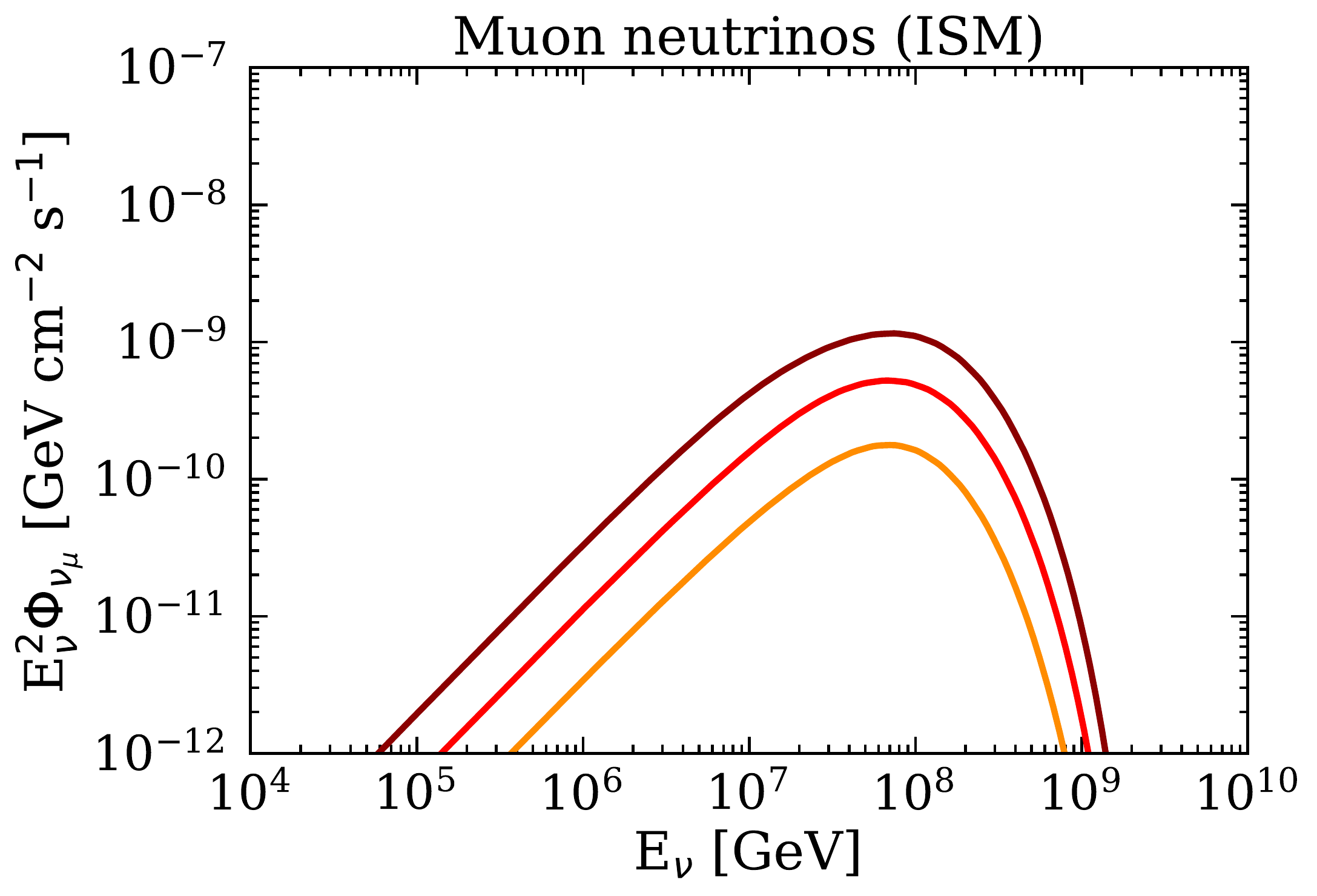}
 \includegraphics[scale=0.36]{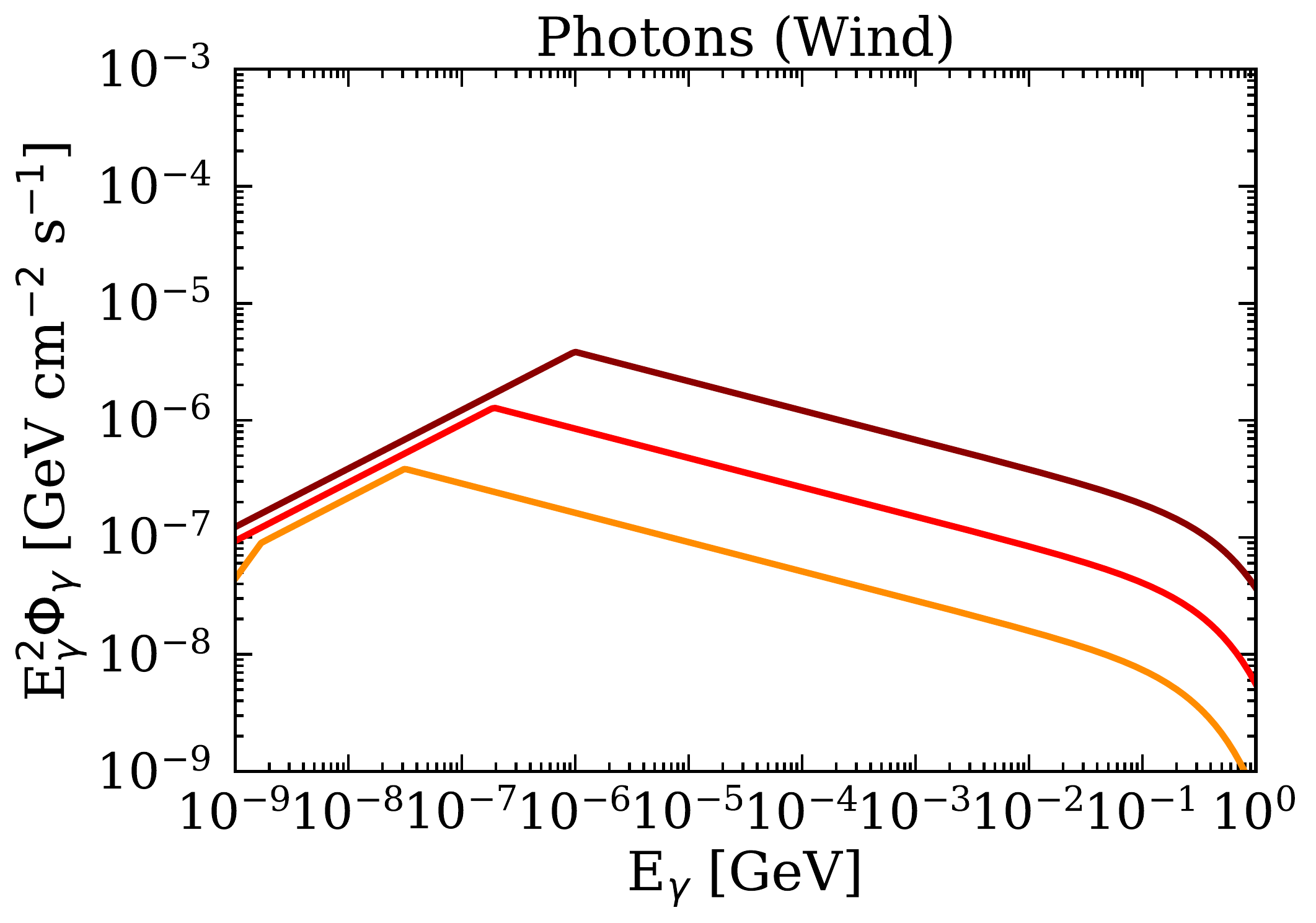}
 \includegraphics[scale=0.36]{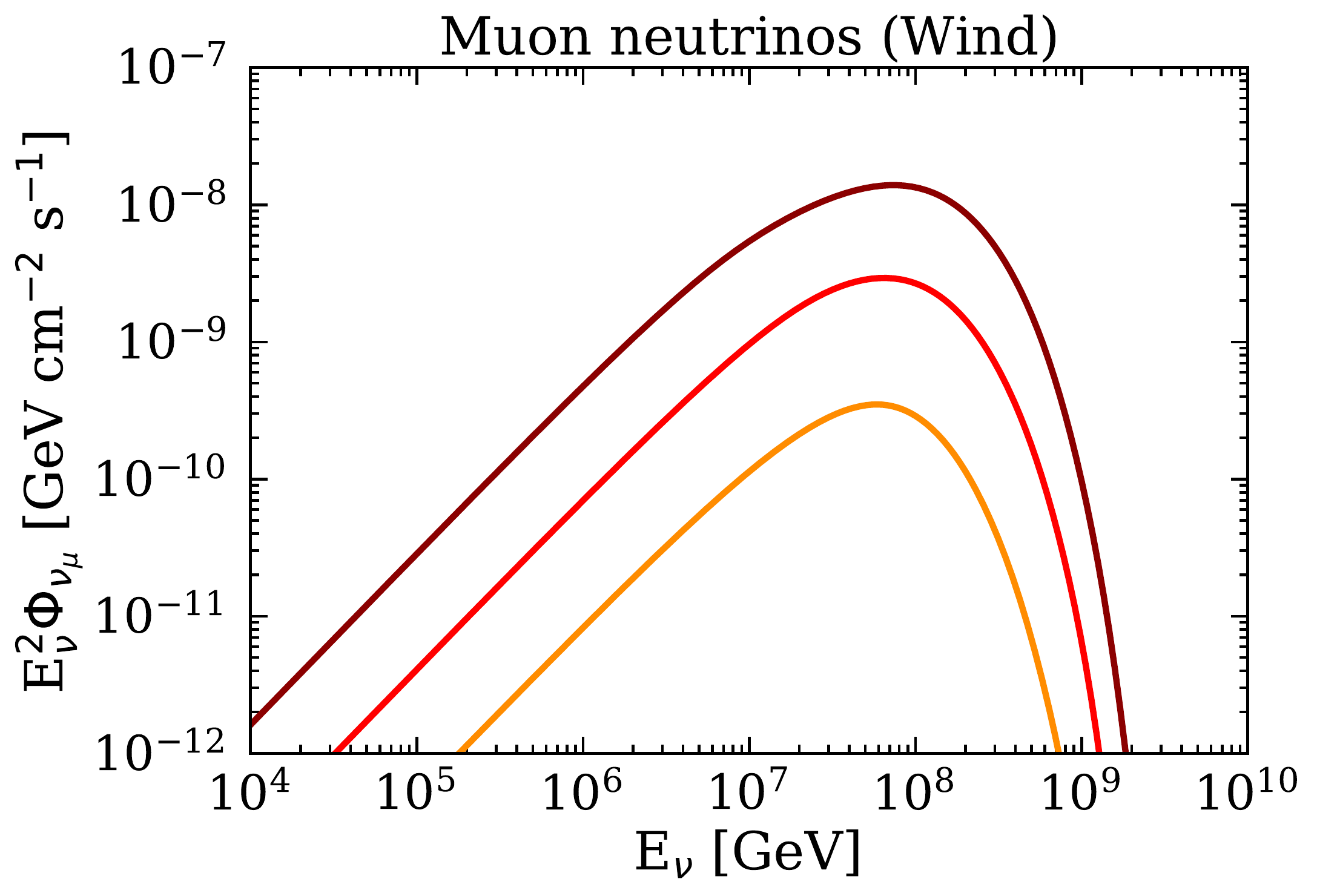}
   \caption{{\it Left:} Synchrotron photon flux expected at Earth for the classic afterglow scenario for  $t = T_{\rm{dec}}$, $3 T_{\rm{dec}}$, and $10 T_{\rm{dec}}$ (see the gray vertical lines in Fig.~\ref{fig:light_curve}) for our benchmark GRB in Table~\ref{tab:benchmark_grb} at $z=1$. {\it Right:} Corresponding neutrino flux expected at Earth. Both fluxes for the wind scenario decrease faster than for the ISM scenario. }
\label{fig:photonISM}
\end{figure}

The photon and neutrino fluxes expected at Earth (see Sec.~\ref{sec:neutrinos})  are shown in Fig.~\ref{fig:photonISM} for $t = T_{\rm{dec}}$, $3 T_{\rm{dec}}$, and $10 T_{\rm{dec}}$ (marked with vertical lines in Fig.~\ref{fig:light_curve}) for the ISM and wind scenarios. We refer the interested reader to  Appendix~\ref{sec:cooling} for a discussion on the characteristic cooling times of protons and mesons affecting the neutrino distributions. For both CBM cases, the flux at Earth decreases with time, as expected~\cite{Sari:1997qe}. Moreover, the peak of the photon energy distribution and its energy breaks shift to lower energies as  time increases. This is due to the fact that  the minimum and cooling energies scale with time as  $E_{\gamma, \min} \propto t^{-3/2}$ $E_{\gamma, \rm{cool}} \propto t^{-1/2}$, respectively~\cite{Sari:1997qe}.

In the right panels of Fig.~\ref{fig:photonISM}, we show our results concerning the neutrino flux. 
In the wind scenario, the neutrino flux  peaks at $E_{\nu}^{\rm{peak}} \simeq 8.1 \times 10^7$~GeV for $t = T_{\rm{dec}}$ and then decreases up to $E_{\nu}^{\rm{peak}} \simeq 6.3 \times 10^7$~GeV for  $t=10 T_{\rm{dec}}$. For the ISM scenario,  the neutrino flux  peaks at $E_{\nu}^{\rm{peak}} \simeq 7.7 \times 10^7$~GeV and at $E_{\nu}^{\rm{peak}} \simeq 7.3 \times 10^7$~GeV for  $t=10 T_{\rm{dec}}$.
The effect of kaon cooling is not visible, since as shown in Appendix~\ref{sec:cooling} (see Fig.~\ref{fig:timescale_meson_slowTdec}) kaons cool at  energies larger than  the maximum energy of protons in the blastwave. Note that both the photon and the neutrino fluxes are larger in the wind scenario than in  the ISM one, but they decreas faster in the wind case~\cite{Razzaque:2013dsa}. This is due to the fact that higher densities of the external medium can be initially reached within the wind profile. At such densities, the blastwave decelerates faster, leading to a rapidly decreasing flux~\cite{Razzaque:2013dsa}. 
The higher densities in the wind scenario also allow for higher magnetic fields, which cause the shift of the cooling frequency in the photon spectrum at  energies lower than  the ISM case. Of course, this is a direct consequence of the  value adopted for $\epsilon_B$. 

The standard afterglow scenario has been already investigated in the literature for what concerns neutrino emission. Nevertheless, there are some relevant differences with respect to  the results presented in this section. Our classic afterglow model resembles the one investigated in Refs.~\cite{Razzaque:2013dsa, Thomas:2017dft}. However, by using the  benchmark input parameters of  Refs.~\cite{Razzaque:2013dsa, Thomas:2017dft}, we find a neutrino flux that is almost $5$ orders of magnitude larger, but with an identical  shape. This discrepancy might be caused by several reasons. First, there is a missing factor $(E^\prime_{\gamma, \min}/ E^\prime_{\gamma, \rm{cool}})^{-1/2}$ in the photon distribution in Eq.~11 of Ref.~\cite{Razzaque:2013dsa}; second, in the definition of the proton flux of Ref.~\cite{Razzaque:2013dsa}  there is a factor $1/[4 \pi (1+z)^2 ]$ in excess, which contributes to further lower the corresponding neutrino flux. On the other hand, our results on the photon and neutrino fluxes are in agreement with the ones in Refs.~\cite{Sari:1997qe, Dermer:2000yd}, respectively.  

The afterglow flux produced by the reverse shock has been investigated in Ref.~\cite{Murase:2007yt}, while  we focus on  the contribution from the forward shock. The neutrino flux obtained in Ref.~\cite{Murase:2007yt} strictly depends on the assumptions on the thickness of the shell. For example, in the case of a thin shell with  $\tilde{E}_{\rm{iso}} = 4 \times 10^{52}$~erg and propagating in an ISM with $n_0= 0.5$~cm$^{-3}$, the estimated flux peaks at $E_{\nu}^{\rm{peak}} \simeq 10^{10}$~GeV, where it should reach  about  
$ 10^{- 10}$~GeV~cm$^{-2}$~s$^{-1}$ for a GRB at $z = 1$. This result is comparable with our maximum flux $\simeq 2 \times 10^{-9}$~GeV~cm$^{-2}$~s$^{-1}$, considering that the isotropic energy adopted in Ref.~\cite{Murase:2007yt}   is one order of magnitude smaller than the one we  adopt in this work.
Nevertheless, the neutrino flux peaks at  energies higher than ours in Ref.~\cite{Murase:2007yt}. Indeed, our fluxes peak at $E_{\nu}\simeq 10^8$~GeV, in contrast with the peak at $\simeq 10^{10}$~GeV in~\cite{Murase:2007yt}, probably because of the different  initial $\Gamma_0$ and because protons are expected to be accelerated at higher energies at the reverse shock. The most optimistic case considered in Ref.~\cite{Murase:2007yt} is a thick shell propagating in a wind environment. In the latter scenario, the afterglow flux reaches an amplitude about $\sim 2$ orders of magnitude larger than ours at the peak energy $E_{\nu} \sim 10^{9}$~GeV, which is shifted by $\sim 1$ order of magnitude with respect to ours.
Also for the wind scenario, the differences are mainly due to the energy of the ejecta, assumed to be $\sim 4$ times larger than ours, and the density of the environment up to $10$ times larger than our benchmark value. Moreover, we rely on the  thin shell assumption rather than the thick one, hence the results are not directly comparable. Finally, note that the emission from the reverse shock lasts longer than the emission from the forward shock.

\subsection{Particle emission in the presence of a late shell collision}\label{sec:afterglow_bump}
In the presence of an optical jump, we model the afterglow light curve through the late collision of two relativistic shells. 
At $t=T_{\rm{coll}}$, we compute the neutrino flux as described in Sec.~\ref{sec:neutrinos} and by using the photon distribution introduced in Sec.~\ref{sec:photon_merger}.
After the merger, the resulting merged shell starts to be decelerated by the external medium, emitting radiation with the standard features expected during the afterglow, as discussed in Sec.~\ref{sec:classic_afterglow}, but with the parameters characteristic of the merged shell. Since energy has been injected in the slow shell during the merger, the merged shell is more energetic than the slow one. Thus, we obtain a higher photon flux as shown in Fig.~\ref{fig:light_curve} (brown continuous line). 
The star denotes the flux  at $t=T_{\rm{coll}}$, given by the sum of the afterglow radiation (see Sec.~\ref{sec:afterglow_nobump}) generated by the slow shell and the radiation from the shocks developing at the collision. For our choice of parameters, electrons accelerated at the collision are in the slow cooling regime both in the ISM and wind scenario (see Appendix~\ref{Theoretical modeling} and figures therein). Therefore, the appropriate photon distribution is given by Eq.~\ref{eq:lum_slowCOLL}.

Since it is assumed that the merger occurs instantaneously at the collision time, we are not taking into account the time needed by the merged shell to relax before being decelerated to the BM solution. Because of this approximation, we neglect the neutrinos produced for $T_{\rm{coll}} < t < T_{{\rm{dec,  m}}}$, since an analytic treatment in this transition phase is not feasible, as already mentioned in Sec. \ref{sec:light_curves}. The time window excluded from our calculations of the neutrino signal corresponds to the gray shadowed area in  Fig.~\ref{fig:light_curve}. Note that, for most of the initial configurations of the slow shell, we find $T_{{\rm{dec,  m}}} \simeq 2 \; T_{\rm{coll}}$. The exclusion of such a time window in our calculation negligibly  affects  the overall time-integrated neutrino signal, which is the main goal of this work (see Sec.~\ref{sec:detection}).

\begin{figure}[]
\centering
\includegraphics[scale=0.31]{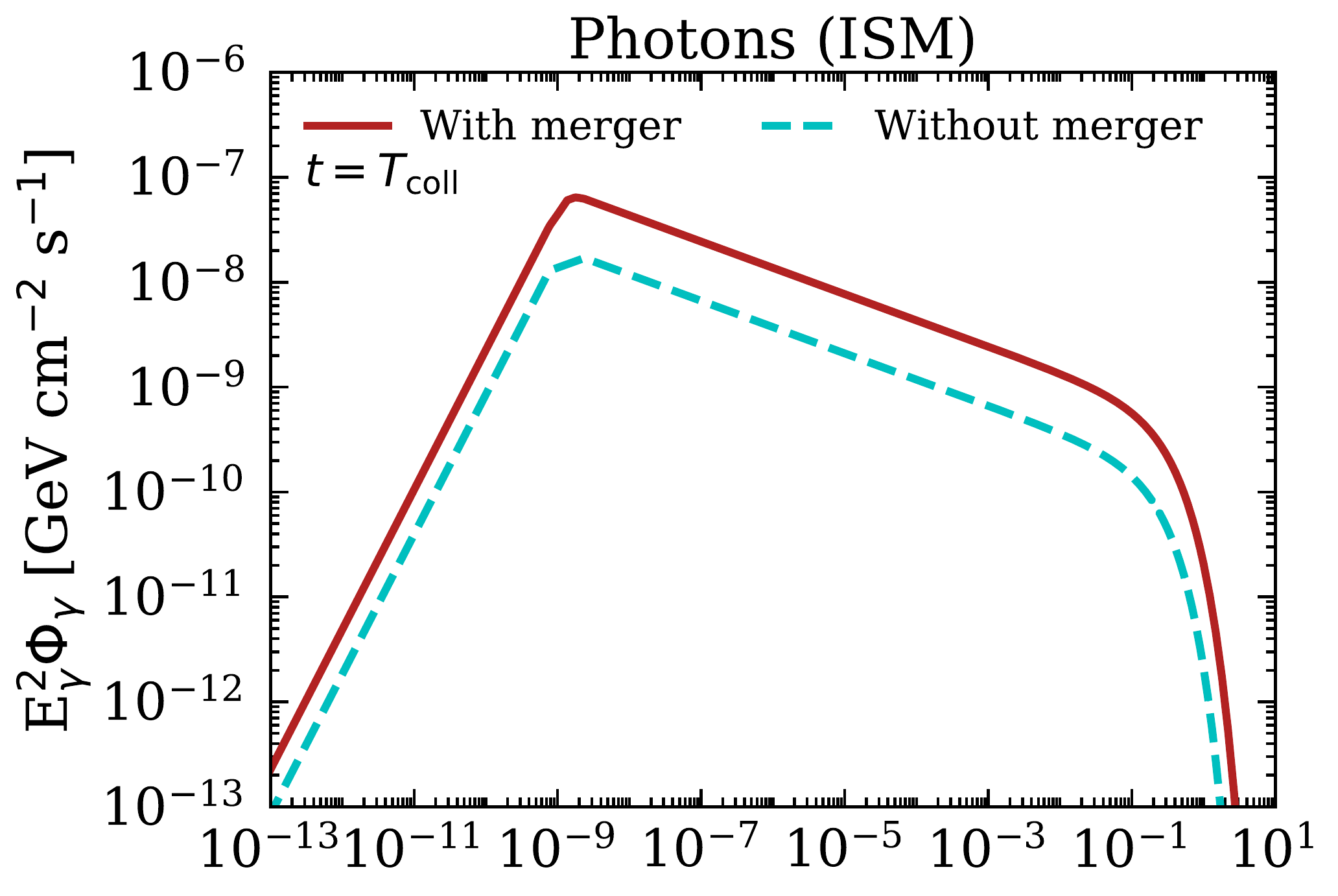}
\includegraphics[scale=0.31]{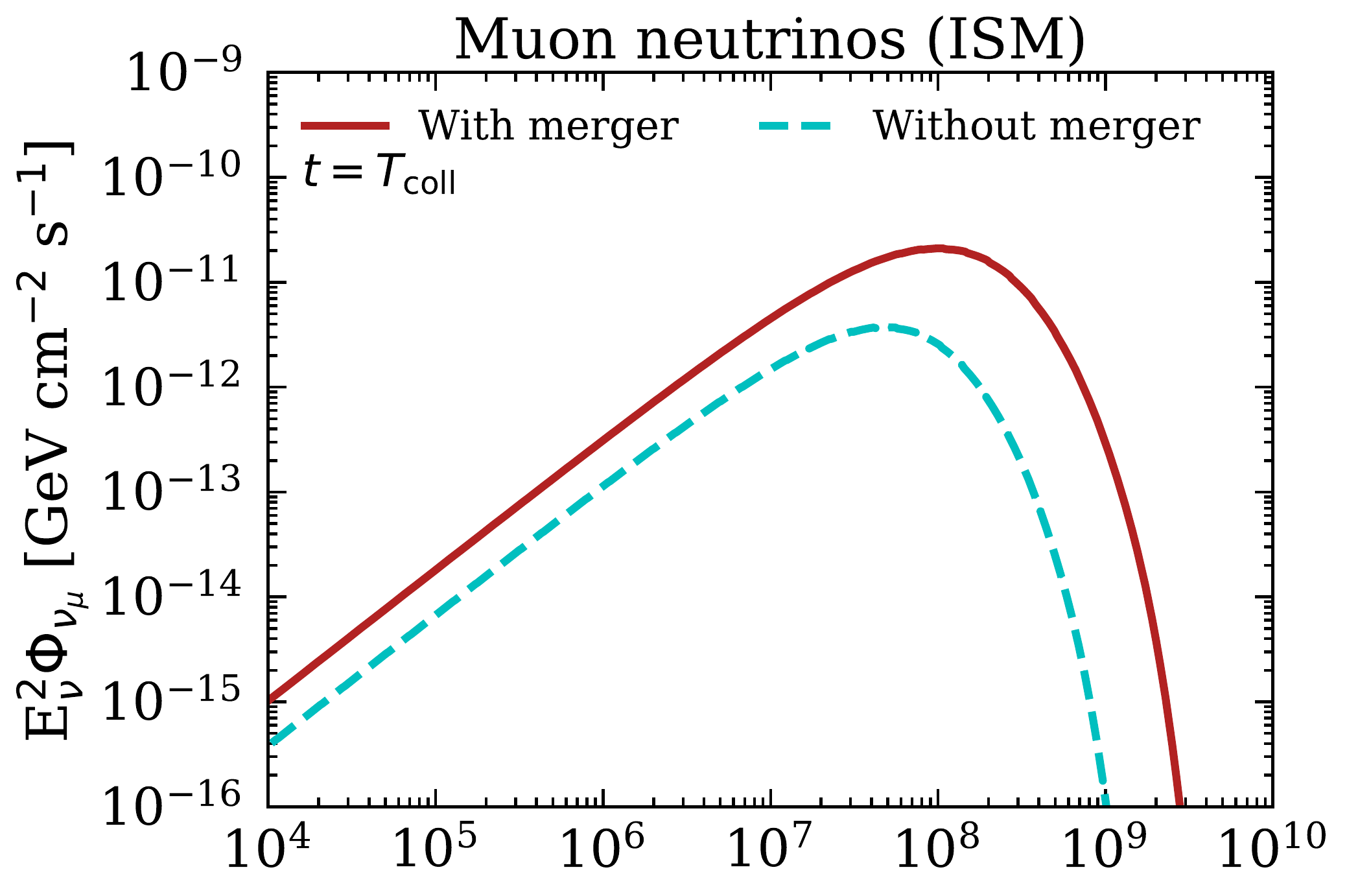}
\includegraphics[scale=0.31]{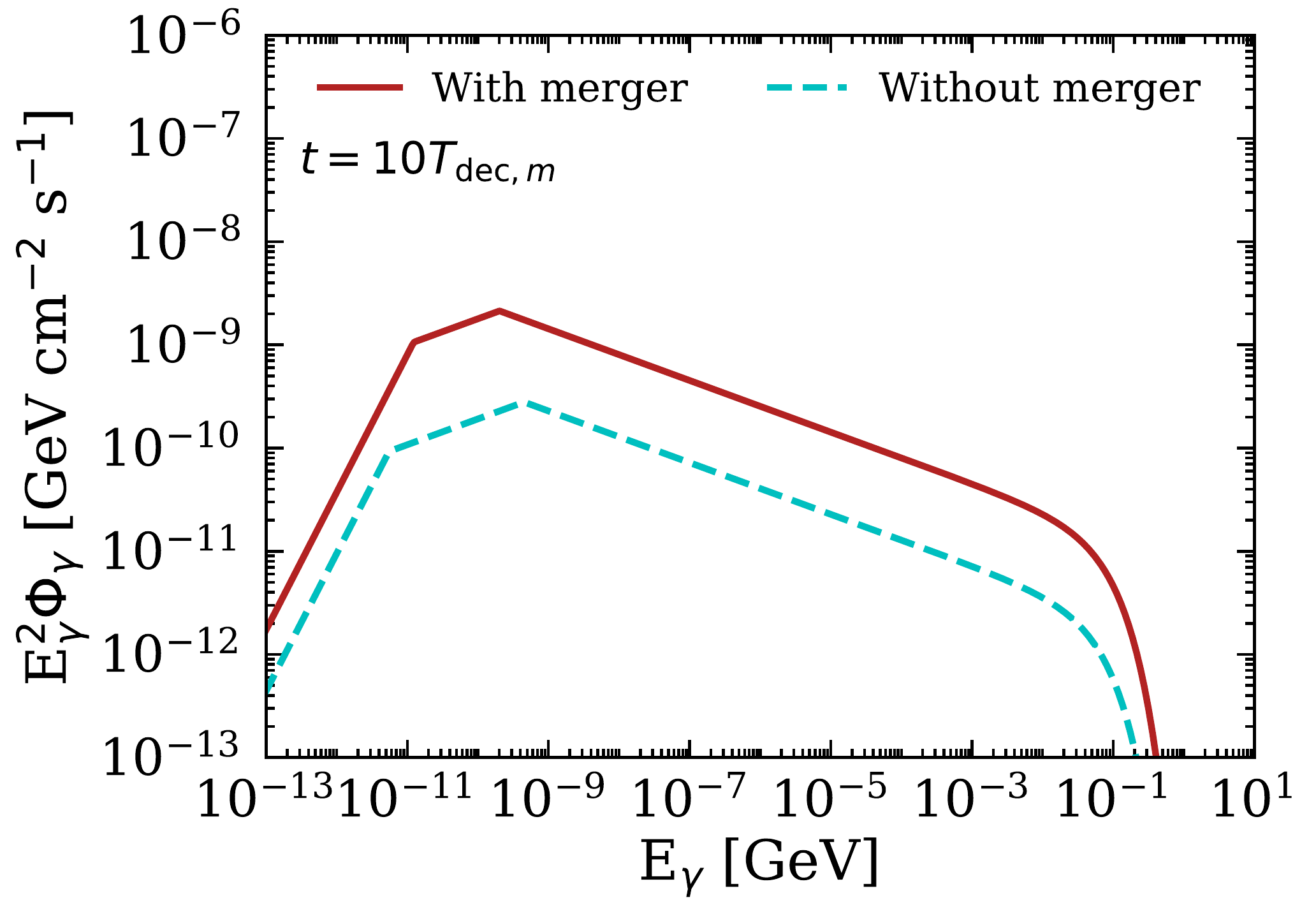}
\includegraphics[scale=0.31]{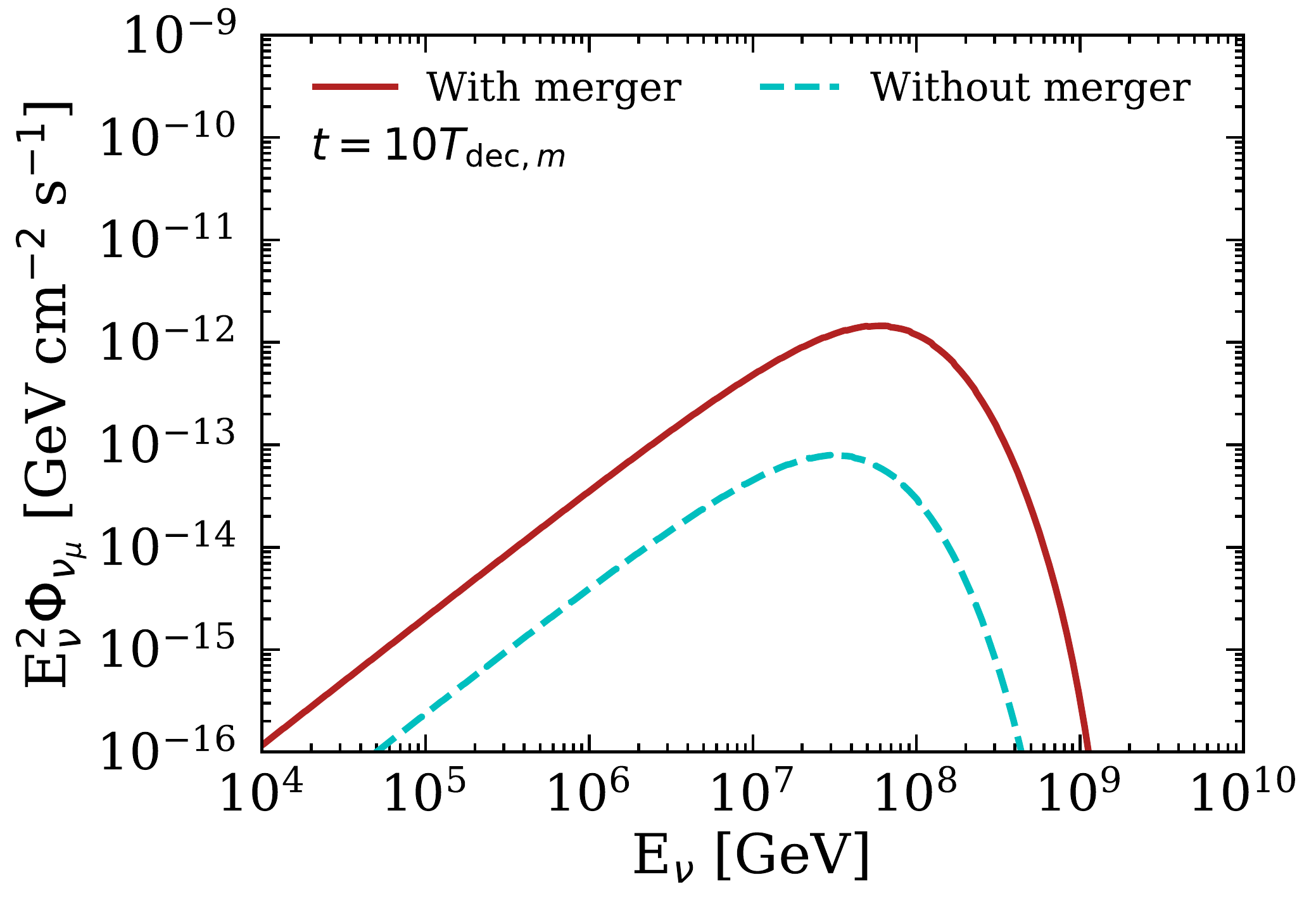}
\includegraphics[scale=0.31]{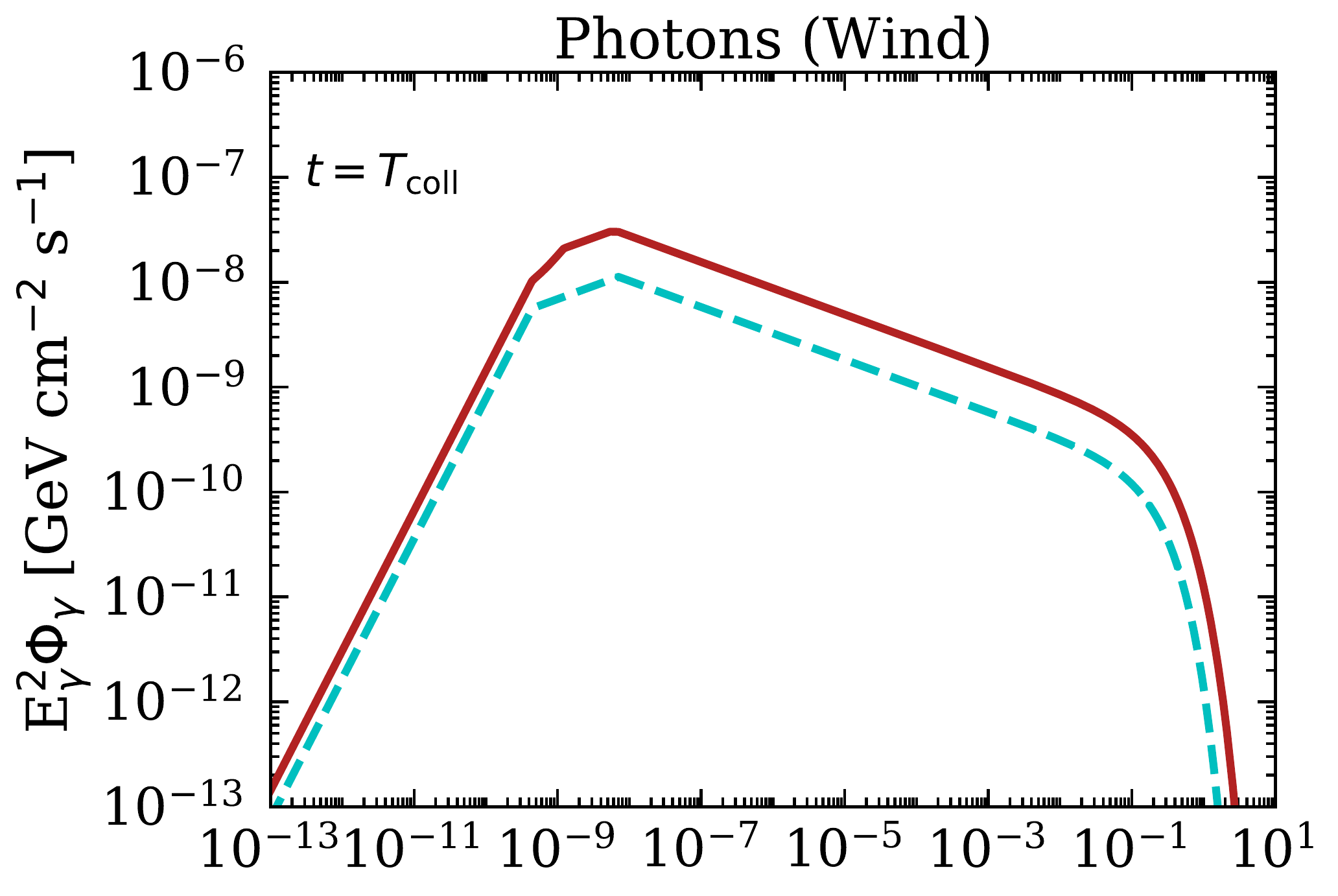}
\includegraphics[scale=0.31]{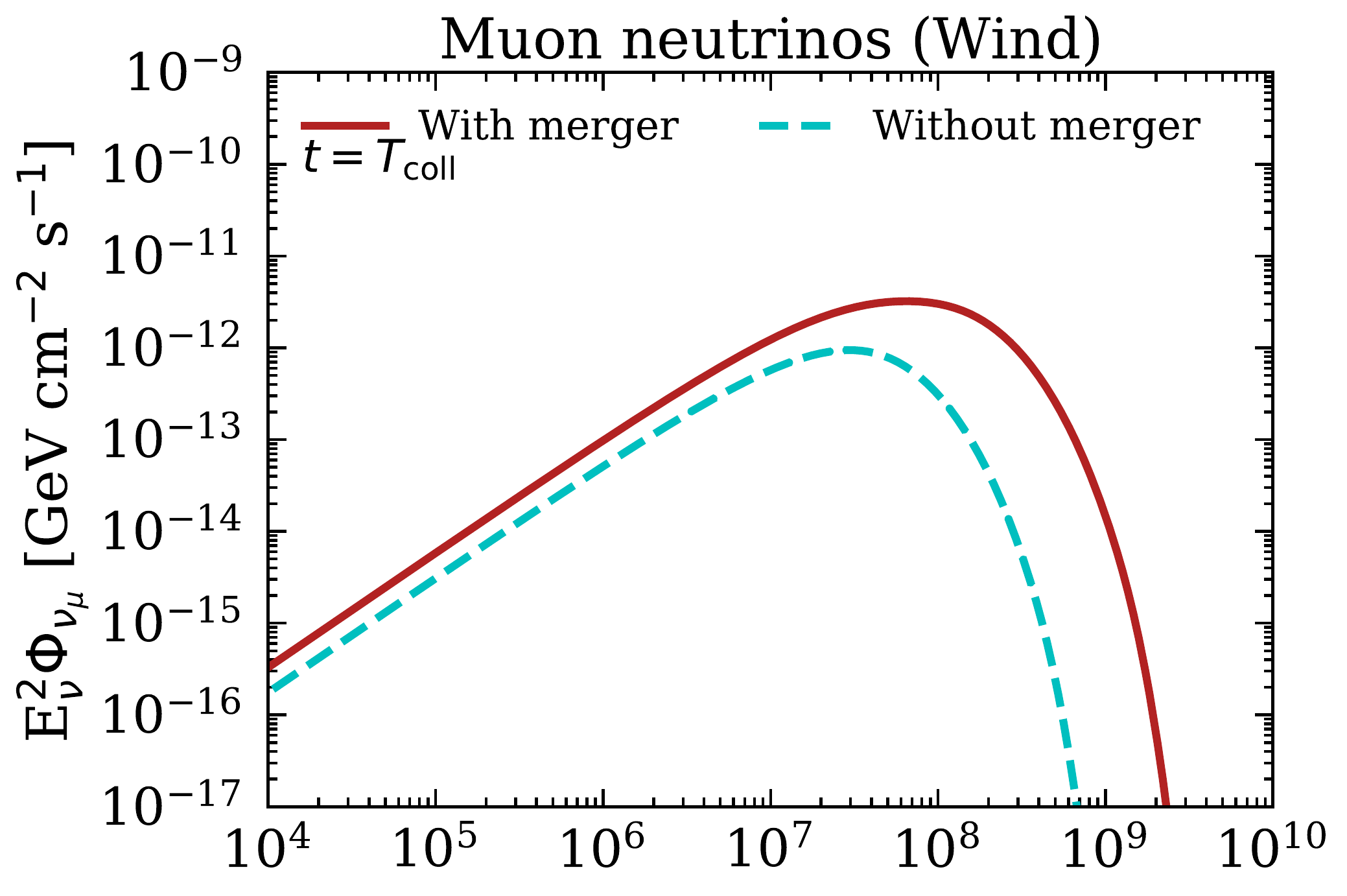}
\includegraphics[scale=0.31]{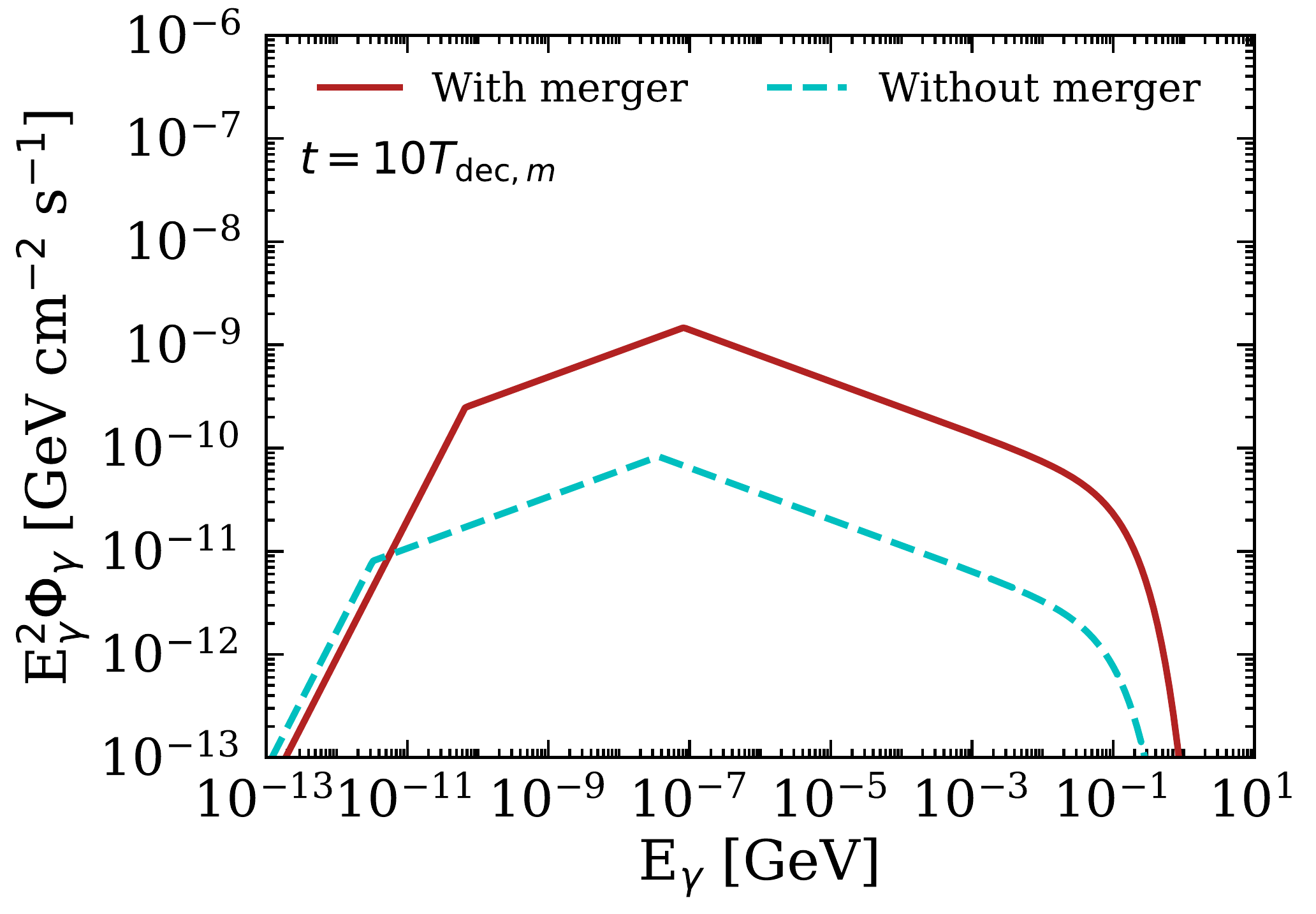}
\includegraphics[scale=0.31]{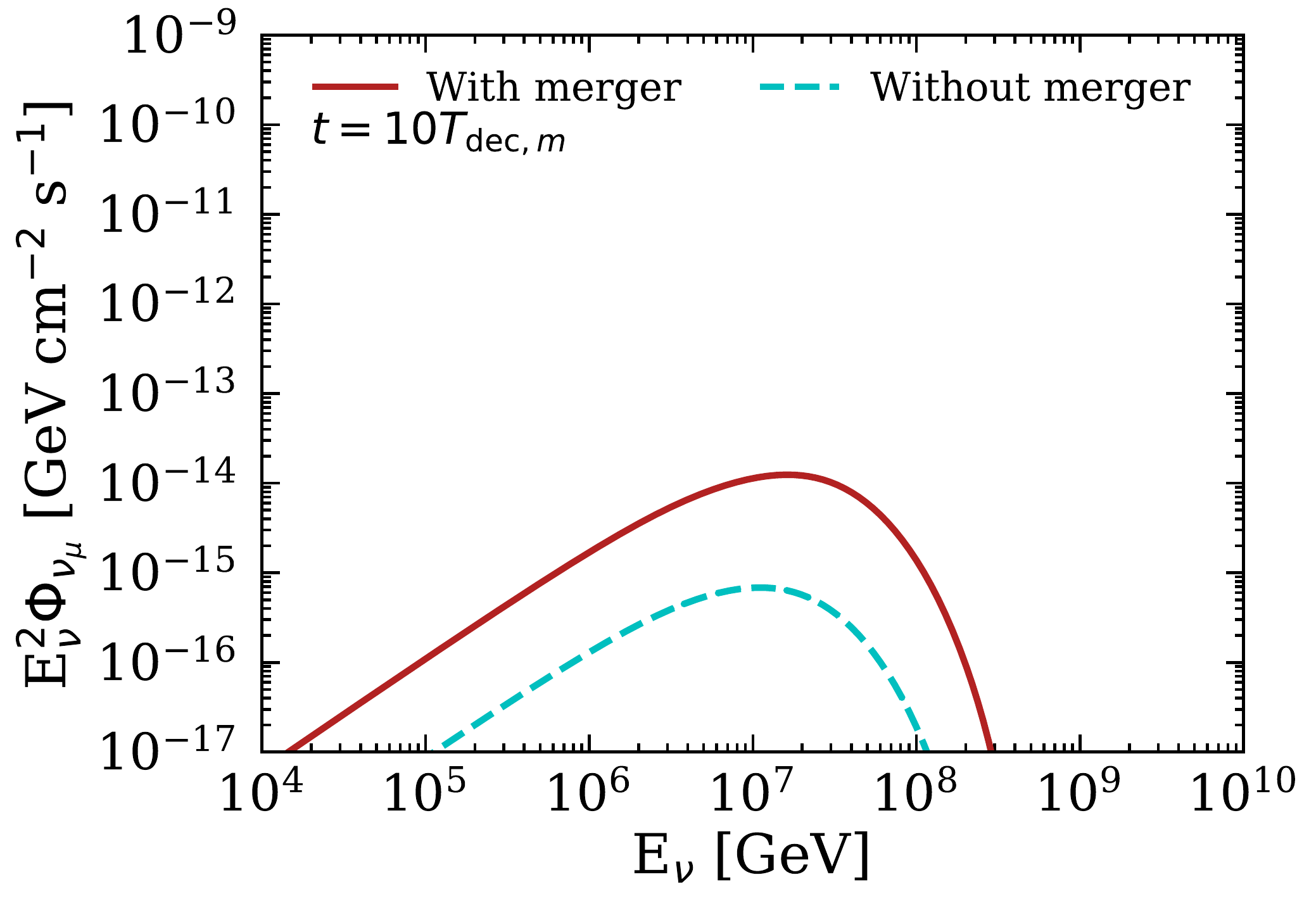}
\caption{Photon (on the left) and neutrino (on the right) fluxes expected at Earth as  functions of the particle energy from the afterglow when the merger of two relativistic shells occurs for the ISM (top two panels) and wind (bottom two panels) scenarios for our benchmark GRB in Table~\ref{tab:benchmark_grb} at $z=1$.  For each CBM scenario, the fluxes are shown  at $t=T_{\rm{coll}}$ and $10\ T_{{\rm{dec}}, m}$ (see vertical lines in Fig.~\ref{fig:light_curve}). The brown lines display the total expected flux in the presence of a merger, while the cyan lines represent the flux  that would be observed in the absence of a jump. The late shell merger enhances the photon and neutrino fluxes compared to the standard afterglow scenario and shifts the peak of the energy distributions at larger energies.}
\label{fig:merger_snapshots}
\end{figure}
\begin{figure}[]
    \centering
    \includegraphics[scale=0.36]{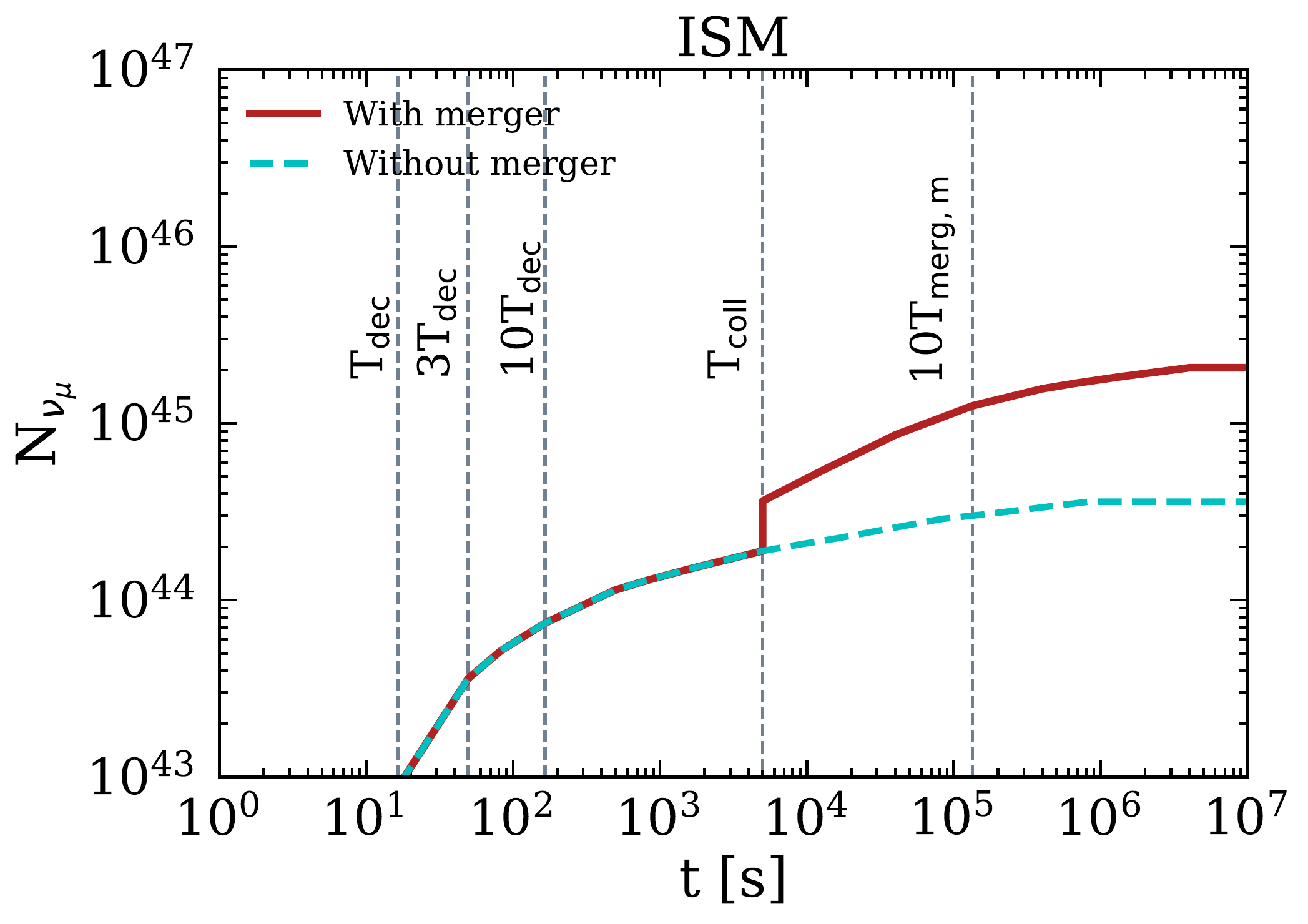}
    \includegraphics[scale=0.36]{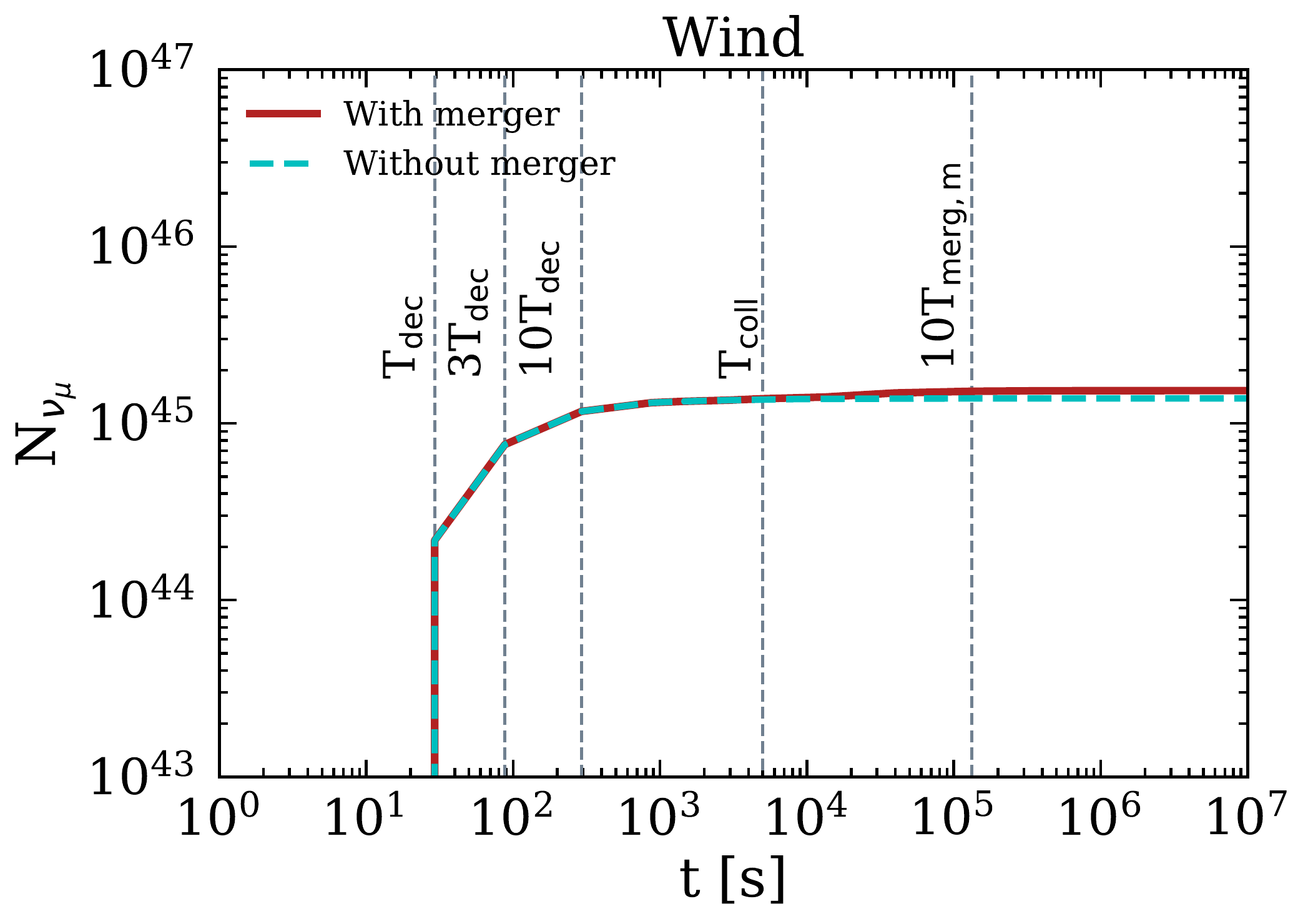}
    \caption{Cumulative number of muon neutrinos expected at Earth for  the ISM  (left panel) and wind  (right panel) scenarios as a function of time for our benchmark GRB (Table~\ref{tab:benchmark_grb})  at $z=1$. The brown solid line represents the number of muon neutrinos produced  when the shell merger occurs, while the cyan dashed line corresponds to the case of the classic afterglow. In order to guide the eye, the gray vertical lines mark the times at which we show the neutrino flux at Earth for the classic afterglow scenario (Fig.~\ref{fig:photonISM}) and when  a jump occurs (Fig.~\ref{fig:merger_snapshots}). In the ISM environment, the jump significantly increases the cumulative number of neutrinos, while  the difference between the two scenarios is negligible in the wind case.}
    \label{fig:cumulative_neutrino}
\end{figure}

Figure~\ref{fig:merger_snapshots} shows the photon and muon neutrino fluxes at $t = T_{\rm{coll}}$ and after the merger at $t = 10 T_{{\rm{dec}}, m}$ for the ISM and wind scenarios. These times are marked in Fig.~\ref{fig:light_curve} by vertical lines. For comparison, we also show the photon and neutrino fluxes that would be generated at $t = 10 T_{{\rm{dec}}, m}$ if no merger occurred. 
In both CBM scenarios, the neutrino flux  increases in the presence of a jump, as expected, due to the  denser photon field leading to more efficient $p \gamma$ interactions (see also the cumulative number of muon neutrinos plotted as a function of time in Fig.~\ref{fig:cumulative_neutrino}).

The peak of the neutrino distribution at late times in Fig.~\ref{fig:merger_snapshots} is shifted at higher energies compared to the case without merger. This is explained because the energy density content of the merged shell is larger than the one of the slow shell, thus the corresponding magnetic field is larger as well. This results into a greater maximum energy of protons in the merged blastwave since $E^{\prime}_{p, \rm{max}}$ depends linearly on the magnetic field; indeed, the acceleration time (see Eq.~\eqref{eq:acceleration_time}) limits the maximum energy of protons. Finally, the quantities entering  the Lorentz transformation of the flux at Earth (e.g.~the Lorentz factor) are larger for the merged shell than for the slow one.

From Fig.~\ref{fig:cumulative_neutrino}, we can see that  the number of neutrinos at $T_{\rm{coll}}$ is given by the sum of the neutrinos produced at the shock front between the slow shell and the CBM and the neutrinos produced at the collision between the slow and fast shells. After the merger, the only contribution comes from the afterglow of the merged shell. 
By comparing the left and right panels of Fig.~\ref{fig:cumulative_neutrino}, we note that a larger efficiency in the neutrino production is achieved in the  ISM scenario in the presence of  shell merger. 
In particular, for the ISM scenario the number of neutrinos increases by a factor of $6$. This result is justified in light of the fact that the neutrino flux rapidly decreases for a wind-like CBM. Thus, the early time emission  dominates the time-integrated neutrino flux. Motivated by these results, in the next section we discuss the detection prospects for neutrinos produced during the GRB afterglow when a jump occurs in the light curve.

\section{Neutrino detection perspectives}
\label{sec:detection}
In this section, we explore  whether the increase in the number of neutrinos expected in the presence of an optical jump could reflect  improved detection perspectives at ongoing and future generation neutrino telescopes. We explore the detection prospects for the  all-sky quasi-diffuse flux as well as point source searches. Finally, we forecast the expected neutrino fluence from  GRB 100621A and for a second hypothetical GRB with  parameters inspired by the bright GRB 130427A.

\subsection{All-sky quasi-diffuse flux}
 The average isotropic kinetic energy from the catalogue of the Gehrels Swift Observatory is $\tilde{E}_{k, \rm{iso}} \simeq 10^{53}$~erg~\cite{Liang:2007rn} and the redshift distribution peaks at $z \simeq 2$~\cite{Jakobsson+2012}.
 Hence, we compute the all-sky quasi-diffuse flux by placing our benchmark GRB at $z=2$ and  assuming that its flux is representative of the  GRB population.
For a GRB rate of $\dot{\rm{N}} \sim 700 \; \rm{yr}^{-1}$~\cite{IceCube:2021qyj,IceCube:2017amx,Paciesas:1999tp} and an isotropic distribution of all the sources in the sky, the all-sky quasi-diffuse flux is: 
\begin{equation}
\label{eq:stacking}
F_{\nu_\mu}(E_\nu) = \frac{\dot{\rm{N}}}{4 \pi} \int dt\ \Phi_{\nu_\mu }(E_\nu, z = 2)\ , 
\end{equation}
being $\Phi_{\nu_\mu}$ defined as in Eq.~\ref{eq:neutrino_flux}. In the case of the afterglow generated by the slow and the merged shells, we perform the time integration for $t \in [T_{\rm{dec}}, T_{\rm{Sedov}}]$, where $T_{\rm{Sedov}}$ marks the Sedov time when the blastwave becomes non-relativistic and enters the Newtonian regime. At $T_{\rm{coll}}$ the integration over time is replaced by the product with $t_{\rm{dyn}, m}^0= t^{\prime 0}_{{\rm{dyn}}, m}(1+z)/\Gamma_m^0$, where $t^{\prime 0}_{{\rm{dyn}}, m}$ is given by Eq.~\eqref{eq:dynamical_merg}, since the collision is considered to be an instantaneous process.

\begin{figure}[]
\centering
\includegraphics[scale=0.33]{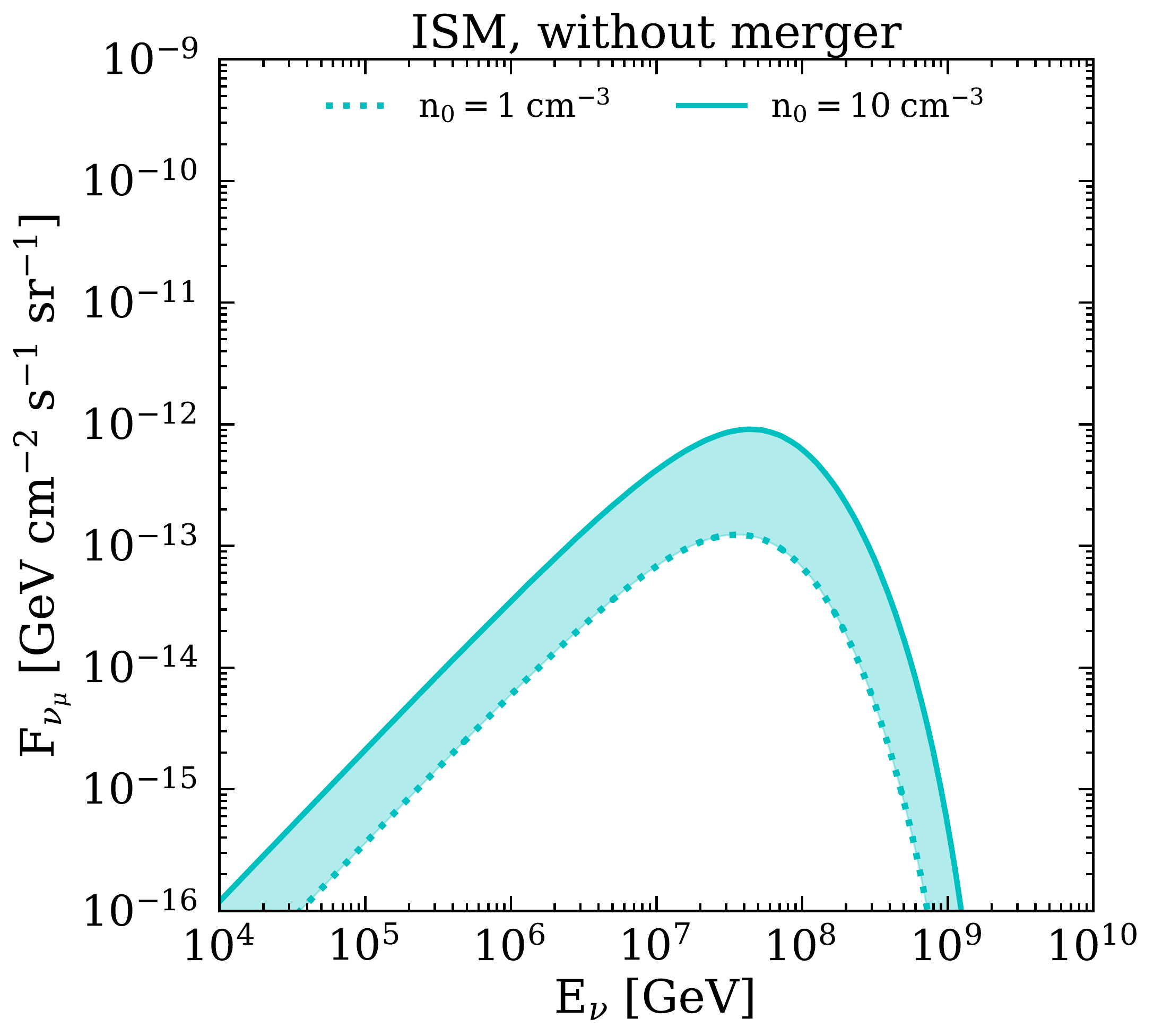}\label{ism_stacked_nomerg}
\includegraphics[scale=0.33]{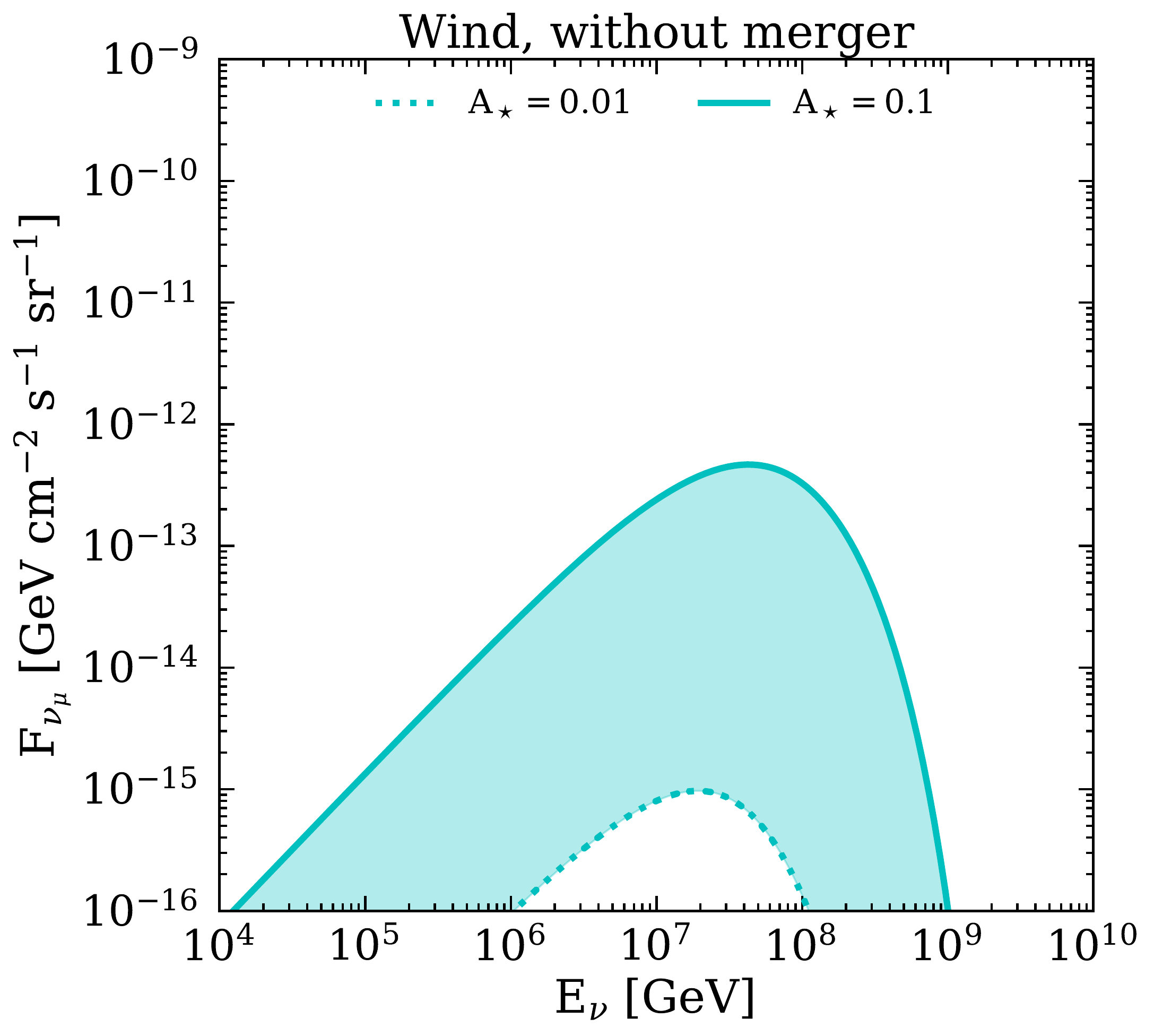}\label{wind_stacked_nomerg}\\
\includegraphics[scale=0.33]{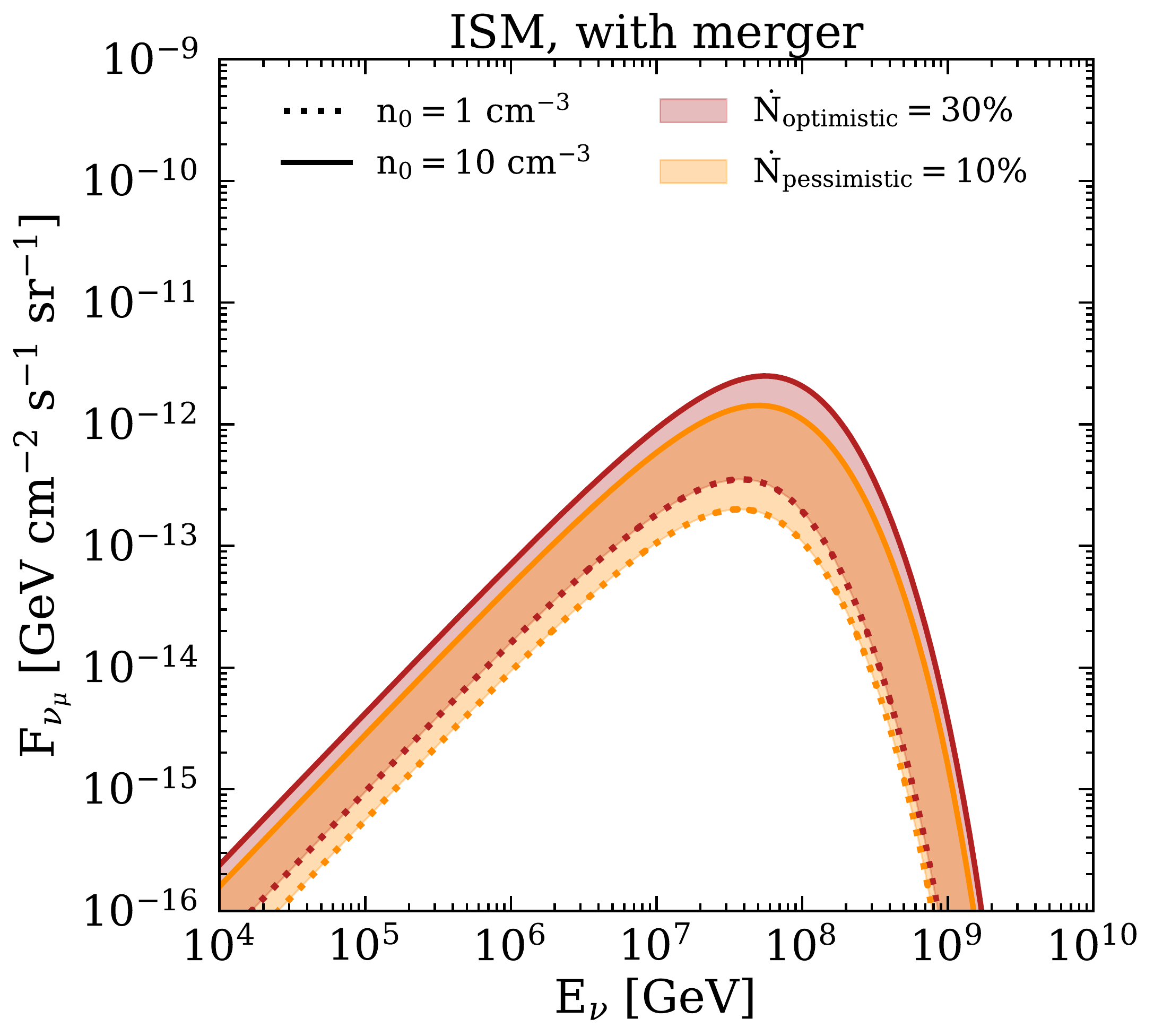}\label{ism_stacked_merg}
\includegraphics[scale=0.33]{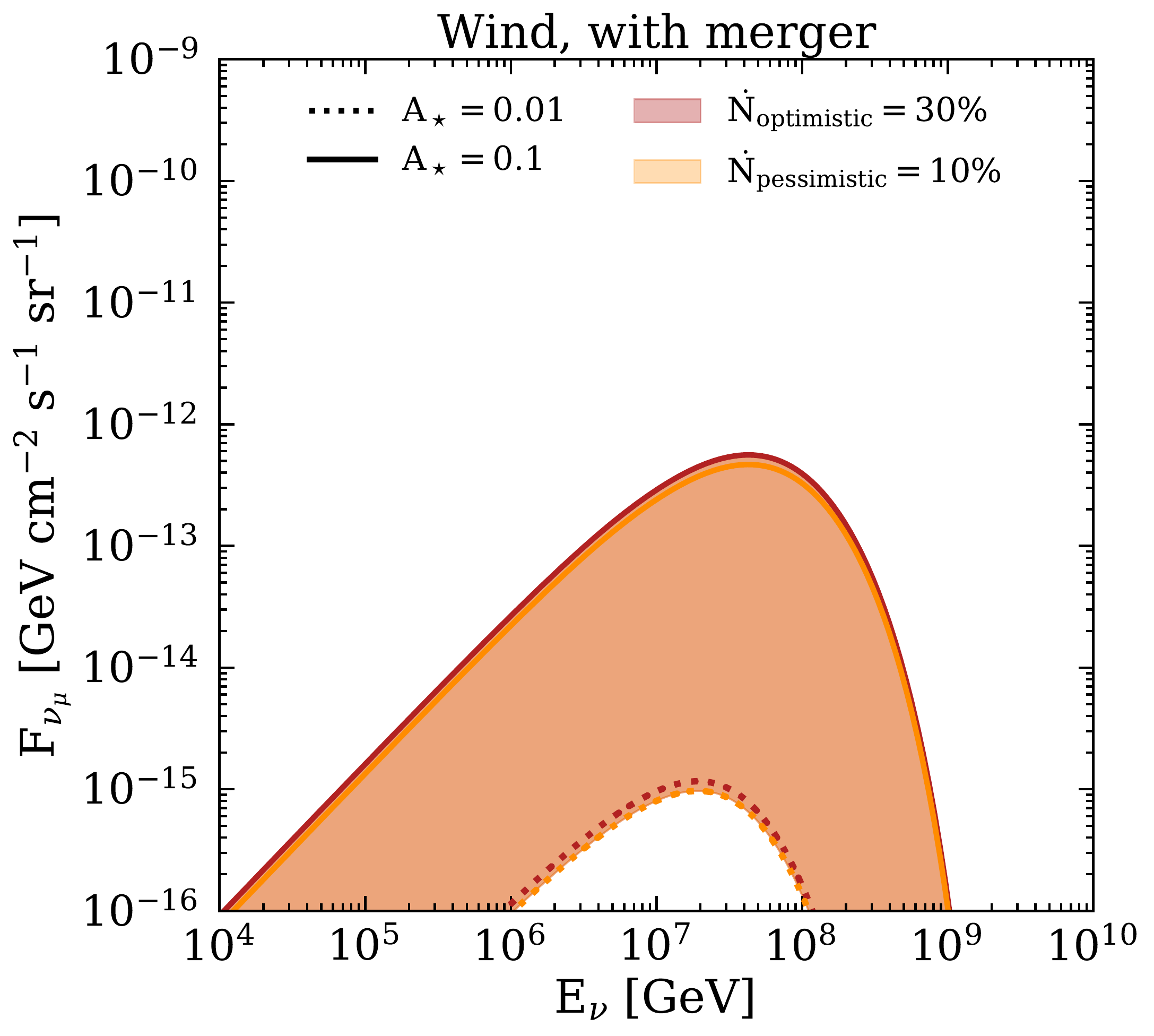}\label{wind_stacked_merg}
\caption{All-sky quasi-diffuse muon neutrino flux from GRB afterglows for the constant density (left panels) and wind (right panels) CBM scenarios, for the standard GRB afterglow  (top panels) and the case with optical jumps (bottom panels). For the  ISM scenario, the band is defined by  $1 \lesssim n_0 \lesssim 10$~cm$^{-3}$  (dotted and solid lines, respectively). For the  wind  scenario, the  band is defined by $0.01 \lesssim A_\star \lesssim 0.1$. For the bottom panels, the quasi-diffuse neutrino flux is computed  for the optimistic scenario with $\dot{\rm{N}}_{\rm{optimistic}} = 30\% \dot{\rm{N}}$ (brown shadowed region) and  $\dot{\rm{N}}_{\rm{pessimistic}} = 10\% \dot{\rm{N}}$ (orange shadowed region). In the presence of optical jumps, the all-sky quasi-diffuse flux slightly increases for the ISM scenario, while negligible changes occur for the wind case. For the wind environment, there is no difference between the optimistic and pessimistic cases since the classic afterglow always dominates the neutrino fluence.  In the cases with and without shell merger, the all-sky quasi-diffuse neutrino flux is in agreement with the results on GRB  afterglow neutrino searches reported in Ref.~\cite{IceCube:2021qyj} and it does not overshoot the  IceCube limits on the GRB prompt emission~\cite{IceCube:2017amx}, as well as the limits placed by the ANTARES collaboration~\cite{ANTARES:2020vzs} and the expected ones for KM3NeT~\cite{KM3Net:2016zxf}.}
\label{stacked_analysis}
\end{figure}

The top panels of Fig.~\ref{stacked_analysis} show the all-sky quasi-diffuse neutrino flux in the absence of shell merger, i.e.~if the light curve resembles the  standard afterglow scenario, for the ISM and wind scenarios. For the ISM scenario, the  band corresponds to $1 \lesssim n_0 \lesssim 10$~cm$^{-3}$; while for the wind scenario, the band includes $0.01 \lesssim A_\star \lesssim 0.1$.

So far, the IceCube Neutrino Observatory has  detected neutrinos with energies up to a few PeV~\cite{Ahlers:2018fkn, Ahlers_2015, IceCube:2020wum, IceCube:2020abv}. 
Even though several sources have been proposed to explain the origin of high-energy neutrinos~\cite{Meszaros:2015krr, Waxman:2015ues, Murase:2015ndr,Anchordoqui:2013dnh,Vitagliano:2019yzm},  only a handful of possible associations have been presented between neutrinos and active galactic nuclei, tidal distruption events (TDEs), and superluminous supernovae~\cite{IceCube:2018dnn,Giommi:2020viy,Franckowiak:2020qrq,Fermi-LAT:2019hte,Krauss:2018tpa,Kadler:2016ygj,Stein:2020xhk,Reusch:2021ztx,Pitik:2021dyf}. In particular,  limits on the quasi-diffuse neutrino flux from GRBs have been placed by the IceCube Collaboration by  taking into account the prompt emission~\cite{IceCube:2017amx}, while a similar analysis on the afterglow phase is missing. A statistical analysis aiming to look for temporal and spatial coincidences between GRB afterglows and neutrinos detected by the IceCube Neutrino Observatory has been carried out in Ref.~\cite{IceCube:2021qyj}. In agreement with the findings of Ref.~\cite{IceCube:2021qyj}, our quasi-diffuse flux does not overshoot existing upper limits on the prompt emission reported by IceCube~\cite{IceCube:2017amx} and by the ANTARES collaboration~\cite{ANTARES:2020vzs}, as well as the ones expected for KM3NeT~\cite{KM3Net:2016zxf}.
Despite differences in the theoretical modeling of the expected signal, our conclusions are also consistent with the detection prospects for the GRB afterglow neutrinos outlined in Ref.~\cite{Razzaque:2014ola}. 
Assuming that jumps occur in the  afterglow light curve, the corresponding all-sky quasi-diffuse muon neutrino flux is shown in the  bottom panels of Fig.~\ref{stacked_analysis} for the two CBM scenarios. Since  the fraction of  GRB afterglows having optical jumps is largely uncertain~\cite{2012ApJ.758.27L, 2013ApJ.774.13L}, we consider an optimistic (pessimistic) case such that the rate of GRBs per year with  jumps is $30\%$ ($10\%$) of $\dot{\rm{N}}$ (see Eq.~\ref{eq:stacking}). The ``pessimistic'' fraction of GRBs with optical jumps  is extrapolated by the analysis carried out in Ref.~\cite{2013ApJ.774.13L}, where they estimate that $10$ out of  $146$ GRBs with well resolved optical light curves displayed a jump. The ``optimistic'' fraction of GRBs with optical jumps is obtained by considering that the actual fraction of GRBs with optical jumps is not known and existing constraints may be plagued by observational biases,
most notably the missing complete coverage over the first few hours.
Therefore, we assume an  upper limit of $\sim 30 \%$ of the GRB population displaying a jump in the light curve.

The all-sky quasi-diffuse neutrino flux for the ISM scenario is enhanced by a factor $\sim 3$ by assuming that  $30 \%$ of the GRB afterglows shows jumps. On the contrary, for the wind scenario,  the variation is basically null since the neutrino fluence is dominated by the early-time flux, i.e.~the neutrino emission expected from the standard afterglow. This is due to the fact that, as mentioned in 
Sec.~\ref{sec:neutrinos}, the flux quickly decreases for the wind profile. Thus, at the time of the shell collision, the flux is already small and does not contribute  to the quasi-diffuse emission substantially. Even though the presence of optical jumps slightly enhances the all-sky quasi-diffuse flux, the latter is still below the limit for the prompt phase of IceCube and is  consistent with the results of Ref.~\cite{IceCube:2021qyj}.

The neutrino diffuse emission associated with late optical jumps has been investigated in \cite{Guo:2019ljp} for optical flares occurring after 1 day from the onset of the prompt emission, thus at times larger than the ones considered in this work. Moreover, Ref.~\cite{Guo:2019ljp} carries out an approximated theoretical modeling of the jump and uses fixed values for the radius of the outflow and its Lorentz factor, while we embed the temporal evolution  of the blastwave and consistently model the shell merger.   In Ref.~\cite{Guo:2019ljp}, a distance of $R \simeq 10^{13}$~cm with Lorentz factor $\Gamma \simeq 10$ at $t = 1$~day is assumed. Through our approach and for the same luminosity, we obtain for the ejecta (that we assumed to be the slow shell) $R \simeq 10^{17}$~cm for $\Gamma \simeq 4$. In the light of these differences, we conclude that our results are not directly comparable to the ones in Ref.~\cite{Guo:2019ljp}. 
Furthermore, the estimation reported in Ref.~\cite{Guo:2019ljp} is  based on Ref.~\cite{Murase:2006dr}, where the expected neutrino signal from the X-ray flares is computed by assuming the late internal shock scenario of Ref.~\cite{Fan:2006ij}. This model assumes that  shock heated electrons in the BM shell are cooled through external inverse Compton scattering. On the other hand, in this work, we  only consider synchrotron emission. Despite the major differences in the modeling with respect to this work,   Ref.~\cite{Guo:2019ljp} also concludes that the optical jump leads to an  increase in  the expected  number of neutrinos.

\subsection{Point source searches}
\label{sec:singleGRB}

\begin{figure}[]
\centering
\includegraphics[scale=0.35]{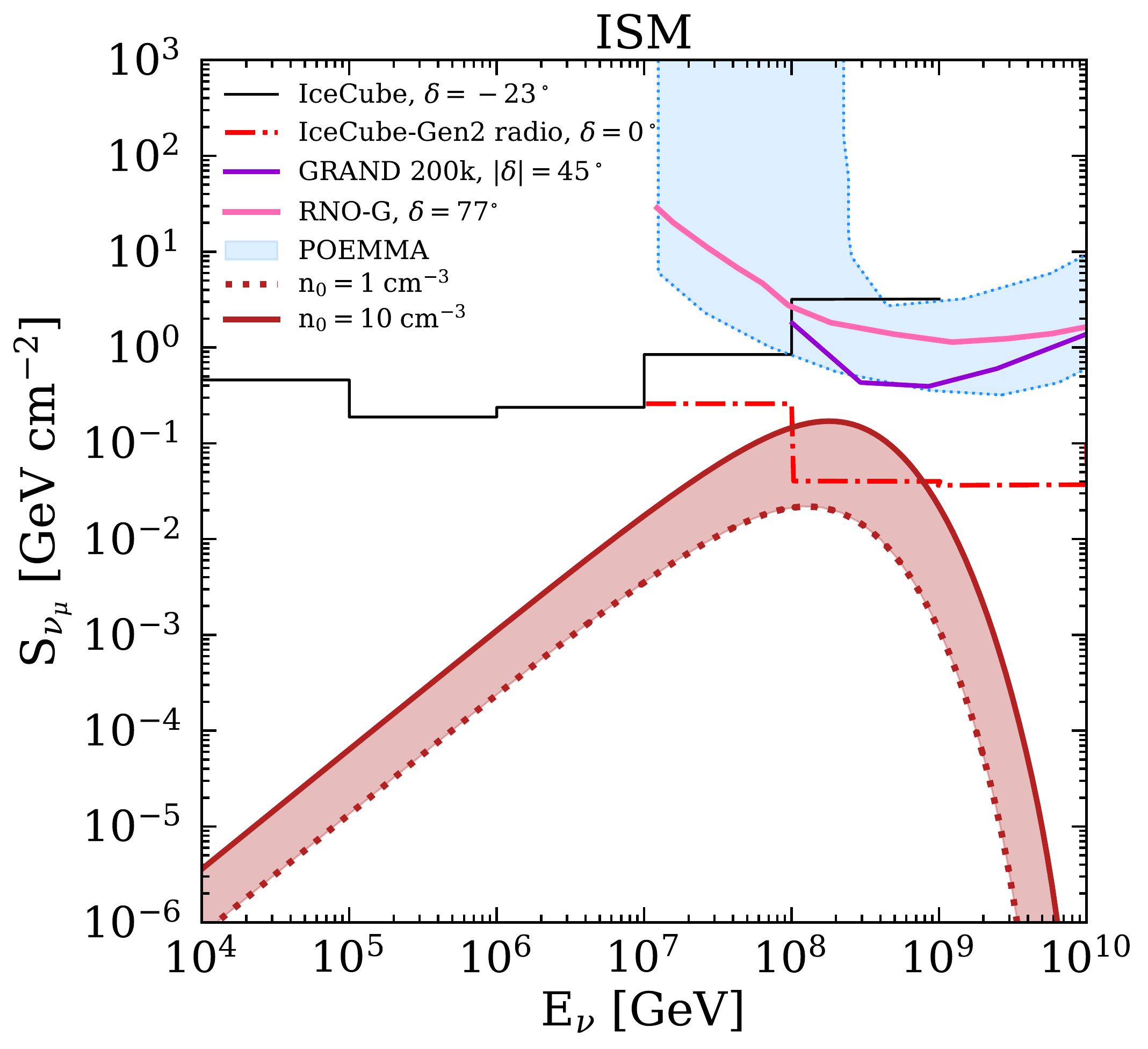} \label{ism_close}
\includegraphics[scale=0.35]{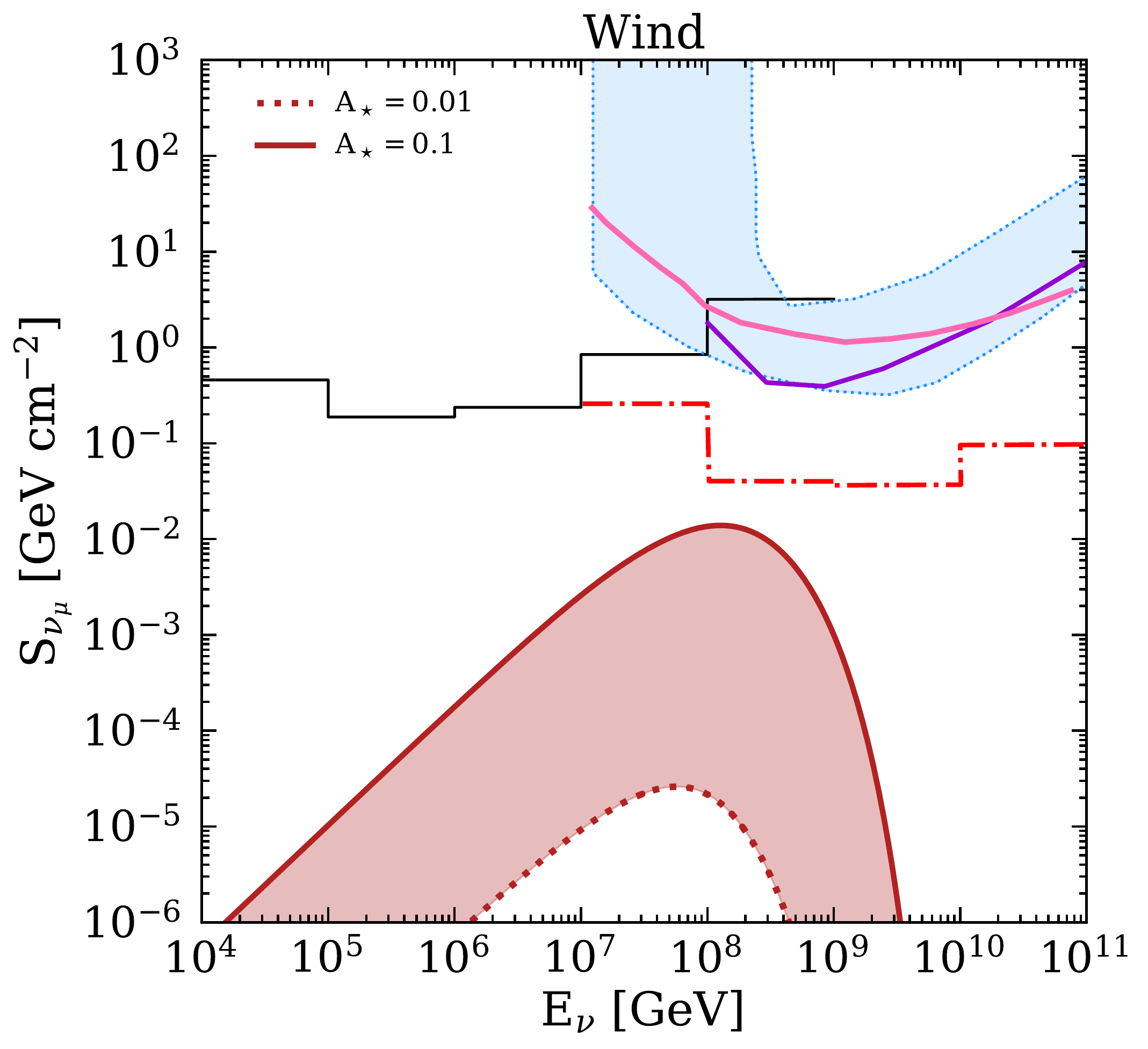}\label{ism_40}
\caption{Muon neutrino fluence for our benchmark GRB afterglow with an optical jump at $d_L=40$~Mpc (brown shadowed region) for the ISM (left panel) and wind (right panel) scenarios. 
The fluence bands correspond to $1 \lesssim n_0 \lesssim 10$~cm$^{-3}$ and $0.1 \lesssim A_\star \lesssim 0.01$ (dotted and solid  lines for the lower and upper bounds, respectively). 
The expected fluence is compared with the estimated sensitivities of IceCube-Gen2 radio for a source at $\delta=0^\circ$~\cite{IceCube-Gen2:2021rkf},  IceCube for  a source located at  $\delta=-23^\circ$~\cite{IceCube-Gen2:2021rkf, IceCube:2020xks}, RNO-G for a source at $\delta=77^\circ$~\cite{RNO-G:2020rmc},  GRAND200k for a source  at $\left| \delta \right| = 45^\circ$~\cite{GRAND:2018iaj}, and the full range time-averaged sensitivity of POEMMA~\cite{Venters:2019xwi}.
For the  ISM scenario, IceCube-Gen2 radio shows promising detection prospects.}
\label{fig:FLUENCE}
\end{figure}

Figure~\ref{fig:FLUENCE} shows the  fluence $S_{\nu_\mu}$ for our benchmark GRB (Table~\ref{tab:benchmark_grb}) with an optical jump assuming a distance of $40$~Mpc (brown-shadowed region) for the ISM  (on the left) and wind (on the right) scenarios. We also assume a band for $1 \lesssim n_0 \lesssim 10$~cm$^{-3}$ (ISM density) and $0.1 \lesssim A_\star \lesssim 0.01$ (wind).
We compare the expected muon neutrino fluence with the most optimistic  sensitivity of IceCube-Gen2 radio expected for the declination angle of the source in the sky  ($\delta=0^\circ$)~\cite{IceCube-Gen2:2021rkf}, the sensitivity of IceCube for a source located at  $\delta=-23^\circ$~\cite{IceCube-Gen2:2021rkf, IceCube:2020xks}, the sensitivity of RNO-G for a source at $\delta=77^\circ$~\cite{RNO-G:2020rmc}, the sensitivity of GRAND200k for  $\left| \delta \right| = 45^\circ$~\cite{GRAND:2018iaj}, and the full range time-averaged sensitivity of POEMMA~\cite{Venters:2019xwi}~\footnote{The declination angles for the detectors are not the same for all instruments since they have been chosen to guarantee the most optimistic conditions for detection.
In addition, GRAND200k and POEMMA are designed to be sensitive to showers initiated by tau neutrinos. Nevertheless, the following flavor composition $(\nu_e : \nu_\mu : \nu_{\tau}) \simeq (1 : 1 : 1)$~\cite{Farzan:2008eg} is expected at detection. Thus, the fluence of tau neutrinos expected at Earth is comparable to the one of muon neutrinos.} .  

Other radio neutrinos detectors have already been operating in the past years, such as the Askaryan Radio Array (ARA)~\cite{ARA:2015wxq, Allison:2011wk}, the Antarctic Ross Ice-Shelf ANtenna Neutrino Array (ARIANNA)~\cite{ARIANNA:2014fsk, Barwick:2014rca} and the Antartic Impulsive Transient Antenna (ANITA)~\cite{ANITA:2008mzi, ANITA:2010hzc}. Nevertheless, in the energy region where the afterglow fluence peaks these detectors have worse sensitivity compared to the ones displayed in Fig.~\ref{fig:FLUENCE}  and thus we did not consider them in our analysis. Note also that, at these energies, the neutrino background could also be populated by cosmogenic  neutrinos~\cite{Kotera:2010yn,Moller:2018isk,vanVliet:2019nse}, neutrinos from TDE~\cite{Guepin:2017abw}, newborn pulsars and millisecond magnetars~\cite{Fang:2017tla,Fang:2013vla}, in addition to GRB afterglow neutrinos~\cite{Razzaque:2013dsa,Murase:2007yt}. 

For a source at $d_L = 40$~Mpc, no detection of neutrinos is expected neither at IceCube---consistently with  current non-observations---nor at GRAND 200k and RNO-G for both CBM scenarios. On the contrary, a successful detection could  be possible with the radio extension of IceCube-Gen2 for the ISM scenario. In principle, in this case through the detection of neutrinos with IceCube-Gen2 radio, we could be able to constrain the CBM through neutrinos as well as probe the mechanism powering the optical jump. Indeed, the results presented in this paper are based on the assumption of a late collision between two shells, but other mechanisms may lead to different signatures in the neutrino signal. 
Furthermore, if no neutrino event is detected in temporal and spatial coincidence with the GRB event, constraints could be set on the parameters describing the jump in the afterglow light curve.

\subsection{Detection prospects for GRB 100621A and a GRB 130427A-like burst}

We now explore the neutrino detection prospects for  GRB 100621A, whose optical jump~\cite{Greiner:2013dma} has been detected in seven channels simultaneously with GROND~\cite{Greiner:2008ms}. We also investigate the detection prospects for a second GRB whose parameters are inspired by the bright and nearby GRB 130427A~\cite{Panaitescu:2013pga, Perley:2013kwa, DePasquale:2016vau}. An optical jump has not been observed for GRB 130427A, however we assume that it has one (hereafter GRB 130427A-like). The model parameters inferred for these two GRB afterglows and related uncertainties are summarized in Table ~\ref{TableGRB1}.
We fix $T_{\rm{coll}}=5 \times 10^3$~s for GRB 100621A, according to  observations. As for  GRB 130427A-like, we choose $T_{\rm{coll}} = 1 \times 10^4$~s for the ISM and wind scenarios, in order to have the light curves decreasing at $T_{\rm{coll}}$ in both  scenarios.

For  GRB 100621A, we fix $\epsilon_{e, m}^{0}$ and $\epsilon_{B, m}^{0}$ by matching the amplitude of the jump in the light curve. For GRB 130427A-like,  we fix $\epsilon_{e, m}^{0} = \epsilon_e$ and we choose $\epsilon_{B, m}^{0}$ in order to get the same rebrightening both for the ISM and wind scenarios. We note that there is a substantial freedom in the choice of $\epsilon_{e, m}^{0}$ and $\epsilon_{B, m}^{0}$.
\begin{table}[]
\centering
\caption{Parameters characteristic of  GRB 100621A~\cite{Greiner:2013dma} (second column) and GRB 130427A-like (inspired by GRB 130427A~\cite{Panaitescu:2013pga, Perley:2013kwa, DePasquale:2016vau}, third and fourth columns). For GRB 100621A, only the wind  scenario is considered, while both CBM scenarios are investigated for GRB 130427A-like, see main text for more details.}
\begin{tabular}{cccc}
\\
\toprule \toprule
& GRB 100621A (ISM) & GRB 130427A-like  (ISM) & GRB 130427A-like  (wind)\\ \midrule
$\tilde{E}_{k, \rm{iso}}$~(erg) & $2.8 \times 10^{53}$ & $3.8 \times 10^{54}$ & $4.2 \times 10^{53}$ \\ 
$z$ & ${0.54}$ & ${0.34}$ & ${0.34}$ \\ 
 $n_0 (\rm{cm}^{-3})$ or $A_\star$ & $1$--$100$ & $(2$--$7) \times 10^{-3}$ & $(1$--$5) \times 10^{-3}$ \\ 
 $\Gamma_0$ & $60$--$104$ & $850$ & $430$ \\ 
 $\epsilon_e$ & $(2$--$6) \times 10^{-2}$ & $0.3$ & $0.3$ \\ 
 $\epsilon_B$ & $6 \times 10^{-6}$--$6 \times 10^{-4}$ & $10^{-4}$ & $3 \times 10^{-2}$ \\ 
 $\epsilon_{e, m}^{0}$ & $0.1$ & $0.3$  & $0.3$  \\ 
 $\epsilon_{B, m}^{0}$ &  $10^{-4}$ -- $10^{-3}$ & $10^{-4}$ & $0.1$\\
\bottomrule \bottomrule 
\end{tabular}
\label{TableGRB1}
\end{table}

The wind  scenario has been excluded  for GRB 1000621A, thus we perform the calculations only for the  ISM case. For our GRB 130427A-like, instead, we explore the detection perspectives both for the ISM~\cite{Panaitescu:2013pga} and wind~\cite{Panaitescu:2013pga,Perley:2013kwa} scenarios. 

\begin{figure}[]
\includegraphics[scale=0.37]{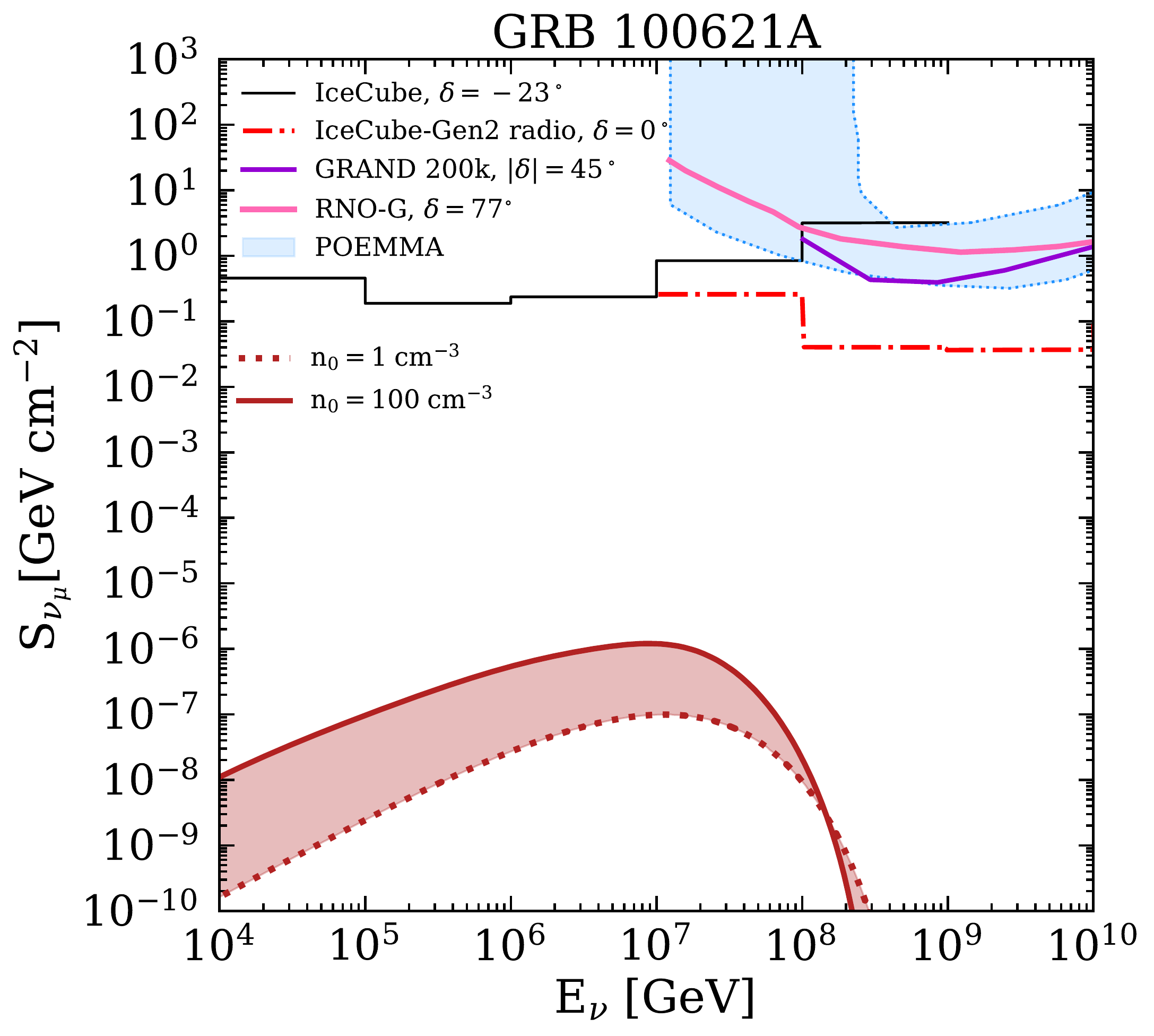}
\includegraphics[scale=0.37]{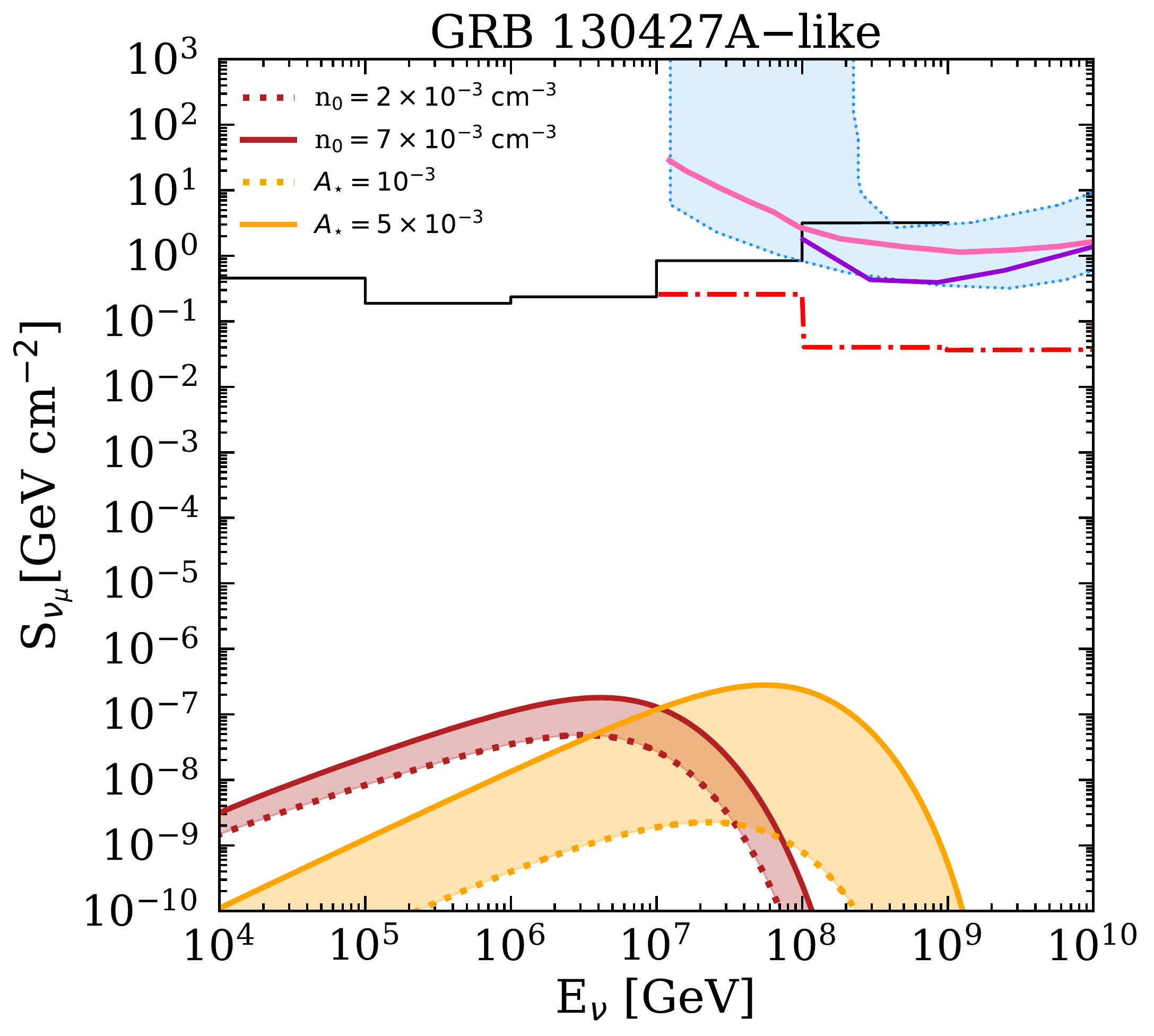} 
\caption{Neutrino fluence for GRB 100621A (left panel) and GRB 130427A-like  (right panel) for the parameters in Table~\ref{TableGRB1}. The brown (orange) bands represent the ISM (wind)  scenario. For GRB 100621A, the dotted (solid) line corresponds to $n_0=1$~cm$^{-3}$ ($n_0=100$~cm$^{-3}$). For GRB 130427A-like, the dotted  lines correspond to  $n_0= 2 \times 10^{-3}$~cm$^{-3}$ (ISM) and $A_\star = 10^{-3}$ (wind), while the solid lines refer to $n_0=7 \times  10^{-3}$~cm$^{-3}$ (ISM) and $A_\star = 5 \times 10^{-3}$ (wind). The expected fluence is compared with the estimated sensitivities of IceCube-Gen2 radio for a source  at $\delta=0^\circ$~\cite{IceCube-Gen2:2021rkf},   IceCube for a source at  $\delta=-23^\circ$~\cite{IceCube-Gen2:2021rkf, IceCube:2020xks},  RNO-G for a source at $\delta=77^\circ$~\cite{RNO-G:2020rmc},  GRAND200k for a source located at $\left| \delta \right| = 45^\circ$~\cite{GRAND:2018iaj}, and the full range time-averaged sensitivity of POEMMA~\cite{Venters:2019xwi}. For both  GRBs, the neutrino fluence lies below the point source sensitivities for all detectors.}
\label{fig:fluenceGRB1}
\end{figure}

The expected neutrino fluences are shown in Fig.~\ref{fig:fluenceGRB1}.
For both GRBs,  the detection of neutrinos seems unlikely. 
Thus, if GRBs with properties similar to GRB 100621A or  GRB 130427A-like  should be  observed, no associated neutrino signal should be expected, unless the burst propagates in an ISM with  $n_0$ larger than the one inferred for GRB 130427A~\cite{Panaitescu:2013pga} or the bursts occur at smaller distances.

\section{Conclusions} \label{sec:conclusions}
The light curve of some gamma-ray burst afterglows  exhibits a sudden intensity jump in the optical band between about one hour and one day after the prompt emission. The origin of this peculiar emission is not known yet, nor the fraction of GRBs displaying this feature. In this paper, we assume that the  optical jump results from the late collision of two relativistic shells, as proposed in Ref.~\cite{Vlasis:2011yp}. 

After modeling  the shell merger analytically, we compute the neutrino emission from the GRB afterglow within a multi-messenger framework by considering two scenarios for the circumburst medium: a constant density case (ISM) and a stellar wind profile. We find that the presence of an optical jump can increase the number of produced neutrinos by about an order of magnitude.

The expected quasi-diffuse flux of afterglow neutrinos falls below the upper limits placed by the non-detection of neutrinos during the GRB prompt phase. 
IceCube-Gen2 radio shows the most promising detection prospects for point source searches, potentially being able to  constrain the mechanism powering the optical jump as well as the properties of the circumburst medium  through neutrinos; for a source at  $d_L = 40$~Mpc,  a successful detection could  be possible with  IceCube-Gen2 radio for the ISM scenario. 

We also explore the neutrino emission  from GRB 100621A and a burst similar to GRB 130427A but with an optical jump, assuming both these GRBs as benchmark cases given their respective luminosity and redshift. However, because of their distance, the neutrino detection prospects from the afterglow of  GRBs similar to these ones could not be  probed with next generation neutrino telescopes. 

This work shows that the (non)-detection of neutrinos from GRB afterglows could offer an independent way to explore the mechanism powering the jump as well as the properties of the circumburst medium, if a GRB occurs relatively nearby or is especially bright.

\acknowledgments
We are thankful to Sargis Gasparyan for useful discussions. We thanks the anonymous referee for insightful comments that significantly improved this manuscript. This project has received funding from the  Villum Foundation (Project No.~37358), the Carlsberg Foundation (CF18-0183),  the Deutsche Forschungsgemeinschaft through Sonderforschungsbereich
SFB~1258 ``Neutrinos and Dark Matter in Astro- and Particle Physics'' (NDM), and the European Research Council via the ERC Consolidator Grant No.~773062 (acronym O.M.J.).

\appendix
\section{A model for the late collision and merger of two relativistic shells} \label{sec:merger_model}
\label{Theoretical modeling}
In this appendix, we  revisit the relativistic shock jump conditions. We then model the dynamical merger of two relativistic shells. In the following, we rely on the assumption of thin shells, for which the reverse shock is at most mildly relativistic. We further assume that the reverse shock has already crossed the ejecta, hence  we focus on the forward shock only.

The first ultrarelativistic, isotropic shell launched by the central engine starts to be decelerated by the ambient medium when it acquires a mass  comparable to $m_0/ \Gamma_0$, with  $m_0$ being the initial mass of the jet and $\Gamma_0$ its initial Lorentz 
factor~\cite{zhang_2018}. 
The number of particles, momentum and energy are conserved across the forward shock; this leads to the Rankine-Hugoniot jump conditions at the shock front, see e.g.~\cite{Taub:1948zz, Blandford:1976uq}, which in the fluid rest frame read as: 
\begin{eqnarray}
w^{\prime} &=& (\Gamma_{u} - 1) n^\prime \frac{h_u}{n_u} \ ,\\ 
n^{\prime} &=& \frac{2 \Gamma_{{\rm {sh}}, u}^2}{\Gamma_u} n_u  \ ,\\ 
\Gamma_{{u}}^2 &=& \frac{1}{2} \Gamma_{{\rm {sh}}, u}^2 \label{eq:gamma} \ ;
\end{eqnarray}
where $u$ refers to the upstream quantities, while the primed quantities are measured in the  downstream region. Here $h_u$ is the enthalpy density of the upstream medium, which  corresponds to the cold CBM, therefore $h_u \equiv \rho_u = n_u m_p c^2$. The quantities $w^\prime$, $\rho^\prime$, and  $n^\prime$ denote the comoving pressure, internal energy density, density and number of particles, respectively. $\Gamma_{{\rm{sh}}, u}$ is the Lorentz factor of the shock in the frame of the unshocked external medium and $\Gamma_{u}$ is the Lorentz factor of the shocked region measured in the same frame. Since the upstream medium is the unshocked CBM, assumed to be at rest in the stellar reference frame, the Lorentz factors in these frames satisfy the equivalence $\tilde{\Gamma} \equiv \Gamma_{u}$. Therefore, as for the Lorentz factors, hereafter we do not distinguish between the stellar and the unshocked CBM frames and simply denote them as $\Gamma$. The shock heats the matter, so that the region behind the shock is a hot plasma for which the equation of state $p^\prime = (\hat{\gamma} - 1) w^\prime$ holds, being $\hat{\gamma}$ the adiabatic index of the fluid and $p^\prime$ its comoving pressure . For a hot fluid $\hat{\gamma} = 4/3$, therefore the equation of state reads $p^\prime = w^\prime/3$. 
Using the shock jump conditions, we can rewrite the first equation in Eq.~\ref{eq:gamma}:
\begin{equation}
w^\prime = 4 \Gamma (\Gamma - 1) \rho_u = 4 \Gamma (\Gamma - 1) n_u m_p c^2 \ ,
\label{eq:en_density}
\end{equation}
which corresponds to Eq.~\ref{eq:densityBlastwave}.
Note that from Eq.~\ref{eq:gamma} one obtains that the plasma behind the shock  moves with a velocity $\Gamma = \Gamma_{\rm sh}/\sqrt{2}$. This region of hot plasma corresponds to a blastwave decelerated to the BM solution~\cite{Blandford:1976uq}, i.e. our slow shell.

The details of the  collision between the slow and the fast shells depend on the hydrodynamical modeling (see, e.g., \cite{Vlasis:2011yp}) and are beyond the scope of this paper. We refer the interested reader to Ref.~\cite{Kumar:1999gi} for a semi-analytic treatment of the shell collision including the reverse shock. Here, we rely on a simplified collision model, expanding on the one adopted to model the internal shock~\cite{Kobayashi:1997jk, Bustamante:2016wpu}. The main difference with respect to the internal shock prescription is that our  slow shell is hot and continuously sweeps  up material from the CBM. As a result, we need to include its internal energy at the collision time~\cite{2013MNRAS.433.2107N} as well as  the mass swept up by the slow shell from the CBM until the time of the collision. 

In the following, we focus on a merger whose duration is smaller than the dynamical time, considering that the jet has an opening angle $\theta_j$ small enough such that the merger process can be approximated as planar. Hence, the comoving volume of the shells can be expressed as $V^\prime \simeq \pi \theta_j^2 R^2 l^\prime$, where $l^\prime$ is the width of the  shell. This assumption is valid 
as long as the Lorentz factor $ \Gamma \gg 1$.

In order to obtain the total energy and momentum of the slow shell at a fixed time $t$, we introduce the energy-momentum tensor of a relativistic fluid in the laboratory frame \cite{2013MNRAS.433.2107N, zhang_2018}: 
\begin{equation}
\tilde{T}^{\mu \nu} = (\rho^\prime c^2 + w^\prime+ p^\prime) \tilde{u}^\mu \tilde{u}^\nu + p^\prime\eta^{\mu \nu} \ ,
\label{eq:tensor_energymom}
\end{equation}
where $\tilde{u}^\mu=\Gamma (1, \vec{\tilde{v}}/c)$ is the adimensional 4-velocity of the fluid in the laboratory frame, $\eta^{\mu \nu} = \mathrm{diag}(-1, 1, 1, 1)$ the Minkowski flat space-time and $p^\prime = w^\prime/ 3$, since we only consider the relativistic shock case. 
From the component with $\mu = \nu = 0$  in Eq.~\ref{eq:tensor_energymom}, we obtain the energy density  in the blastwave at the fixed time $t$:
\begin{equation}
\tilde{T}^{00}= \Gamma^2 ( \rho^\prime c^2 +  w^\prime + p^\prime) - p^\prime = \Gamma^2 \rho^\prime c^2 + (\hat{\gamma} \Gamma^2 - \hat{\gamma} + 1) w^\prime\ .
\label{eq:en_densitySlow}
\end{equation}
The total energy of the slow shell in the laboratory frame is computed by integrating Eq.~\ref{eq:en_densitySlow} over  $\tilde{V} = V^\prime/\Gamma$, where $V^\prime$ is defined as described previously. By denoting the total internal energy of the slow shell in the comoving frame as $W^\prime=w^\prime V^\prime$, its total energy is~\cite{2013MNRAS.433.2107N}: 
\begin{equation}
\tilde{E} = \Gamma c^2 m + \frac{\hat{\gamma}\Gamma^2 - \hat{\gamma} + 1}{\Gamma} W^\prime = \Gamma c^2 m + \Gamma_{\rm{eff}} W^\prime\ ,
\label{eq:en_slow}
\end{equation}
where $m$ is the mass of the slow shell given by Eq.~\ref{eq:mass_slow}.
Furthermore,  the effective Lorentz factor of the slow shell $\Gamma_{\rm{eff}}$ in Eq.~\ref{eq:en_slow} is
\begin{equation}
\Gamma_{\rm{eff}} =  \frac{\hat{\gamma} \Gamma^2 - \hat{\gamma} + 1}{\Gamma} \simeq \hat{\gamma} \Gamma = \frac{4}{3} \Gamma \ ,
\label{eq:gamma_eff_slow}
\end{equation}
where we have done the approximation $\Gamma \gg 1$ (valid in the time window we are looking at) and the relativistic $\hat{\gamma} = 4/3$ has been adopted. 

Similarly, taking  the  ($\mu=1, \nu=0$) component  in Eq.~\ref{eq:tensor_energymom},  the 4-momentum density of the slow shell at a fixed time $t$ is
\begin{equation}
\tilde{T}^{10} = \Gamma^2 \frac{\tilde{v}^1}{c}( \rho^\prime c^2 + w^\prime + p^\prime) \ ,
\label{eq:momentum}
\end{equation}
where $\eta^{10} = 0$. The $1$-st component of the total momentum of the slow shell is
    $\tilde{P}^1= {1}/{c} \int_{\tilde{V}} \tilde{T}^{10}  d \tilde{V}$,
from which:
\begin{equation}
\tilde{P} = c \Gamma {{\beta}} \left( m + \hat{\gamma} \frac{W^\prime}{c^2} \right) = c \sqrt{\Gamma^2 - 1} \left(m + \frac{\hat{\gamma}W^\prime}{c^2} \right)\ .
\label{eq:momentum_slow}
\end{equation}
Equations~\ref{eq:en_slow} and \ref{eq:momentum_slow} represent the energy and  momentum of the slow shell. 

If the second shell is emitted with energy $\tilde{E}_f$ and Lorentz factor $\Gamma_f = \rm{const.}$, its mass is
$m_f = {\tilde{E}_f}/{(\Gamma_f c^2)}$. The fast shell is cold, since it propagates freely, thus its energy and momentum are
\begin{eqnarray}
\tilde{E}_f = \Gamma_f m_f c^2 \label{eq:en_fast} \ \  \mathrm{and}\ \ 
\tilde{P}_f = c m_f \sqrt{\Gamma_f^2 - 1} \label{eq:momentum_fast} \ . 
\end{eqnarray}

In order to obtain the Lorentz factor and  energy of the resulting merged shell right after the collision, we impose energy and momentum conservation:
\begin{eqnarray}
\label{eq:en_merg}
\tilde{T}^{00}_f \tilde{V}_f + \tilde{T}^{00} \tilde{V} & = & \tilde{T}^{00}_m \tilde{V}^{0}_m \ ; \\
\tilde{T}^{i0}_f \tilde{V}_f + \tilde{T}^{10} \tilde{V} & = & \tilde{T}^{10}_m \tilde{V}^0_m\ , \label{eq:momentum_merg} 
\end{eqnarray}
being $\tilde{T}^{\mu \nu}_m$ the energy-momentum tensor of the merged shell and $\tilde{V}_m^0$ its volume, both evaluated at the collision time.
Hereafter, we  denote  all relevant quantities of the merged shell computed at the collision time with the apex ``0,'' in order to distinguish them from the ones describing its deceleration phase.
Plugging Eqs.~\ref{eq:en_slow}-\ref{eq:en_fast} in Eq.~\ref{eq:en_merg}, we obtain:
\begin{eqnarray}
 \Gamma_f m_f+  \Gamma m + \frac{\Gamma_{ \rm{eff}} W^\prime}{c^2} &=& \Gamma_m^0 m^0_m + \frac{\Gamma_{m, \rm{eff}}^0 W^{\prime 0}_m}{c^2} \ , \label{eq:first_conservation}\\ 
 \sqrt{\Gamma_f^2 - 1} m_f +  \sqrt{\Gamma^2 - 1} \left(m + \frac{\hat{\gamma} W^\prime}{c^2}\right) &=& \sqrt{{\Gamma_m^0}^2-1} \left(m_m^0 + \frac{\hat{\gamma} W^{\prime 0}_m}{c^2}\right)\ , \label{eq:conservation_second}
\end{eqnarray}
where $\Gamma_m^0$, $m_m^0 \equiv m + m_f$, $W^{\prime 0}_m$ are the initial Lorentz factor, the initial mass and the comoving internal energy of the merged shell, respectively. $\Gamma_{m, \rm{eff}}^0$ is the effective Lorentz factor of the merged shell and is defined as in Eq.~\ref{eq:gamma_eff_slow} by replacing $\Gamma \rightarrow \Gamma_m^0$. Note that all the physical quantities of the merged shell are evaluated at the collision time, thus they describe its initial setup.
Equations~\ref{eq:first_conservation} and \ref{eq:conservation_second} have a simple solution in the relativistic case, i.e.~for $\Gamma_f \gg 1$ and $\Gamma \gg 1$, which also implies $\Gamma_m^0 \gg 1$. Indeed, in this case $\Gamma_{\rm{eff}} \approx \hat{\gamma} \Gamma$ and $\Gamma_{m, \rm{eff}}^0 \approx \hat{\gamma} \Gamma_m^0$ so that we can rewrite Eqs.~\ref{eq:first_conservation} and \ref{eq:conservation_second} as follows:
\begin{eqnarray}
m_f \Gamma_f + \Gamma m_{\rm{eff}} &=& \Gamma_m^0 m_{m, \rm{eff}}^0  \label{eq:rewritten_first} \ ; \\ 
m_f \sqrt{\Gamma_f^2-1} + m_{\rm{eff}} \sqrt{\Gamma^2-1} & = & m_{m, \rm{eff}}^0 \sqrt{{\Gamma_m^0}^2-1}\ , \label{eq:rewritten_second}
\end{eqnarray}
where we have introduced the effective masses of the slow and merged shells:
$m_{\rm{eff}} = m + {\hat{\gamma}} W^\prime/{c^2}$ and 
$m_{m, \rm{eff}}^0 = m_m^0 + { \hat{\gamma}W^{\prime 0}_m}/{c^2}$.
After performing a Taylor expansion around $1/\Gamma_f$, $1/\Gamma$ and $1/\Gamma_m^0$ in Eq.~\ref{eq:rewritten_first}, we  obtain the initial Lorentz factor of the merged shell:
\begin{equation}
\Gamma_m^0 \approx \sqrt{\frac{m_f \Gamma_f + m_{\rm{eff}} \Gamma }{m_f/ \Gamma_f + m_{ \rm{eff}}/ \Gamma }} \ .
\label{eq:gamma_m}
\end{equation}
From energy conservation (Eq.~\ref{eq:rewritten_first}), we obtain the internal energy $\tilde{W}^0_m$ of the merged shell in the laboratory frame: 
\begin{equation}
\tilde{W}^0_m \equiv \Gamma_m^0 W^{\prime 0}_m = \frac{1}{\hat{\gamma}} \left[(m_f \Gamma_f  + m \Gamma ) c^2 - (m + m_f) \Gamma c^2 \right] + \Gamma W^\prime \ .
\label{eq:e_m}
\end{equation}
Equations~\ref{eq:gamma_m} and \ref{eq:e_m}  describe the initial conditions of the merged shell. 

We  assume that the shocks immediately cross the plasma. During the crossing, the resulting shell will be compressed, so that the correct expression of the initial width of the resulting merged shell is the one in Eq.~7 of Ref.~\cite{Kobayashi:1997jk}. In this paper, we make the simple assumption that its width is given by the sum of the widths of the slow shell $\tilde{l}$ and the fast shell $\tilde{l}_f$: 
\begin{equation}
\tilde{l}_m^0 \simeq \tilde{l} + \tilde{l}_f\ .
\label{eq:width_merg}
\end{equation}
This result differs from the one in Ref.~\cite{Kobayashi:1997jk}  for a small numerical correction factor.

\begin{figure}
\includegraphics[scale=0.35]{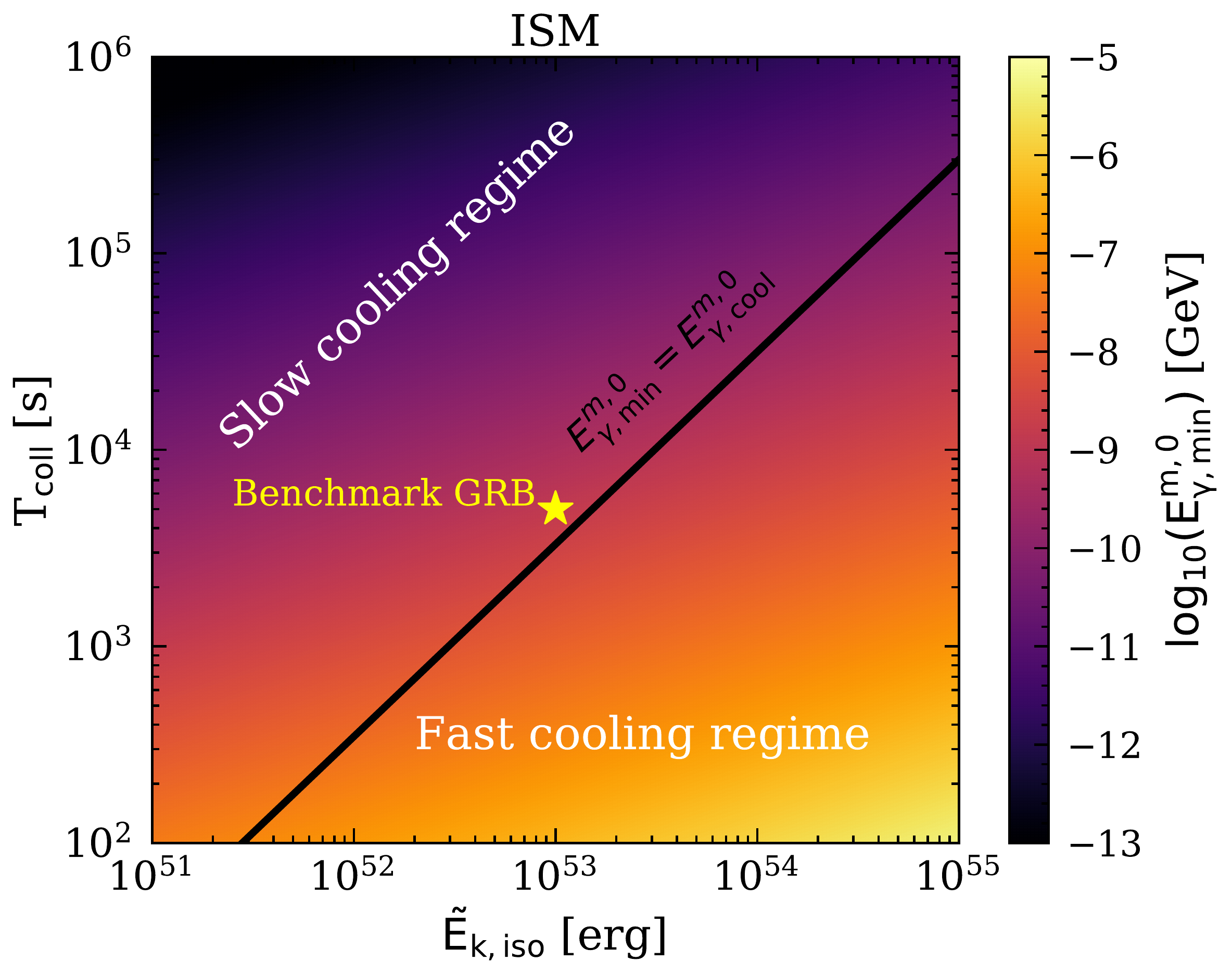}
\includegraphics[scale=0.35]{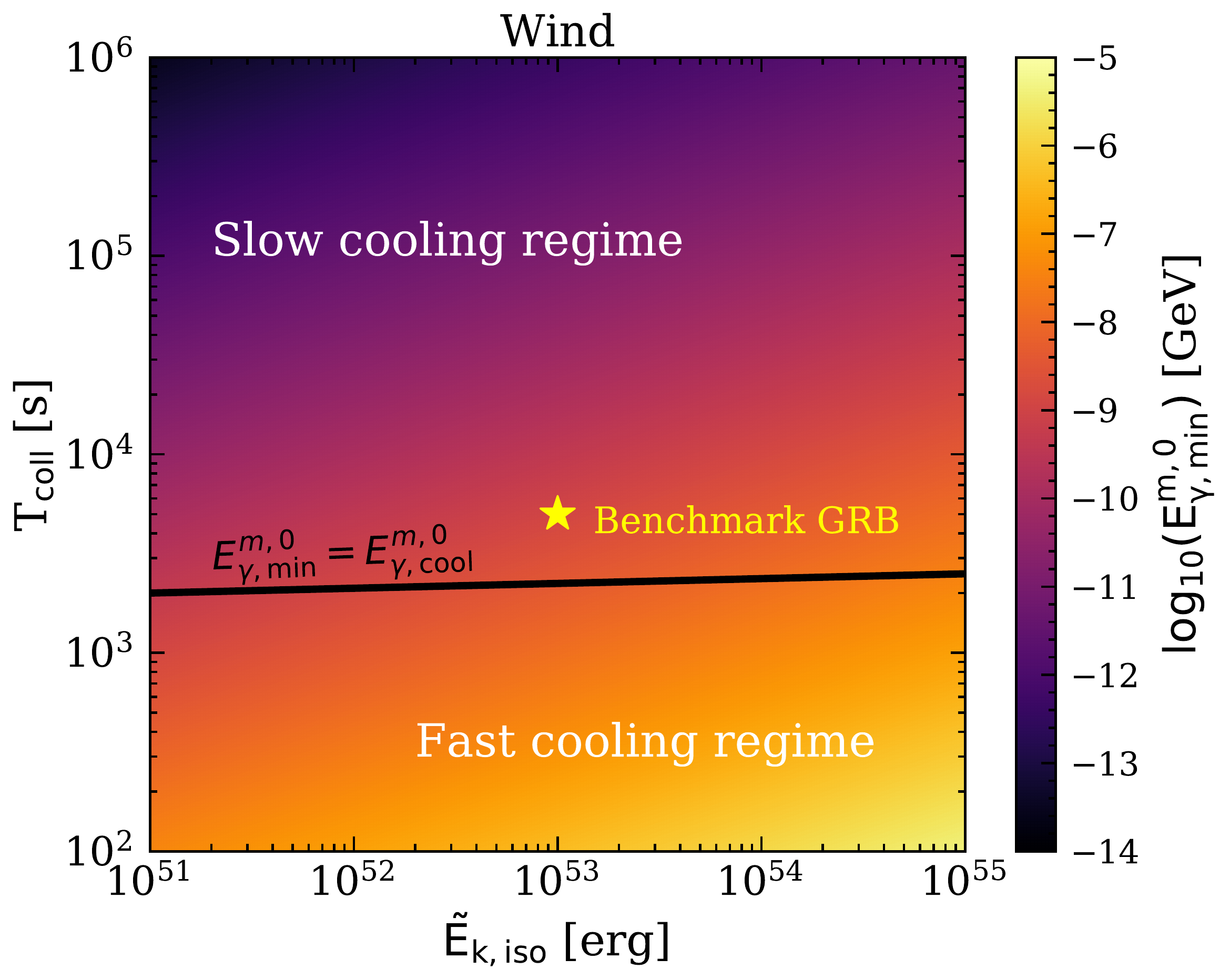}
\caption{Contour plot of the minimum energy  of the synchrotron photons emitted at the collision ($E_{\gamma, \rm{min}}^{m, 0}$) in the plane spanned by  $(\tilde{E}_{k, \rm{iso}}, T_{\rm{coll}})$, where $\tilde{E}_{k, \rm{iso}}$ is the isotropic kinetic energy of the slow shell and $T_{\rm{coll}}$ the collision time. The  ISM (wind) scenario is shown on the left (right).  The black solid line marks $E_{\gamma, \rm{min}}^{m, 0} = E_{\gamma, \rm{cool}}^{m, 0}$. For our set of parameters, electrons accelerated at the collision are in the slow cooling regime for the ISM and wind CBM scenarios.}
\label{fig:slowfast}
\end{figure}

After the merged shell forms, it interacts with the CBM.
Even though in our model the merged shell is expected to produce a standard afterglow flux through its interaction with the CBM, its dynamics is slightly different from the one of the  slow shell. This is because the merged shell is already hot and thus already has internal energy. Moreover, it also contains the matter material previously swept up by the slower shell. 
The total initial energy of the merged shell is: 
\begin{equation}
    \tilde{E}^0_{{\rm{tot}}, m} \simeq \frac{4}{3} \tilde{W}^0_m+ \Gamma_m^0 m_m^0 c^2\ .
    \label{eq:tot_merg}
\end{equation}
At the collision, a fraction $\epsilon_{e, m}^{0}$ of the internal energy $\tilde{W}^0_m$ goes into electrons and a fraction $\epsilon_{B, m}^{ 0}$ to the magnetic field. For our choice of parameters, electrons accelerated at the collision are in the slow cooling regime both in the ISM and wind scenarios, as shown in Fig.~\ref{fig:slowfast}. Hence only a small fraction of electrons efficiently radiates, and all the internal energy $\tilde{W}^0_m$ stays in the merged shell. Nevertheless, it is worth stressing that even if the fast cooling condition should be satisfied and all the electrons should cool through synchrotron radiation, the fraction of  energy carried away by photons is rather small ($ \simeq 10$--$30 \%$ of the internal energy, depending on the assumptions on the microphysical parameter $\epsilon_e$). Therefore, also in the fast cooling regime, most of the internal energy released at the collision stays in the merged shell as it is carried by protons which predominantly lose their energy via adiabatic cooling. Thus, the isotropic kinetic energy of the merged shell at the beginning of its deceleration is $\tilde{E}_{k, m} = \tilde{E}^0_{{\rm{tot}}, m}$.  

When the mass $m_{m, \rm{swept}}$ is swept up from the CBM  by the expanding blastwave, conservation of energy reads as: 
\begin{equation}
\Gamma_m^0 \biggl( \frac{\tilde{E}_{k, m}}{\Gamma_m^0 c^2} \biggr) + m_{m, \rm{swept}} = \Gamma_m \biggl[ \biggl( \frac{\tilde{E}_{k, m}}{\Gamma_m c^2} \biggr) + \hat{\gamma} \Gamma_m m_{m, \rm{swept}} \biggr] \ ,
\label{eq:conservation_energy}
\end{equation}
where $\Gamma_m$ is the Lorentz factor of the merged shell after the interaction with the medium and $m_{m, \rm{swept}}$ is the swept up mass dependent on the density profile of the external medium.
The shell starts to be decelerated when the two terms on  the right side of Eq.~\ref{eq:conservation_energy} become comparable~\cite{zhang_2018}: 
\begin{equation}
        m_{m, \rm{swept}}\simeq \frac{1}{\hat{\gamma}\Gamma_{m}} \biggl(   \frac{\tilde{E}_{k, m}}{\Gamma_m^0 c^2} \biggr) \simeq \frac{1}{\Gamma_m^0} \biggr(   \frac{\tilde{E}_{k, m}}{\Gamma_m^0 c^2} \biggr) \ ,
\label{eq:swept_merg}
\end{equation}
where we have considered that the Lorentz factor of the merged shell at the deceleration onset has been reduced to half of its initial value ($\Gamma_m \simeq \Gamma_0^m /2$) and we have neglected the numerical correction factor $2/3$. 

By integrating the density profile between $R_{\rm{coll}}\equiv R(T_{\rm{coll}})$ and $R_{{\rm{dec}}, m}$ and equating with Eq.~\ref{eq:swept_merg}, we finally obtain:
\begin{eqnarray}
\frac{4}{3} \pi n_0 m_p c^2 ({R_{ {\rm{dec}}, m}^{\rm{ISM}}}^3 - R_{\rm{coll}}^3) &\simeq&  \frac{\tilde{E}_{k, m} }{{\Gamma_m^0}^2}\  , \\
4 \pi A (R_{{\rm{dec}}, m}^{\rm{wind}} - R_{\rm{coll}}) m_p c^2 &\simeq& \frac{ \tilde{E}_{k,m}}{{\Gamma_m^0}^2}\ ,
\end{eqnarray}
for the ISM  and  wind  scenarios, respectively. Thus the deceleration radius for the merged shell is
\begin{eqnarray}
R_{{\rm{dec}}, m}^{\rm{ISM}} &\simeq& \biggl(\frac{3 \tilde{E}_{k, m} }{8 \pi n_0 m_p c^2 {\Gamma_m^0}^2} + R_{\rm{coll}}^3\biggr)^{1/3}\ ,\label{eq:dec_mergISM} \\ 
R_{{\rm{dec}}, m}^{\rm{wind}} &\simeq& \frac{\tilde{E}_{k, m}}{4 \pi A m_p c^2 {\Gamma_m^0}^2} + R_{\rm{coll}}\ . \label{eq:dec_mergWIND}
\end{eqnarray}
Finally, the deceleration time of the merged shell is 
\begin{equation}
T_{{\rm{dec}}, m}^{\rm{ISM, wind}} \simeq \frac{R_{{\rm{dec}}, m}^{\rm{ISM, wind}} (1+z)}{2 {\Gamma_m^0}^2 c}\ .
\label{eq:dec_merg}
\end{equation}
From $T_{{\rm{dec}}, m}^{\rm{ISM, wind}}$ on, the merged shell follows the standard BM solution. In particular, the temporal evolution of its Lorentz factor $\Gamma_m$ is described by Eq.~\ref{eq:ad_i}, by considering Eq.~\ref{eq:dec_merg} for the deceleration time and replacing $\Gamma_0 \rightarrow \Gamma_m^0$.

\section{Degeneracies among the parameters characteristic of the merging shells}
\label{A}

The two shells in our model collide when their position relative to the central engine coincides, i.e.~when $R(T_{\rm{coll}}) = R_f(T_{\rm{coll}})$ (see Eqs.~\ref{eq:radius_blastwave} and \ref{eq:radius_fast})~\cite{Laskar:2017qrq}:
\begin{equation}
    \frac{8 \Gamma^2 T_{\rm{coll}} c}{(1+z)}= \frac{2 \Gamma_f^2 (T_{\rm{coll}}- \Delta_T) c}{(1+z)} \ .
    \label{eq:collision_condition}
\end{equation} 

\begin{figure}[]
\includegraphics[scale=0.36]{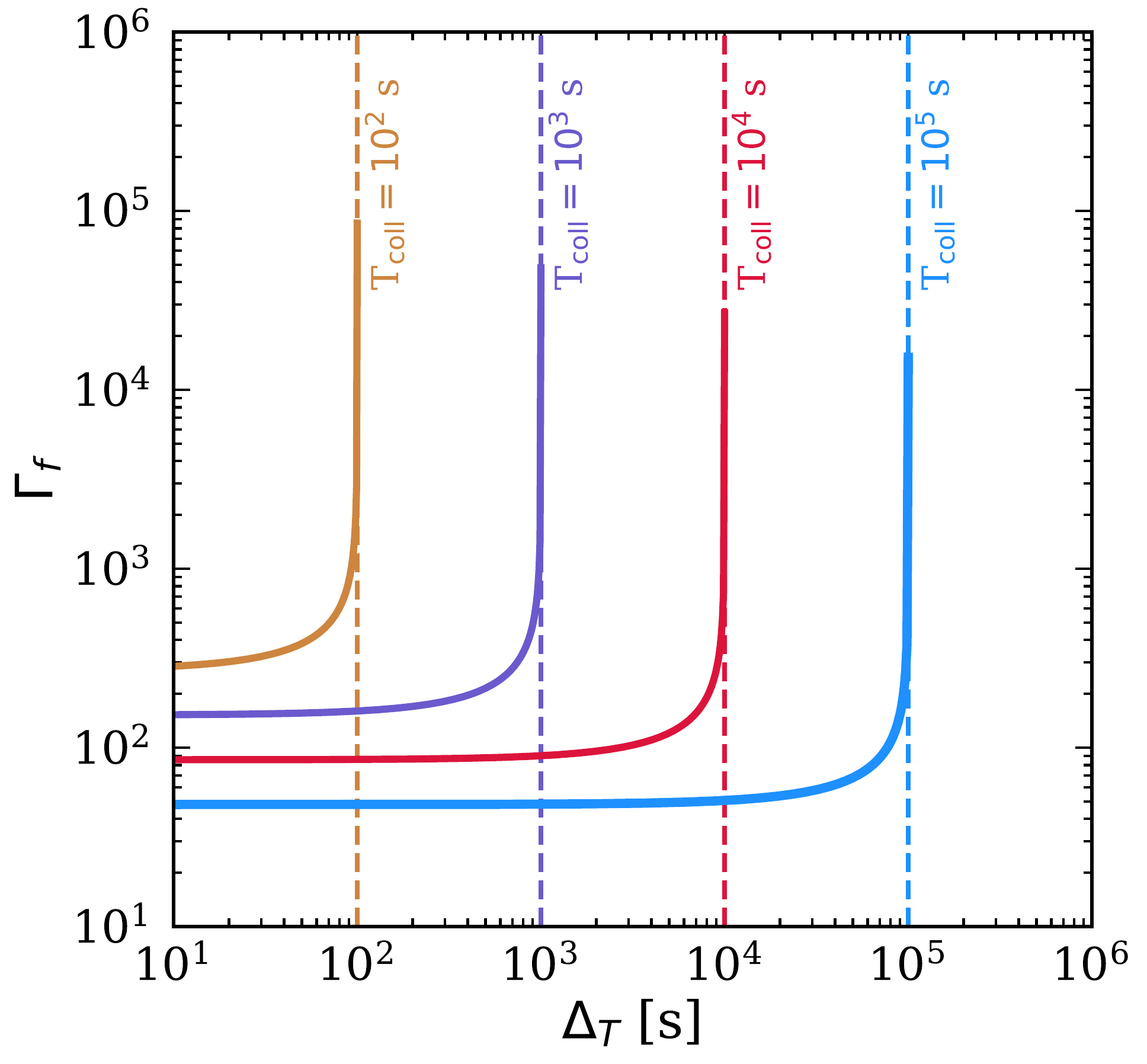}
\includegraphics[scale=0.355]{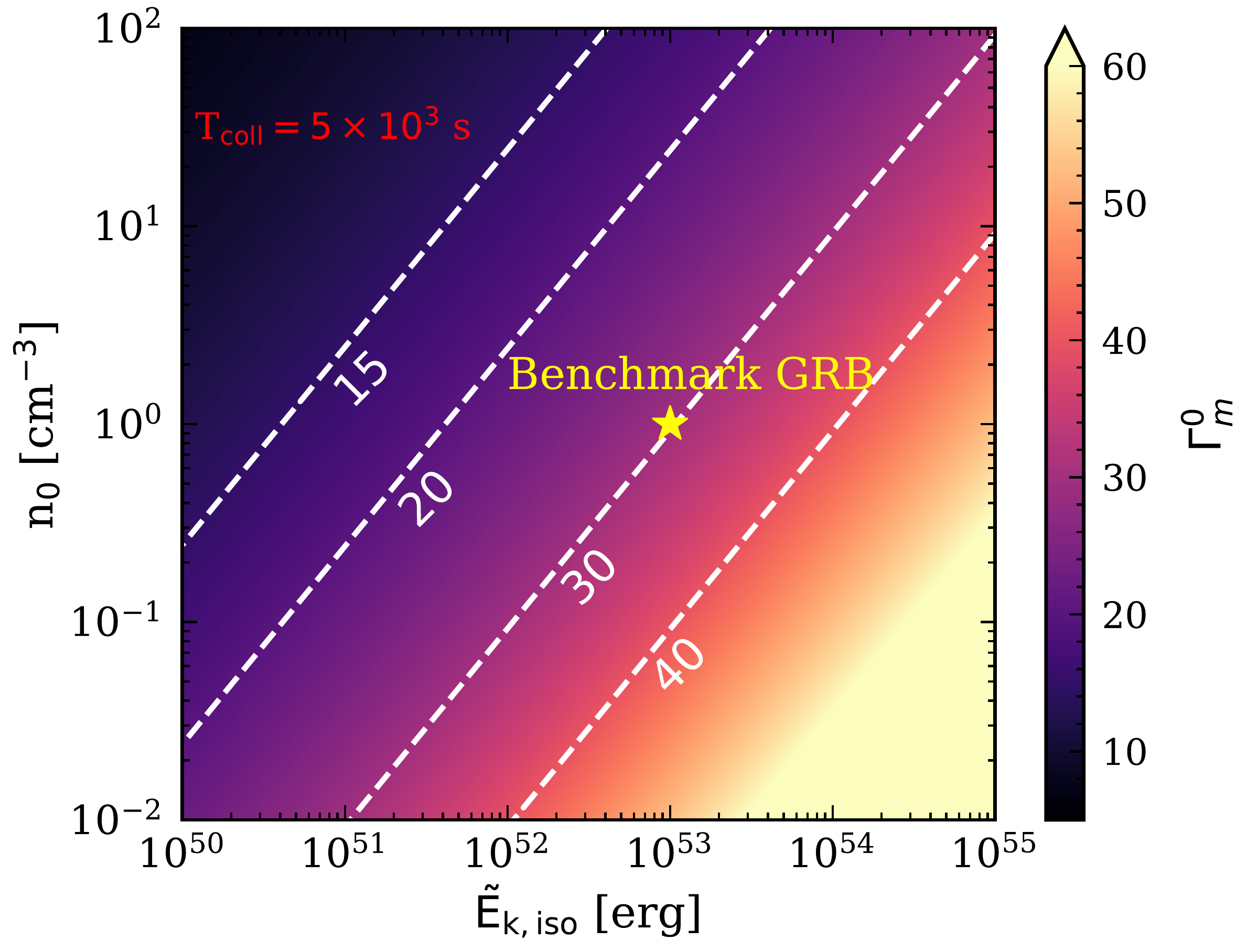}
\caption{{\it Left:} Isocontours of $T_{\rm{coll}}$ in the plane spanned by $\Delta_T$ and $\Gamma_f$ in the   ISM scenario.  The  function $\Gamma_f$ diverges when $\Delta_T \rightarrow T_{\rm{coll}}$.  {\it Right:} Contour plot of $\Gamma_m^0$ in the plane spanned by $\tilde{E}_{k, \rm{iso}}$ and $n_0$ (ISM scenario) for  $T_{\rm{coll}} = 5 \times 10^4$~s. The red dashed lines denote  $\Gamma_m^0 = 15$, $20$, $30$, and $40$. The yellow stars mark  our benchmark GRB (Table~\ref{tab:benchmark_grb}). Similar results also hold  for the wind case, both for the degeneracy between $\Gamma_f$ and $\Delta_T$ and for  $\Gamma_m^0$, by replacing $n_0 \rightarrow A_\star$.}
\label{fig:deg_gammaf}
\end{figure}

The collision of the two shells entails degeneracies among the parameters characteristic of the merging shells. One of these degeneracies occurs  between the Lorentz factor of the fast shell $\Gamma_f$ and the time delay $\Delta_T$ relative to the emission time of the first shell. Indeed, from Eq. \ref{eq:collision_condition}:
\begin{equation}
\Gamma_f = 2 \Gamma(T_{\rm{coll}}) \biggl(1 - \frac{\Delta_T}{T_{\rm{coll}}} \biggr)^{-1/2}\ ,
\end{equation}
i.e.~a shell launched with a large $\Delta_T$ can reach the first slow shell at the same collision time $T_{\rm{coll}}$ of a shell launched with a smaller delay and smaller $\Gamma_f$. This degeneracy can be better understood by looking at  the left panel of Fig.~\ref{fig:deg_gammaf} for our benchmark GRB (see Table~\ref{tab:benchmark_grb}).
A shortcoming  of our model is that it is not possible to distinguish between $\Gamma_f$ and $\Delta_T$, if no other information is available except for the amplitude of the optical jump. Perhaps, an analysis of the reverse shock may break this degeneracy, but it is out of the scope of this paper. Hence, in this work, we take $\Delta_T/ T_{\rm{coll}} \ll 1 $, meaning that the emission of the second shell would occur shortly after the explosion. 

Another  degeneracy in our model is in the definition of $\Gamma$ (see Eqs.~\ref{eq:ad_i}). The same value of $\Gamma$ can be obtained for different  $(\tilde{E}_{k, \rm{iso}}, n_0)$ pairs for the ISM scenario or  $(\tilde{E}_{k}, A_\star)$ for the wind scenario.  Once the collision time has been fixed, this results in the same value of  $\Gamma_m^0$, as displayed in the right panel of Fig.~\ref{fig:deg_gammaf} for the ISM  case. Similar results are obtained in the case of a wind environment, by replacing $n_0 \rightarrow A_\star$ (results not shown here). We do not
exclude any region of the parameter space in Fig.~\ref{fig:deg_gammaf}, since there are not observational constraints for the jump component. In principle, $\tilde{E}_{k,\rm{iso}}$ can be estimated from modeling the afterglow or by assuming that it is in the same order as $\tilde{E}_{\gamma,\rm{iso}}$, see e.g.~\cite{2011ApJ...732...29C}.

\begin{figure}[]
\centering
\includegraphics[scale=0.45]{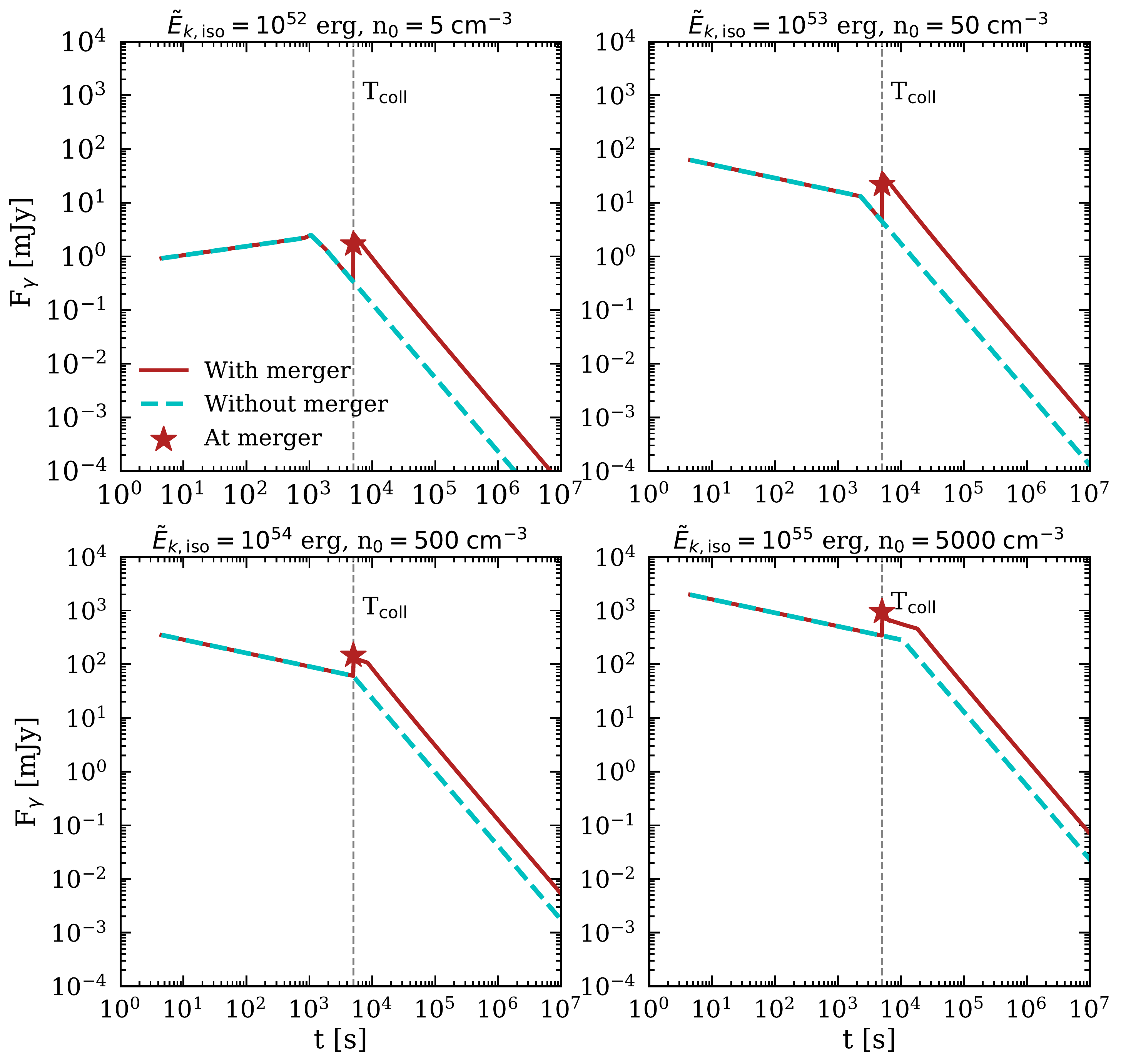}
\caption{Light curves, generated from different  $(\tilde{\rm{E}}_{k, \rm{iso}}, n_0)$ pairs in the ISM scenario, with the same  $\Gamma_m^0$ ($z = 1$ for all the panels). Each pair leads to  a different light curve, both in the absence (cyan dashed line) and in the presence (brown solid line) of the shell merger. The flux at the optical jump (marked by a brown star) is different for each $(\tilde{\rm{E}}_{k, \rm{iso}}, n_0)$ pair. Similar conclusions hold for the wind scenario.}
\label{fig:nodeg_spectrum}
\end{figure}

Even though the same $\Gamma$  and $\Gamma_m^0$ can be obtained at a fixed time for different values of the  energy and density of the external environment, the degeneracy is not observable in the resulting spectrum. Indeed, there are other parameters (e.g.~the break frequencies and magnetic field) that strictly depend on the density of the environment and thus allow to break this degeneracy---see Fig.~\ref{fig:nodeg_spectrum} for the ISM scenario (similar conclusions hold for the wind scenario, results not shown here).  

\section{Cooling timescales of protons and mesons}
\label{sec:cooling}
\begin{figure}[]
 \centering
   \includegraphics[scale=0.41]{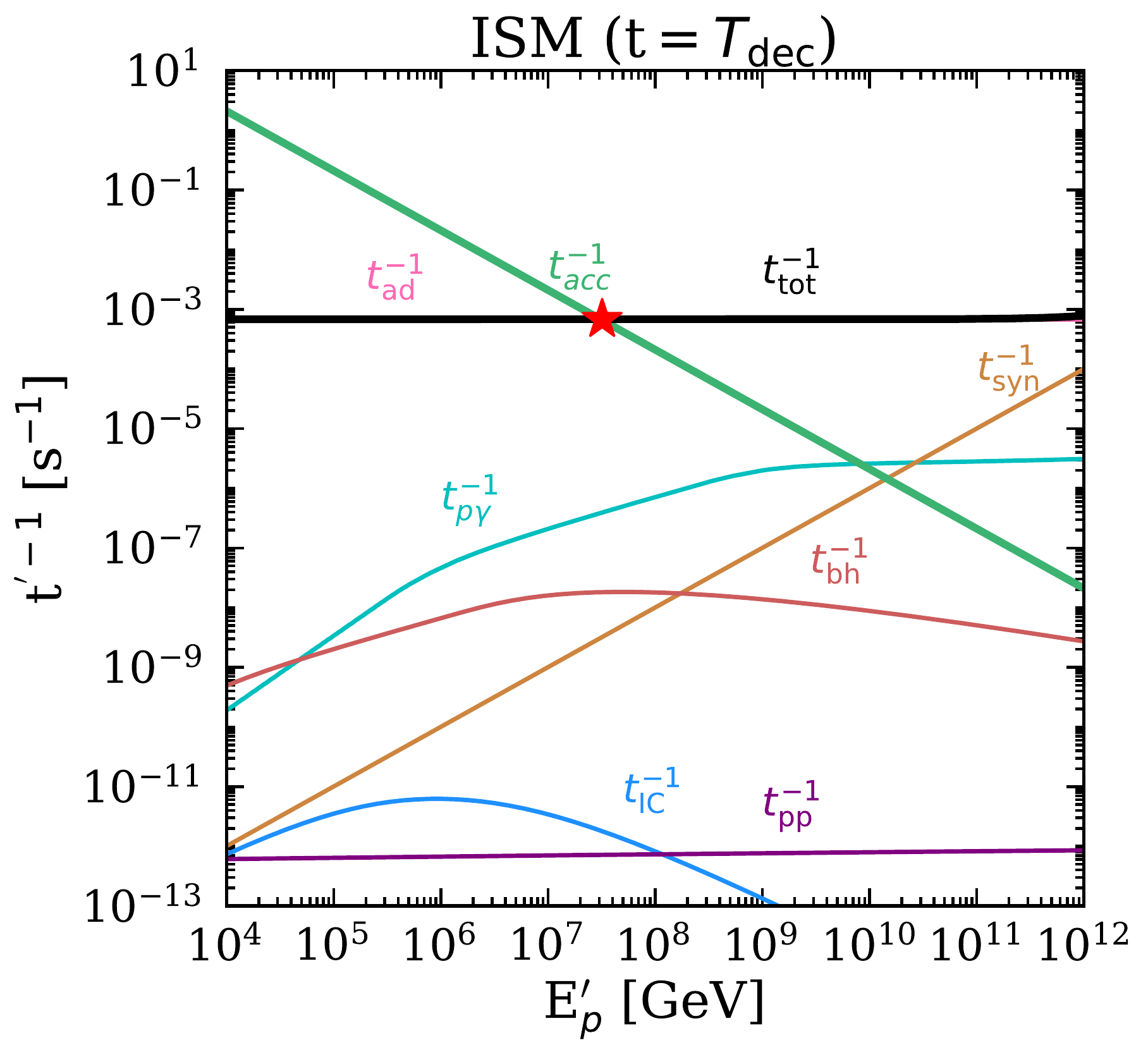}
   \includegraphics[scale=0.41]{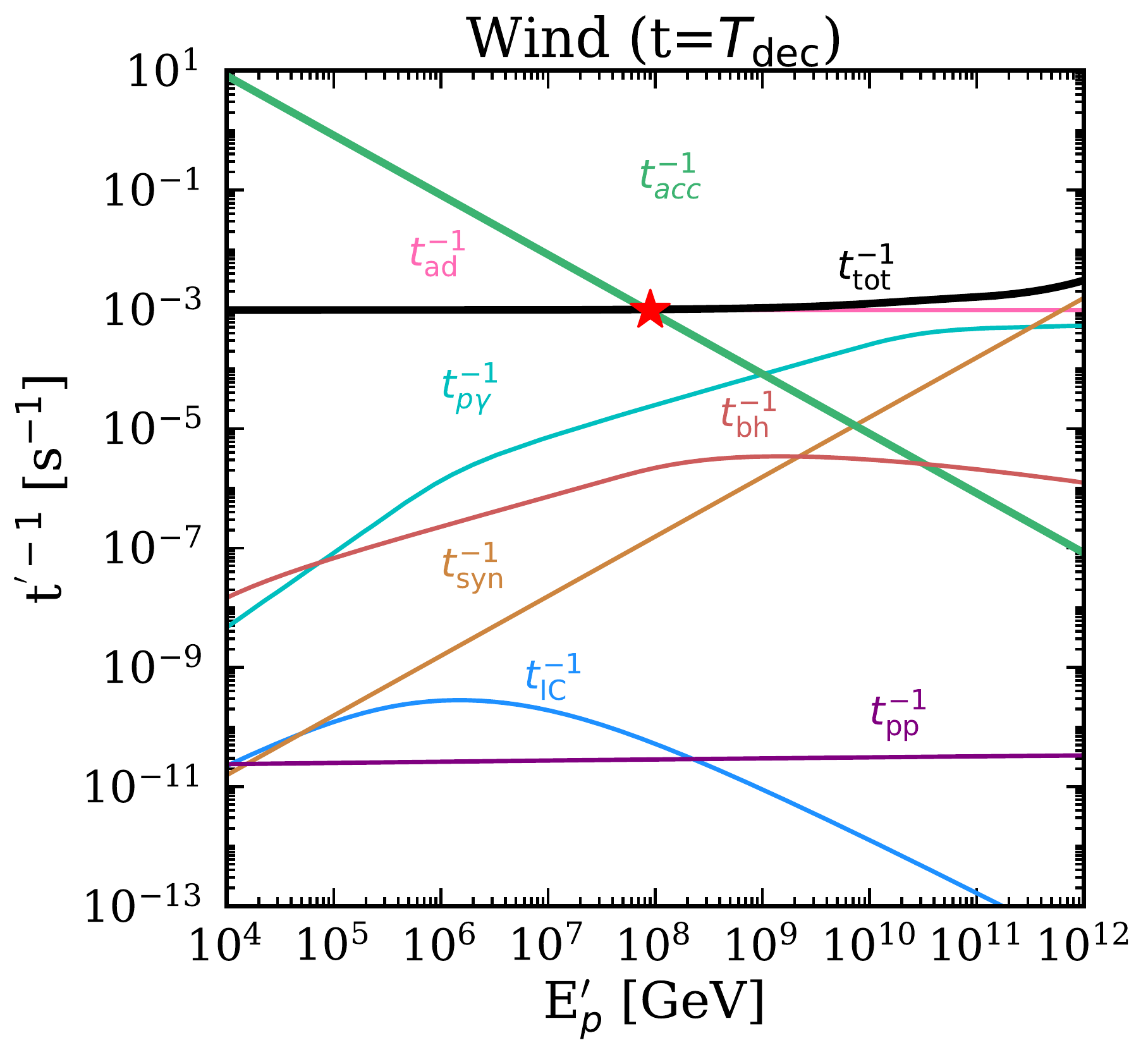}
   \caption{Inverse cooling timescales of protons  as functions of the comoving proton energy at the deceleration time $T_{\rm{dec}}$ for our benchmark GRB (Table~\ref{tab:benchmark_grb}) placed at $z = 1$ in the ISM  (left panel) and  wind  (right panel) scenarios . The red star marks the maximum energy up to which protons can be accelerated. The main cooling process for the  ISM scenario is the adiabatic one; for the  wind scenario, adiabatic cooling dominates at lower energies, while   synchrotron and the $p\gamma$ interactions become important at higher energies.}
    \label{fig:timescale_slowTdec}
\end{figure}

In order to compute the neutrino energy distributions, we need to take into account  the main cooling processes for accelerated protons, $\pi^{\pm}$, $\mu^{\pm}$, and $K^{\pm}$. The proton inverse cooling timescales for our benchmark GRB (see Table~\ref{tab:benchmark_grb}) are shown in Fig.~\ref{fig:timescale_slowTdec} at $t=T_{\rm{dec}}$ for the ISM and wind scenarios. Both in the ISM scenario (left panel) and in the wind scenario (right panel), the main cooling process for protons is the adiabatic one, that defines $E^\prime_{p, \max}$. The adiabatic timescale decreases with time, as a consequence  of the fact that $\Gamma$ of the shell decreases, while its radius increases. 

\begin{figure}[]
\centering
\includegraphics[scale=0.41]{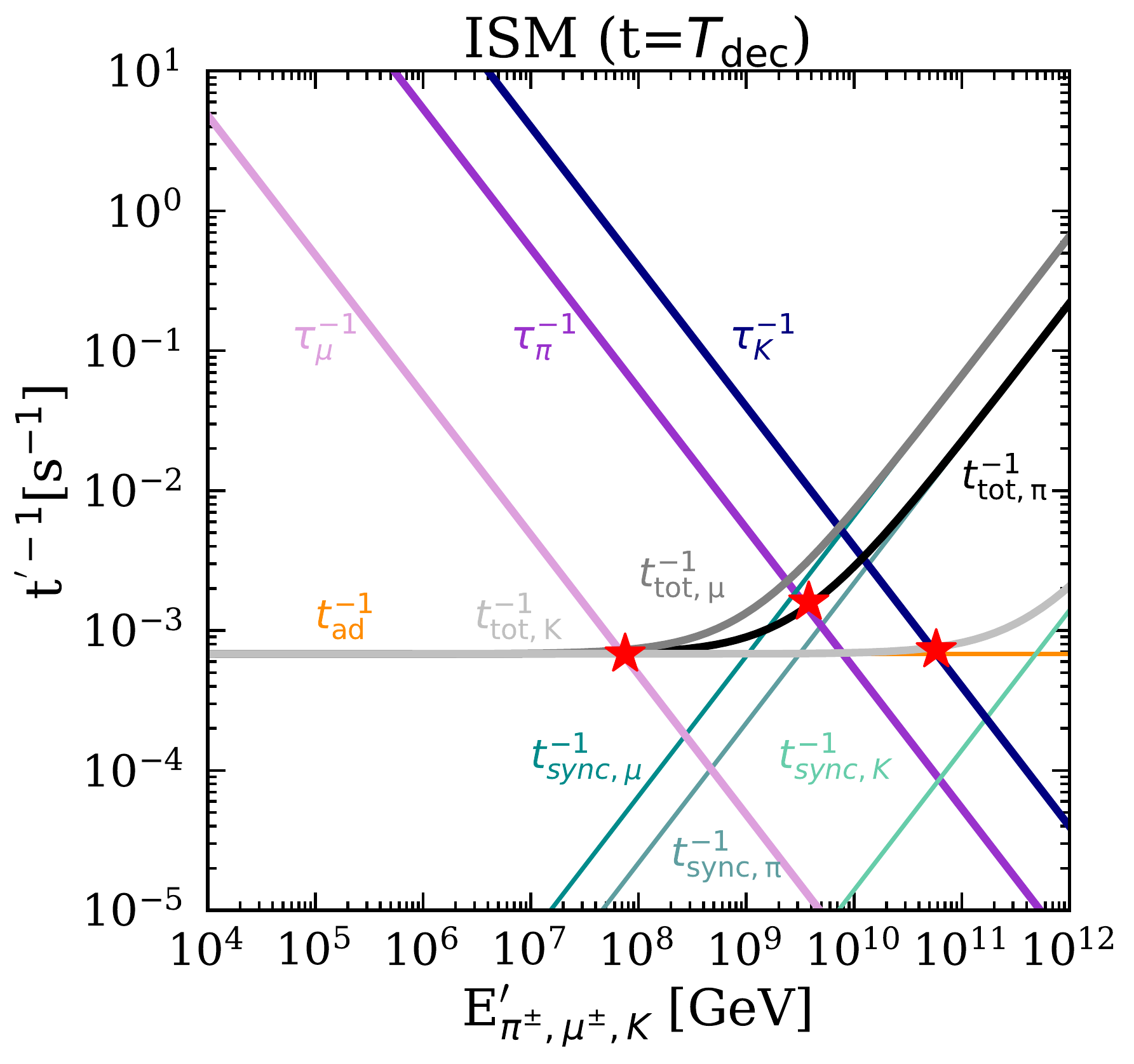}
  \includegraphics[scale=0.41]{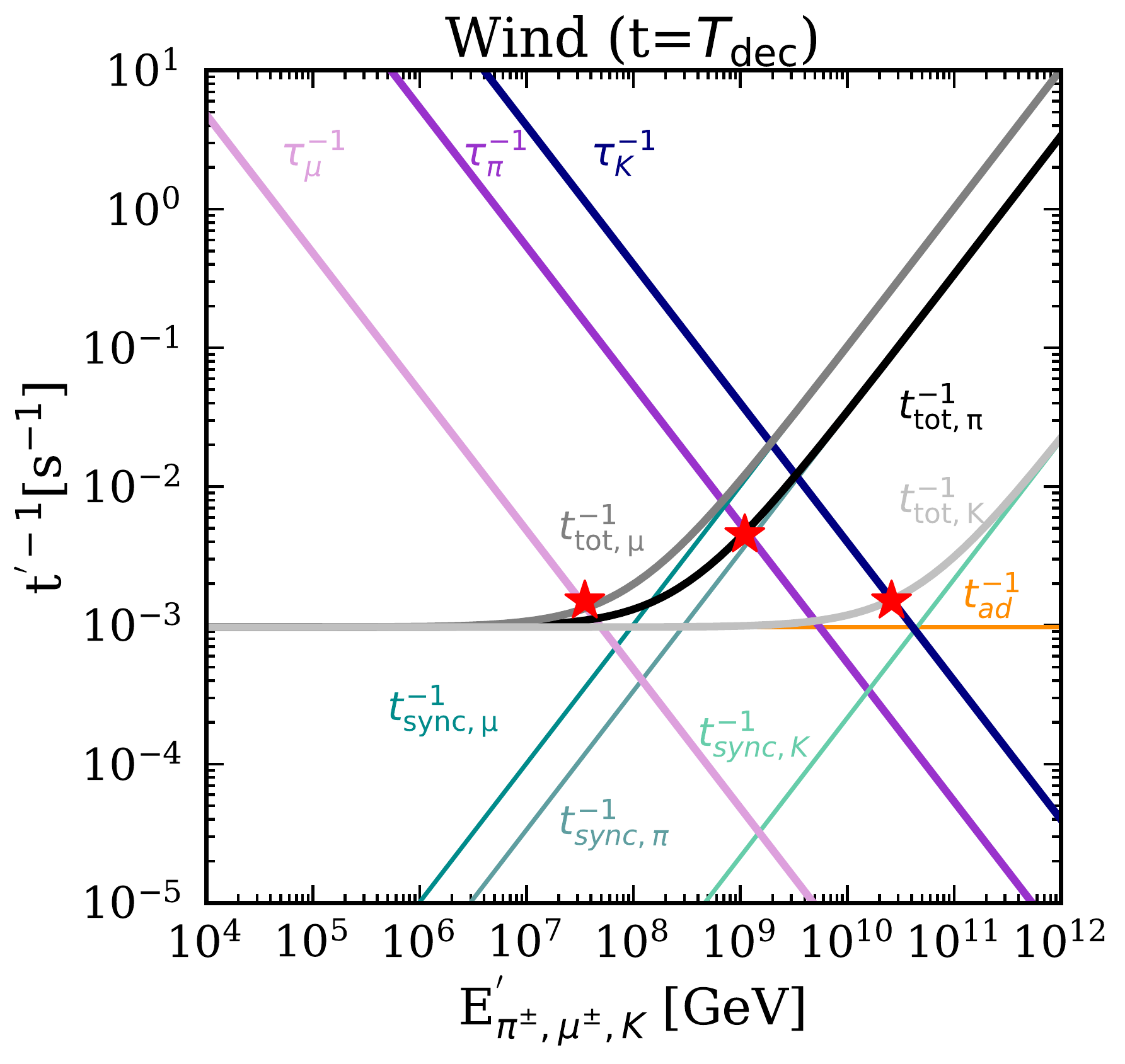}
   \caption{Same as Fig.~\ref{fig:timescale_slowTdec} but for  $\pi^{\pm}$, $\mu^{\pm}$, and $K^{\pm}$. For the  ISM scenario, adiabatic cooling is the most important process for kaons and muons, while synchrotron losses are important for pions. For the wind  scenario, both synchrotron and adiabatic cooling  are relevant for pions and muons. In both scenarios, the cooling of kaons occurs at energies larger than the maximum proton energy $E^\prime_{p, \rm{max}}$; thus, their cooling is negligible.}
    \label{fig:timescale_meson_slowTdec}
\end{figure}

Concerning the $\pi^{\pm}, \mu^{\pm}, K^{\pm}$, the cooling time scales for the slow shell at $t=T_{\rm{dec}}$ are shown  in Fig.\ref{fig:timescale_meson_slowTdec}. For the ISM scenario, adiabatic cooling can be important, yet  not relevant, for muons at the onset of the deceleration. Pions and kaons, instead, are expected to cool at energies larger than the maximum proton energy. Thus, their cooling does not affect the resulting neutrino energy distribution. 
For the wind case, the cooling timescales of mesons at $t=T_{\rm{dec}}$ are shown in Fig.~\ref{fig:timescale_meson_slowTdec}. In this scenario, muons cool at  energies lower than  the maximum energy of protons, affecting the  neutrino energy distribution. For our benchmark GRB, kaons always cool  at energies that are higher than the maximum proton energy. Thus, their contribution is negligible. 
In both  scenarios, the cooling of secondary particles becomes less relevant at larger times and it does not affect the shape of the resulting neutrino distribution.
\clearpage

\bibliographystyle{JHEP}
\bibliography{references}

\end{document}